\documentclass[12pt,english]{article}
\usepackage{lmodern}
\usepackage[T1]{fontenc}
\usepackage[latin9]{inputenc}
\usepackage{amsmath}
\usepackage{amssymb}
\usepackage{esint}

\usepackage{geometry}
\makeatletter
\usepackage{float} 
\usepackage{hyperref}
\usepackage{cite}
\usepackage{babel}
\usepackage{mathdots}
\usepackage{amsfonts}
\usepackage{mathrsfs}
\usepackage{amsmath}
\usepackage{esint}
\usepackage{graphicx}
\usepackage{youngtab}
\usepackage{multicol}
\usepackage{multirow}
\usepackage{slashed}
\usepackage{bm}
\usepackage{relsize}
\usepackage[dvipsnames,table,xcdraw]{xcolor}
\usepackage{subfigure}
\usepackage{feynmp}
\usepackage{cancel}
\allowdisplaybreaks

\makeatother

\usepackage[dvipsnames]{xcolor}

\begin{document}
\author{Wan-Zhe Feng\footnote{Email: vicf@tju.edu.cn},~~Zi-Hui
Zhang\footnote{Email: zhangzh\_@tju.edu.cn},~~Kai-Yu Zhang\footnote{Email: kaiyu\_zhang@tju.edu.cn}\\
 \textit{\small{}Center for Joint Quantum Studies and Department of
Physics,}{\small{}}\\
{\small{} }\textit{\small{}School of Science, Tianjin University,
Tianjin 300350, PR. China}}
\title{Sub-GeV millicharge dark matter from the $U(1)_X$ hidden sector}

\date{}

\maketitle

\begin{abstract}
We conduct a comprehensive study on the sub-GeV millicharge dark matter
produced through the freeze-in mechanism.
We discuss in general the mixing mechanism,
encompassing both kinetic mixing and mass mixing,
between the $U(1)_X$ hidden sector and the standard model,
which can generate millicharge carried by the dark fermions from the hidden sector.
We discuss in depth how such millicharge is generated,
and clarify several misunderstandings regarding this subject in the literature.
Without employing an effective field theory approach,
where the photon field directly mixed with the additional $U(1)$,
we analyze a general renormalizable model and
investigate the complete evolution of the hidden sector particles.
Due to the substantial self-interactions among hidden sector particles,
the evolution of the hidden sector temperature plays a crucial role,
which is addressed concurrently with the number densities of hidden sector particles
by solving a set of coupled Boltzmann equations.
We thoroughly examine eight benchmark models from six distinct cases.
Some of our key findings from the analysis of these benchmark models may be generalizable
and applicable to broader freeze-in scenarios.
We also explore the possibility that the $\mathcal{O}$(keV) $U(1)_X$ dark photon is a viable dark matter candidate,
even though it can contribute at most $\sim 5\%$  to the total observed dark matter relic density.
\end{abstract}
\newpage{}

\tableofcontents{}

\section{Introduction}\label{sec:Intro}

The Dirac quantization condition explains the observed quantization of the electric charge,
although the magnetic monopole is not yet discovered.
From an experimental standpoint, however,
the pursuit of discovering new phenomena and making new discoveries continues to be the foremost goal.
The search for new particle carrying tiny electric charge, normally referred to as the
millicharge,\footnote{From here and the following, we will use the term millicharge referring to ``milli-{\it electric} charge''.
The tiny amount of other quantum charges will be otherwise specified.}
is one of those interesting directions for the new physics exploration.
If millicharge particles exist,
the lightest millicharge particle must be stable and can serve as the dark matter candidate.
Experiments have searched for years and established stringent bounds for the millicharge dark matter particle,
for reviews see~\cite{Davidson:2000hf,Krnjaic:2022ozp}.
In this paper, we concentrate on millicharge dark matter in the sub-GeV mass range,
which has been extensively investigated by experimental searches~\cite{Mohapatra:1990vq,Davidson:1991si,Davidson:1993sj,Prinz:1998ua,
Jaeckel:2012yz,Diamond:2013oda,Vogel:2013raa,Adshead:2022ovo,Fung:2023euv,Dvorkin:2013cea,Tashiro:2014tsa,
Planck:2015fie,Vinyoles:2015khy,Gluscevic:2017ywp,Xu:2018efh,Chang:2018rso,
Liu:2018jdi,Kelly:2018brz,Liu:2019knx,Liang:2019zkb,Ball:2020dnx,Foroughi-Abari:2020qar,
Budker:2021quh,Du:2022hms,Berlin:2023gvx}.
From the theoretical perspective,
the exploration of how a consistent theory can generate such a millicharge,
represents an intriguing avenue of research.

The connection between kinetic mixing and the millicharge can be traced back to~\cite{Holdom:1985ag,Goldberg:1986nk}.
The kinetic mixing between a massless $U(1)_a$ and a massless $U(1)_b$ can generate a milli-$U(1)_a$ charge
for a $U(1)_b$-charged fermion, proportional to the kinetic mixing parameter.
In a massive amount of literature, the generation of millicharge particle
boils down to a simplified effective model that
an extra $U(1)_X$ mixed with the photon field through gauge kinetic terms,
i.e., via the effective operator
$\mathcal{L} \sim \frac{\boldsymbol{\delta}}{2}F_{\mu\nu}^{\rm em}F_{X}^{\mu\nu}$.
After the mixing, fermions charged under the extra $U(1)_X$ will carry a millicharge.
However, as we will review in Section~\ref{sec:mixing2U10M}, when one of the $U(1)$ fields gets massive, say $U(1)_b$,
the kinetic mixing with the massless $U(1)_a$ can {\it no longer} generate a milli-$U(1)_a$ charge for a $U(1)_b$-charged fermion,
no matter how tiny the massive $U(1)_b$ is.

In the full electroweak theory, the photon field is a linear combination of
the hypercharge $B$ and the neutral $SU(2)$ gauge fields $A_3$.
Thus the kinetic mixing between the extra $U(1)$ and the hypercharge gauge fields
should be treated simultaneously with the $\{ B, A_3\}$ electroweak mass mixing matrix~\cite{Foot:1991kb}.
It has been shown that pure kinetic mixing between an extra $U(1)$ and the hypercharge gauge field
{\it cannot} generate a millicharge~\cite{Feldman:2007wj}.
More precisely, for a general discussion of the Standard Model (SM) extended by a $U(1)_X$ hidden sector,
the millicharge carried by the $U(1)_X$ dark fermions can be obtained from:
(1) The dark particle carries a tiny amount of hypercharge as a prior.
(2) A kinetic mixing between a massless $U(1)$ and the hypercharge gauge field,
and the generated millicharge is proportional to the kinetic mixing parameter.
(3) The mass mixing between a massive $U(1)$ with the hypercharge gauge field~\cite{Cheung:2007ut,Feldman:2007wj}.
In the last case the kinetic mixing does not play any role in generating the millicharge,
and the millicharge generated is proportional to the mass mixing parameter.
In this paper we will review all these possibilities
and will focus on the third case, exploring details in a realistic model
involving viable freeze-in millicharge dark matter from a $U(1)_X$ hidden sector.
For a realistic model,
it is essential to ensure that in the final mass eigenbasis,
the photon is precisely massless.

The nature of dark matter remains to be a puzzle in modern physics.
In particle physics, dark matter is typically assumed to reside in the hidden sector,
comprising new hypothetical particles and/or new interactions.
Hidden sectors are common in various grand unified theory models and in string theory.
In general, the Universe may encompass a large amount of hidden sectors
that can have direct or indirect, strong or weak connections with the SM.
With direct and indirect detections of dark matter imposing increasingly stringent constraints
on freeze-out scenarios,
researchers have turned their attention to the freeze-in mechanism~\cite{Hall:2009bx,Chu:2011be,Bernal:2017kxu,Hambye:2019dwd,Aboubrahim:2019kpb,Aboubrahim:2020wah,Du:2020avz,Aboubrahim:2020lnr,Aboubrahim:2021ycj,Aboubrahim:2021dei}.
While the freeze-in mechanism demands a feeble coupling between the SM and the dark matter particle,
posing a puzzle that why such tiny coupling might emerge in the theory.
The kinetic mixing precisely provide such smallness for the coupling constant
between the SM and the hidden sector~\cite{Chu:2013jja,Blennow:2013jba,Dutra:2018gmv,Dvorkin:2019zdi,
Banerjee:2019asa,Koren:2019iuv,Chang:2019xva,An:2020tcg,Banerjee:2022fiw,Compagnin:2022elr,Gan:2023wnp}.

If a hidden sector is connected with the SM with only feeble couplings,
the hidden sector will evolve almost independently and will have its own temperature.
The temperature of the hidden sector may differ significantly from the temperature of the observed Universe,
and this can be analyzed using the formalism established in~\cite{Aboubrahim:2020lnr}.
Consequently, our Universe could contain numerous hidden sectors
with temperatures identical or different from the temperature of the Universe~\cite{Aboubrahim:2021ycj,Aboubrahim:2021dei}.

When the hidden sector possesses strong interactions among its particles,
calculating the evolution of hidden sector particles becomes challenging,
making it difficult to determine the final relic density of dark matter.
In the literature, numerous assumptions and approximations are employed,
and such calculations cannot be carried out convincingly.
The full evolution of a $U(1)_X$ hidden sector with feeble connection to the SM,
is first analyzed in~\cite{Aboubrahim:2020lnr},
where the number densities of dark particles as well as the hidden sector temperature can be accurately computed,
by solving a set of coupled Boltzmann equations.
In this paper,
we concentrate on the extension of the SM with an extra $U(1)_X$ hidden sector,
interacting feebly through tiny kinetic and mass mixings with the SM,
which may produce sub-GeV millicharge dark matter (some models also involve dark photon dark matter).
We follow the approach represented in~\cite{Aboubrahim:2020lnr},
calculating the complete evolution of the $U(1)_X$ hidden sector
and determining the final relic abundance of the dark matter candidate(s).
We discover several interesting novel results in the computations,
which could provide valuable insights for future research
on hidden sectors feebly interacting with the SM.

The paper is organized as follows.
In Section~\ref{sec:mixing}, we discuss in general the kinetic and mass mixing effect,
including the simplified model where the photon field mixed directly with an extra $U(1)$,
as well as the general case that the extra $U(1)_X$ mixed with the hypercharge gauge field,
for a complete treatment.
In particular, we discuss the kinetic mixing and mass mixing effects in the full electroweak theory in Section~\ref{sec:mixingF},
with a detailed derivation shown in Appendix~\ref{sec:APPKMMM}
for the case of Stueckelberg mass mixing.
In Section~\ref{sec:BEMT} we review the general formalism of computing the complete evolution of the hidden sector,
and write down the corresponding coupled Boltzmann equations.
A comprehensive investigation on the sub-GeV dark matter from a $U(1)_X$ hidden sector is discussed in Section~\ref{sec:EoM}.
To this end, we explore six distinct cases with different evolution details,
depending on the masses of the dark fermion and the dark photon.
We discussed various experimental constraints and carefully chose eight illustrative benchmark models to show
the evolution of hidden sector particles and the hidden sector temperature.
In some of the models, dark photon can also be a valid dark matter candidate,
although it can only occupy at most $\sim 5\%$ of the total observed dark matter relic density.
For the freeze-in production of millicharge dark particles,
the decay from effective massive photon from hot thermal bath (plasmon)
is one of the important production channels and is reviewed in Appendix~\ref{sec:APPPlasmon}.
All relevant cross-sections are summarized in Appendix~\ref{sec:APPScatter}.
Finally we conclude in Section~\ref{sec:Con}.

\section{The $U(1)$ extension of the SM and the generation of millicharge particle}\label{sec:mixing}

The SM can be extended by a $U(1)_{X}$ hidden sector
mixed with the SM via gauge kinetic terms and mass terms. In this
section, we discuss in general the effects of both the kinetic mixing
and the mass mixing. The simplest $U(1)_{X}$ extension consists of a
dark $U(1)_{X}$ gauge boson $C_{\mu}$ and the dark fermion $\chi$
carrying $U(1)_{X}$ charge $Q_{\chi}=+1$ but no SM quantum number.
In the final mass eigenbasis the dark $U(1)_{X}$ gauge boson is referred
to as the $Z^{\prime}$ gauge boson and is usually called a ``dark
photon'' if its mass is less than $\mathcal{O}$(GeV). The total
Lagrangian of the $U(1)_{X}$ extended SM is given by
\begin{equation}
\mathcal{L}=\mathcal{L}_{{\rm SM}}+\mathcal{L}_{{\rm hid}}+\mathcal{L}_{{\rm mix}}^{{\rm kin}}+\mathcal{L}_{{\rm mix}}^{{\rm mass}}\,,
\end{equation}
where $\mathcal{L}_{{\rm hid}}$ is given by
\begin{equation}
\mathcal{L}_{{\rm hid}}=-\frac{1}{4}F_{X\mu\nu}F_{X}^{\mu\nu}+g_{X}\bar{\chi}\gamma^{\mu}\chi C_{\mu}+m_{\chi}\bar{\chi}\chi\,,
\end{equation}
where $g_{X}$ is the $U(1)_{X}$ gauge coupling, $C_{\mu}$ is the
$U(1)_{X}$ gauge boson,
$F_{X\mu\nu}$ is the field strength of $U(1)_{X}$,
$\chi$ is the dark fermion with mass $m_{\chi}$.
$\mathcal{L}_{{\rm mix}}^{{\rm kin}}$ is the kinetic mixing term,
written as~\cite{Holdom:1985ag}
\begin{equation}
\mathcal{L}^{{\rm kin}}_{{\rm mix}}=-\frac{\delta}{2}F_{\mu\nu}F_{X}^{\mu\nu}\,,\label{eq:KinMix}
\end{equation}
where $F_{\mu\nu}$ is the hypercharge field strength,
$\delta$ is the kinetic mixing parameter. If the $U(1)_{X}$
gauge boson acquires a mass via Stueckelberg mechanism, $\mathcal{L}_{{\rm mix}}^{{\rm mass}}$
is written as~\cite{Feldman:2007wj}
\begin{equation}
\mathcal{L}_{{\rm mix}}^{{\rm st}}=-\frac{1}{2}(M_{1}C_{\mu}+M_{2}B_{\mu}+\partial_{\mu}\sigma)^{2}\,,\label{eq:StMMix}
\end{equation}
and we define the mass mixing parameter throughout the paper as $\epsilon=M_{2}/M_{1}$,
denoting the mass mixing between the $U(1)_{X}$ and the hypercharge
gauge field. In the Stueckelberg case the gauge field $C_{\mu}$ is
massive without breaking the $U(1)_{X}$ gauge symmetry.

If the $U(1)_{X}$ gauge boson achieves a mass via a dark Higgs $\phi$,
the covariant derivative reads
\begin{equation}
D_{\mu}\phi=(\partial_{\mu}-ig_{X}C_{\mu})\,\phi\,,\label{eq:darkHCD}
\end{equation}
and the hidden $U(1)_{X}$ is broken at some new scale $u$.
However, the mixing between the dark Higgs $\phi$ and the SM Higgs field $H$
via the term $(\phi^{*}\phi)(H^{\dagger}H)$ will not generate a mass
mixing in the gauge sector, i.e., the mixing of $U(1)_{X}$ with either
the hypercharge gauge field or the neutral $SU(2)$ gauge field is absent.
The mass mixing in the gauge sector can be generated
if the hidden $U(1)_{X}$ gauge boson $C_{\mu}$ also receives a tiny SM Higgs mass.\footnote{
Another possibility is that the SM gauge bosons also acquire tiny masses from the dark Higgs.
In this scenario, the mass and couplings of the dark Higgs will be severely constrained.}
One way to realize such scenario is to gauge a linear
combination of the hypercharge and any quantum number $X$,
i.e., $U(1)_{y+X}$ where $y$ is a very small fraction of the hypercharge $Y$~\cite{Zhang:2022nnh}.
Thus the dark particle, in this case, can carry a $U(1)_{y+X}$ charge $y$.
The covariant derivative acting on the SM Higgs doublet $H$
is then written as
\begin{equation}
D_{\mu}H=\Big(\partial_{\mu}-ig_{2}T^{a}A_{\mu}^{a}-\frac{i}{2}g_{Y}YB_{\mu}-\frac{i}{2}g_{Y}yC_{\mu}\Big)\,H\,,\label{eq:HCD}
\end{equation}
where the hidden $U(1)_{X}$ also carries a tiny hypercharge $y$.
After the symmetry breaking and the mixing,
the dark fermion originally carried a tiny hypercharge proportional to $y$ now
carries a millicharge.
However, this millicharge is not achieved by either the kinetic mixing or the mass mixing,
c.f., Section~\ref{sec:mixingEH} for details.
Though there seems to be no evident explanation for the inclusion of milli-hypercharge in the hidden $U(1)_{X}$ as a prior.

We outline our primary conclusion at the beginning that
the millicharge particle can be generated {\it by mixing} only in two ways:
(1) kinetic mixing between two massless $U(1)$ fields~\cite{Holdom:1985ag},
and (2) mass mixing between two massive
$U(1)$ fields (though in the mass eigenbasis there can still be a massless $U(1)$)~\cite{Cheung:2007ut,Feldman:2007wj}.
We emphasize here that except for the two massless $U(1)$ case,
the kinetic mixing does not play any role in generating the
millicharge carried by the dark particle. For the case of two massive
$U(1)$ fields, if kinetic mixing and mass mixing are both evoked, the
millicharge carried by the fermion will only depend on the mass mixing.

Since the photon is a linear combination of the hypercharge gauge field
and the neutral $SU(2)$ gauge field in the electroweak theory,
the dark $U(1)_{X}$ should mix with the hypercharge at high energies
rather than with the photon field.
For the analysis in the early universe,
it is important to explore how the effective kinetic mixing term of
photon and dark photon is generated~\cite{Gherghetta:2019coi,Rueter:2019wdf},
and further, how the millicharge carried by the dark particle is generated at low energies.
It is also intriguing to explore the stringy origin of the kinetic mixing~\cite{Abel:2008ai,Acharya:2018deu,Hebecker:2023qwl}
as well as the production of the millicharge~\cite{Shiu:2013wxa,Feng:2014eja,Feng:2014cla}.
After the discussion of this section,
one will observe that in the case of a {\it massive} $U(1)_X$ mixed with the hypercharge gauge field
in the full electroweak theory, {\it the kinetic mixing cannot generate a millicharge in any case.}
The millicharge can only be generated by
(1) the Stueckelberg mass mixing or (2) the dark particle originally carrying a tiny hypercharge.

In the following we would like to summarize some of the known results regarding
the generation of millicharge particles, and we wish to clarify several
important issues as well as some misunderstandings regarding the generation
of the millicharge with kinetic and/or mass mixing.
We will first review the mixing effect for two $U(1)$ fields and then
discuss the mixing effect of an extra $U(1)_X$ with the full electroweak theory.

\subsection{Kinetic mixing between two massless $U(1)$ fields}\label{sec:mixing2U1}

The simplest case of kinetic mixing arises from two massless $U(1)$ fields~\cite{Holdom:1985ag,Goldberg:1986nk}.
From an effective theory approach, at low energies one can consider
the case that an extra $U(1)_{X}$ mixed with the photon field.
In the gauge eigenbasis, the kinetic part and interaction part of the
Lagrangian read
\begin{align}
\mathcal{L}_{{\rm kin}} & =-\frac{1}{4}F_{\mu\nu}F^{\mu\nu}-\frac{1}{4}F_{X\,\mu\nu}F_{X}^{\mu\nu}-\frac{\boldsymbol{\delta}}{2}F_{\mu\nu}F_{X}^{\mu\nu}\,,\label{eq:2U1kinM}\\
\mathcal{L}_{{\rm int}} & =eA_{\mu}J_{{\rm em}}^{\mu}+g_{X}C_{\mu}J_{{\rm d}}^{\mu}\,,
\label{eq:2U1kinMint}
\end{align}
where $\boldsymbol{\delta}$ is the kinetic mixing parameter,\footnote{
In this paper we use the boldface $\boldsymbol{\delta}$ to denote
the kinetic mixing between the the extra $U(1)_X$ and $U(1)_{\rm em}$ in the simplified model,
and use the regular $\delta$ to denote kinetic mixing between the extra $U(1)_X$ and
the hypercharge gauge field in the full electroweak theory.}
$e$ is the electromagnetic coupling constant, i.e., the electric
charge and $g_{X}$ is the dark $U(1)_{X}$ gauge coupling constant,
$J_{{\rm em}}^{\mu}$ and $J_{{\rm d}}^{\mu}$ are the electromagnetic current and dark current respectively.
To keep the gauge kinetic terms in the canonical form, one performs
a non-unitary transformation
\begin{equation}
K=\left(\begin{array}{cc}
c_{\boldsymbol{\delta}} & 0\\
-s_{\boldsymbol{\delta}} & 1
\end{array}\right)\,,\label{eq:NUT1}
\end{equation}
where $c_{\boldsymbol{\delta}}=1/\sqrt{1-\boldsymbol{\delta}^{2}}$
and $s_{\boldsymbol{\delta}}=\boldsymbol{\delta}/\sqrt{1-\boldsymbol{\delta}^{2}}$.
After the diagonalization, the gauge kinetic terms are in the canonical
form with the mixing term eliminated, and the fields transform from
the gauge eigenbasis $\{A,C\}$ to the physical eigenbasis $\{A_{\gamma},A_{X}\}$
including the physical photon and the dark photon as
\begin{equation}
\left(\begin{array}{c}
A\\
C
\end{array}\right)=\left(\begin{array}{cc}
c_{\boldsymbol{\delta}} & 0\\
-s_{\boldsymbol{\delta}} & 1
\end{array}\right)\left(\begin{array}{c}
A_{\gamma}\\
A_{X}
\end{array}\right)\,.\label{eq:NUTE}
\end{equation}
In the physical eigenbasis, the interactions can be rewritten as
\begin{equation}
\mathcal{L}_{{\rm int}}=ec_{\boldsymbol{\delta}}A_{\gamma}J_{{\rm em}}+g_{X}(-s_{\boldsymbol{\delta}}A_{\gamma}+A_{X})J_{{\rm d}}\,.\label{eq:Int1}
\end{equation}
One can see now the dark current can couple to photon with strength $s_{\boldsymbol{\delta}}g_{X}$,
corresponding to a millicharge $s_{\boldsymbol{\delta}}g_{X}/e$
carried by $\chi$ (we assume $\chi$ carries the $U(1)_{X}$ charge $+1$).
However, the non-unitary transformation Eq.~(\ref{eq:NUT1}) is not symmetric with respect to the fields $A,C$.
Thus instead of the transformation made in Eq.~(\ref{eq:NUTE}), reversing the order of $A,C$ one has
\begin{equation}
\left(\begin{array}{c}
C\\
A
\end{array}\right)=\left(\begin{array}{cc}
c_{\boldsymbol{\delta}} & 0\\
-s_{\boldsymbol{\delta}} & 1
\end{array}\right)\left(\begin{array}{c}
A_{X}\\
A_{\gamma}
\end{array}\right)\,.\label{eq:NUTE2}
\end{equation}
The interactions in this case are then written as
\begin{equation}
\mathcal{L}_{{\rm int}}=g_{X}c_{\boldsymbol{\delta}}A_{X}J_{{\rm d}}+e(-s_{\boldsymbol{\delta}}A_{X}+A_{\gamma})J_{{\rm em}}\,,\label{eq:Int2}
\end{equation}
corresponding to the case where dark particles carry zero electric charge (no millicharge),
while SM fermions carry milli-$U(1)_{X}$ charges $s_{\boldsymbol{\delta}}eQ_{i}/g_{X}$
($Q_{i}$ denotes the electric charges carried by SM fermions in the unit of $e$).

Thus for the kinetic mixing between two massless $U(1)$ fields,
there can be two distinct outcomes:
(1) Dark particles carry millicharge while SM particles carry no $U(1)_X$ charge.
(2) Dark particles carry no millicharge while SM particles carry milli-$U(1)_X$ charge.
This characteristic arises due to the asymmetry in the non-unitary transformation.
These two possibilities are both hypothetical and will be confirmed or falsified through experimentation.
However, to generate a millicharge carried by dark particles, the transformation in Eq.~(\ref{eq:NUTE}) must be applied.

More generally, there is a rotation degree of freedom for the transformation
Eq. (\ref{eq:NUT1}). One can always multiply an orthogonal matrix
\begin{equation}
R=\left(\begin{array}{cc}
\cos\alpha & -\sin\alpha\\
\sin\alpha & \cos\alpha
\end{array}\right)\equiv\left(\begin{array}{cc}
c_{\alpha} & -s_{\alpha}\\
s_{\alpha} & c_{\alpha}
\end{array}\right)\label{eq:2MLR}
\end{equation}
to the non-unitary transformation $K$, which gives rise to the interaction
terms as
\begin{equation}
\mathcal{L}_{{\rm int}}=e(c_{\boldsymbol{\delta}}c_{\alpha}A_{\gamma}-c_{\boldsymbol{\delta}}s_{\alpha}A_{X})J_{{\rm em}}+g_{X}\big[(-c_{\alpha}s_{\boldsymbol{\delta}}+s_{\alpha})A_{\gamma}+(c_{\alpha}+s_{\boldsymbol{\delta}}s_{\alpha})A_{X}\big]J_{{\rm d}}\,.\label{eq:Int3}
\end{equation}
The case $\alpha=0$ corresponds to no rotation, which goes back
to Eq. (\ref{eq:Int1}), the case $\alpha=\pi/2$ corresponds to the
reverse case Eq. (\ref{eq:Int2}) with the redefinition of the physical
fields $A_{X}\leftrightarrow A_{\gamma}$.
In practice, this rotation
degree of freedom is redundant because the physical photon is measured
to be the gauge boson that couples to the EM current $J_{{\rm em}}$.
Namely one should redefine a real (measured) photon field $A_{\gamma}^{{\rm real}}$
and $A_{X}^{{\rm real}}$ as
\begin{align}
A_{\gamma}^{{\rm real}} & \equiv c_{\alpha}A_{\gamma}-s_{\alpha}A_{X}\,,\\
A_{X}^{{\rm real}} & \equiv s_{\alpha}A_{\gamma}+c_{\alpha}A_{X}\,.
\end{align}
Under such field redefinition, one finds the interactions in the physical
eigenbasis as
\begin{equation}
\mathcal{L}_{{\rm int}}=ec_{\boldsymbol{\delta}}A_{\gamma}^{{\rm real}}J_{{\rm em}}+g_{X}(-s_{\boldsymbol{\delta}}A_{\gamma}^{{\rm real}}+A_{X}^{{\rm real}})J_{{\rm d}}\,,
\end{equation}
which is exactly in the same form as Eq. (\ref{eq:Int1}).

In summary, the kinetic mixing of two massless $U(1)$ gauge fields
can generate the dark particle carrying a millicharge proportional
to the kinetic mixing parameter $\boldsymbol{\delta}$.

\subsection{Kinetic mixing between a massless $U(1)$ and a massive $U(1)$}\label{sec:mixing2U10M}

A more realistic case is kinetic mixing between the photon field with
a massive $U(1)_X$. In this case the Lagrangian reads
\begin{align}
\mathcal{L}_{{\rm eff}} & =-\frac{1}{4}F_{\mu\nu}F^{\mu\nu}-\frac{1}{4}F_{X\,\mu\nu}F_{X}^{\mu\nu}-\frac{\boldsymbol{\delta}}{2}F_{\mu\nu}F_{X}^{\mu\nu}-\frac{1}{2}M^{2}C^{2}\,,\\
\mathcal{L}_{{\rm int}} & =eA_{\mu}J_{{\rm em}}^{\mu}+g_{X}C_{\mu}J_{{\rm d}}^{\mu}\,.
\end{align}
To maintain the physical photon field exactly massless,
the only possible way of eliminating the kinetic mixing term is given by the following transformation
\begin{equation}
\left(\begin{array}{c}
C\\
A
\end{array}\right)=\left(\begin{array}{cc}
c_{\boldsymbol{\delta}} & 0\\
-s_{\boldsymbol{\delta}} & 1
\end{array}\right)\left(\begin{array}{c}
A_{X}\\
A_{\gamma}
\end{array}\right)\,.\label{eq:0MU1TR}
\end{equation}
The above transformation gives rise to the interaction in the physical eigenbasis as
\begin{equation}
\mathcal{L}_{{\rm int}}=g_{X}c_{\boldsymbol{\delta}}A_{X}J_{{\rm d}}+e(-s_{\boldsymbol{\delta}}A_{X}+A_{\gamma})J_{{\rm em}}\,.
\end{equation}
It is clear that dark particles carry precisely zero electric charge in this case.

As mentioned in the previous subsection,
the non-unitary transformation Eq.~(\ref{eq:NUT1}) is not symmetric with respect to the fields $A,C$.
For the kinetic mixing between two massless $U(1)$ fields,
there are two ways of taking the transformation Eqs.~(\ref{eq:NUTE}) and (\ref{eq:NUTE2}),
one leads to dark particles carry the millicharge and the other leads to no millicharge.
However, the emergence of the mass term for the massive $U(1)$ determines the transformation to be Eq.~(\ref{eq:0MU1TR}),
ensuring that the physical photon remains precisely massless.
The transformation Eq.~(\ref{eq:0MU1TR}) in the current case is
identical to the transformation in Eq.~(\ref{eq:NUTE2}) for two massless $U(1)$ fields,
which leads to no millicharge for dark particles.

Hence, it is {\it not} appropriate to consider a kinetic mixing between
the photon and a massive dark photon, and still claiming the dark
particles from the dark $U(1)$ sector can achieve a millicharge proportional
to the kinetic mixing parameter $\boldsymbol{\delta}$.
The generation of millicharge from the case of kinetic mixing between two massless $U(1)$ fields
is distinct from the situation where a massless $U(1)$ kinetically mixed with a massive $U(1)_X$,
even when the latter has an extremely small but nonzero mass $M$.
One cannot take the $M\to0$ limit in this case
and use directly the result from the case of kinetic mixing between two massless $U(1)$ fields in Eq.~(\ref{eq:Int1}),
where the millicharge carried by dark particles is generated,
since these two cases used different transformations shown in Eq.~(\ref{eq:0MU1TR}) and Eq.~(\ref{eq:NUTE}).
The $M\to0$ limit brings us to the transformation in Eq.~(\ref{eq:NUTE2}) for two massless $U(1)$ fields,
resulting in no millicharge being produced for the dark particles while SM particles carry milli-$U(1)_X$ charge.

In summary, for a dark photon
with a non-negligible mass (for example, a $\mathcal{O}$(keV--MeV) dark
photon as commonly studied), even though this mass is much smaller than the
experimental scale one considered,
the kinetic mixing between a massless
$U(1)$ and a massive $U(1)$ does {\it not} generate a millicharge carried
by the dark particle.

As long as one of the $U(1)$ fields is massive, the only possible way of
generating the millicharge through mixing is from the mass mixing effect,
and we will review this mechanism in the next subsection.

\subsection{Kinetic and mass mixing between two massive $U(1)$ fields}\label{sec:mixing2U12M}

The mass mixing effect of two massive $U(1)$ is discussed in~\cite{Foot:1991kb,Feldman:2007wj}.
The full Lagrangian reads
\begin{align}
\mathcal{L}_{{\rm kin}} & =-\frac{1}{4}F_{\mu\nu}F^{\mu\nu}-\frac{1}{4}F_{X\,\mu\nu}F_{X}^{\mu\nu}-\frac{\boldsymbol{\delta}}{2}F_{\mu\nu}F_{X}^{\mu\nu}\,,\\
\mathcal{L}_{{\rm mass}} & =-\frac{1}{2}M_{2}^{2}A^{2}-\frac{1}{2}M_{1}^{2}C^{2}-M_{1}M_{2}AC\,, \label{eq:2U1massmix}\\
\mathcal{L}_{{\rm int}} & =eA_{\mu}J_{{\rm em}}^{\mu}+g_{X}C_{\mu}J_{{\rm d}}^{\mu}\,.
\end{align}
To obtain the physical eigenbasis, one needs to diagonalize the kinetic
mixing matrix and mass mixing matrix simultaneously for the gauge
eigenbasis $V^{T}=(C,A)$,
\begin{equation}
\mathcal{K}=\left(\begin{array}{cc}
1 & \boldsymbol{\delta}\\
\boldsymbol{\delta} & 1
\end{array}\right)\,,\qquad M^{2}=\left(\begin{array}{cc}
M_{1}^{2} & M_{1}M_{2}\\
M_{1}M_{2} & M_{2}^{2}
\end{array}\right)=M_{1}^{2}\left(\begin{array}{cc}
1 & \epsilon\\
\epsilon & \epsilon^{2}
\end{array}\right)\,,
\end{equation}
where we define the mass mixing parameter $\epsilon=M_{2}/M_{1}$.

Firstly we know that the transformation $K$ of Eq. (\ref{eq:NUT1})
takes $\mathcal{K}$ to be diagonal,
\begin{equation}
K^{T}\mathcal{K}K=1\,,
\end{equation}
and sets a new basis as
\begin{equation}
KV^{\prime}=V=(C,A)^{T}\,,\qquad{\rm or}\qquad V^{\prime}=K^{-1}V\,.
\end{equation}
The gauge kinetic terms then become canonical with the new basis $V^{\prime}$.
At the same time the new basis $KV^\prime$ also changes the mass terms to be
\begin{equation}
V^{T}M^{2}V\to(KV^{\prime})^{T}M^{2}(KV^{\prime})=V^{\prime T}K^{T}M^{2}KV^{\prime}\,.
\end{equation}
Performing one further orthogonal transformation $O$ to diagonalize
the matrix $K^{T}M^{2}K$ such that
\begin{equation}
O^{T}K^{T}M^{2}KO=D^{2}\,,
\end{equation}
where the matrix $D^2$ is the final diagonalized mas-square matrix. The mass
terms reduce to
\begin{equation}
V^{\prime T}K^{T}M^{2}KV^{\prime}=V^{\prime T}O(O^{T}K^{T}M^{2}KO)O^{T}V^{\prime}=(V^{\prime T}O)D^{2}(O^{T}V^{\prime})\equiv E^{T}D^{2}E\,,
\end{equation}
where $E$ is the final mass eigenbasis. The kinetic terms will still
stay in the canonical form since $V^{\prime T}V^{\prime}=V^{\prime T}OO^{T}V^{\prime}=E^{T}E$.
Thus the transformation $R=KO$ diagonalizes the kinetic and mass
mixing matrices simultaneously, and it transforms the original gauge
eigenbasis $V^{T}=(C,A)$ to the mass eigenbasis $E^{T}=(A_{X},A_{\gamma})$,
which is
\begin{equation}
V=(KO)E \equiv RE\,.
\end{equation}
One should keep in mind that the physical photon must be exactly massless,
i.e., the mass term is zero for $A_{\gamma}$. One finds
\begin{equation}
\tan\theta=\frac{\epsilon\sqrt{1-\boldsymbol{\delta}^{2}}}{1-\epsilon\boldsymbol{\delta}}\,,
\end{equation}
which leads to the following interaction terms
\begin{align}
\mathcal{L}_{{\rm int}} & =(g_{X}J_{{\rm d}}^{\mu},eJ_{{\rm em}}^{\mu})V=(g_{X}J_{{\rm d}}^{\mu},eJ_{{\rm em}}^{\mu})RE\nonumber \\
 & =\frac{1}{\sqrt{1-2\epsilon\boldsymbol{\delta}+\epsilon^{2}}}\frac{1}{\sqrt{1-\boldsymbol{\delta}^{2}}}A_{X}\big[(1-\epsilon\boldsymbol{\delta})g_{X}J_{{\rm d}}+eJ_{{\rm em}}(\epsilon-\boldsymbol{\delta})\big]\nonumber \\
 & +\frac{1}{\sqrt{1-2\epsilon\boldsymbol{\delta}+\epsilon^{2}}}A_{\gamma}(eJ_{{\rm em}}-\epsilon g_{X}J_{{\rm d}})\,.
\end{align}
In this case, one can see clearly that the millicharge can be induced
only when the mass mixing, denoted by $\epsilon$, is non-vanishing,
and is {\it not} related to the kinetic mixing parameter $\boldsymbol{\delta}$ at all.
One can also see, for the case that $M_{2}=0$ or $\epsilon=0$, the above
result reduces to the result in the previous subsection, i.e., a
massless $U(1)$ mixed with a massive $U(1)$, where again no millicharge is produced.

\subsection{Kinetic and mass mixing in the full electroweak theory}\label{sec:mixingF}

In the full electroweak theory, the Higgs mechanism generates the
mass mixing between the hypercharge gauge field $B$ and the neutral $SU(2)$ gauge
field $A_{3}$, and the photon is a linear combination of them.
Hence, the extra $U(1)_{X}$ should mix
with the hypercharge rather than the photon field at high energies.
For a realistic model beyond the SM, the extra $U(1)_{X}$ can mix
with the hypercharge via kinetic and mass terms. To obtain the
$\{A^{\prime},A^{\gamma},Z\}$ mass eigenstates,
where $A^{\prime}$ is the additional gauge boson in the theory,
either called a $Z^\prime$ or a dark photon for small masses,
and the $A_{3}$ gauge
boson is always entangled in the mass matrix after the electroweak
phase transition when the Higgs picks a \textit{vev}.

In this subsection we discuss realist models that can generate a millicharge
carried by the dark particle. To this end, we discuss the kinetic
mixing and mass mixing effects of an extra $U(1)_{X}$ with the full
electroweak sector. This encompasses three different situations,
and we will see that the kinetic mixing between $U(1)_Y$ and a massive $U(1)_X$
alone cannot produce a millicharge for the dark fermion.

\subsubsection{$U(1)_{X}$ gauge boson achieves a mass via the Stueckelberg mechanism}\label{sec:mixingSt}

In this subsection we review the kinetic mixing and mass mixing in
the Stueckelberg extension of the SM, c.f., Eqs. (\ref{eq:KinMix})
and (\ref{eq:StMMix}), first discussed in~\cite{Feldman:2007wj}.
The Stueckelberg mass mixing effect is discussed in details in the context of string
theory in~\cite{Feng:2014eja,Feng:2014cla}. Within the context of string theory, the quantization
of the electric charge, i.e., any charge should  be an integral  (more
accurately, a fractional number)  multiple of the electric charge, can
be explained that the millicharge carried by the dark particles corresponds
to the ratio of integer or half-integer wrapping numbers of D-branes which
wrap on cycles in the internal six dimensional manifold in string theory.

In the gauge eigenbasis of the $U(1)_{X}$, hypercharge and the neutral
$SU(2)$ gauge field $V^{T}=(C,B,A^{3})$, the mixing matrices can
be written as
\begin{equation}
\mathcal{K}=\left(\begin{array}{ccc}
1 & \delta & 0\\
\delta & 1 & 0\\
0 & 0 & 1
\end{array}\right)\,,\qquad M_{{\rm St}}^{2}=\left(\begin{array}{ccc}
M_{1}^{2} & M_{1}M_{2} & 0\\
M_{1}M_{2} & M_{2}^{2}+\frac{1}{4}v^{2}g_{Y}^{2} & -\frac{1}{4}v^{2}g_{2}g_{Y}\\
0 & -\frac{1}{4}v^{2}g_{2}g_{Y} & \frac{1}{4}v^{2}g_{2}^{2}
\end{array}\right)\,.\label{eq:33mixM}
\end{equation}
Similar to the discussion for the two $U(1)$ case in previous subsection,
a simultaneous diagonalization is performed for the above two mixing matrices.
This boils down to the problem of diagonalizing the matrix
$K^{T}M_{{\rm St}}^{2}K$ using an orthogonal transformation $O$
such that $O^{T}K^{T}M_{{\rm St}}^{2}KO$ is diagonal, where
\begin{equation}
K=\left(\begin{array}{ccc}
c_{\delta} & 0 & 0\\
-s_{\delta} & 1 & 0\\
0 & 0 & 1
\end{array}\right)\,,
\label{eq:33KMnuTR}
\end{equation}
with $c_{\delta}=1/\sqrt{1-\delta^{2}}$ and $s_{\delta}=\delta/\sqrt{1-\delta^{2}}$.
The original gauge eigenbasis $V^{T}=(C,B,A^{3})$ then transforms
to the mass eigenbasis $E^{T}=(A^{\prime},A^{\gamma},Z)$ following
the relation
\begin{equation}
V=(KO)E\equiv \mathcal{R}E \,,
\end{equation}
the detailed calculation can be found in Appendix~\ref{sec:APPKMMM}
and the transformation matrix $\mathcal{R}$ is shown in Eq.~(\ref{eq:RSt}).
It will be useful to define the mass mixing parameter $\epsilon=M_{2}/M_{1}$.
The interaction of gauge bosons with fermions
can be obtained from
\begin{equation}
\mathcal{L}_{{\rm int}}=\left(g_{X}J_{X},g_{Y}J_{Y},g_{2}J_{3}\right)V=\left(g_{X}J_{X},g_{Y}J_{Y},g_{2}J_{3}\right)\mathcal{R}E\,.
\end{equation}

Here we summarize the neutral current couplings. The interactions
of $A^{\prime},Z,A^{\gamma}$ to SM fermions can be written as
\begin{equation}
-\mathcal{L}_{A^{\prime},Z,A^{\gamma}}=\frac{1}{2}\bar{f}_{i}\gamma^{\mu}\left[\left(v_{i}^{\prime}-a_{i}^{\prime}\gamma^{5}\right)f_{i}A_{\mu}^{\prime}+\left(v_{i}-a_{i}\gamma^{5}\right)f_{i}Z_{\mu}\right]+eQ_{i}\bar{f}_{i}\gamma^{\mu}f_{i}A_{\mu}^{\gamma}\,,
\label{eq:StNGBc}
\end{equation}
where
\begin{align}
v_{i} & =\left(g_{2}\mathcal{R}_{33}-g_{Y}\mathcal{R}_{23}\right)T_{i}^{3}+2g_{Y}\mathcal{R}_{23}Q_{i}\,,\\
a_{i} & =\left(g_{2}\mathcal{R}_{33}-g_{Y}\mathcal{R}_{23}\right)T_{i}^{3}\,,\\
v_{i}^{\prime} & =\left(g_{2}\mathcal{R}_{31}-g_{Y}\mathcal{R}_{21}\right)T_{i}^{3}+2g_{Y}\mathcal{R}_{21}Q_{i}\,,\\
a_{i}^{\prime} & =\left(g_{2}\mathcal{R}_{31}-g_{Y}\mathcal{R}_{21}\right)T_{i}^{3}\,.
\end{align}
In the above equations, $Q_{i}e$ is the electric charges of the SM
fermions, $g_{Y}$ is the original hypercharge gauge coupling in the
electroweak theory.

The electric charge is altered by the mixing effects and is written
as
\begin{equation}
e=\frac{g_{2}g_{Y}^{\prime}}{\sqrt{g_{2}^{2}+g_{Y}^{\prime2}}}\,,
\end{equation}
where the hypercharge gauge coupling is also modified to be
$g_{Y}^{\prime}=g_{Y}/\sqrt{1+\epsilon^{2}-2\delta\epsilon}\approx g_Y$ for small $\delta,\epsilon$.

\begin{figure}
\begin{center}
\includegraphics[scale=1]{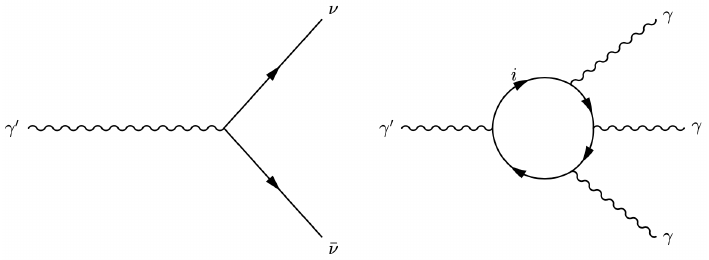}
\caption{
A display of dark photon decay channels for $M_{\gamma^\prime} < 2m_e$ and $M_{\gamma^\prime} < 2m_\chi$,
including the decay to pair of neutrino and anti-neutrino, and to three photons.
Considering various constraints,
for the benchmark models we consider,
the decay of the dark photon to neutrinos due to the mixing effect is always suppressed
compared to the three-photon decay channel.}
\label{fig:FeyZpDecay}
\end{center}
\end{figure}

When the mass of the dark photon is less than twice of dark fermion mass and is also less than twice of electron mass,
the dark photon can only decay to pairs of neutrino and anti-neutrino, and to three photons, as shown in Fig.~\ref{fig:FeyZpDecay}.
The coupling of the dark photon to neutrinos is calculated to be
\begin{align}
-\mathcal{L}_{A^{\prime}\bar{\nu}\nu} & =\frac{1}{2}a_{\nu}^{\prime}\bar{\nu}_{L}\gamma^{\mu}\nu_{L}A_{\mu}^{\prime}\approx-\frac{1}{2}g_{Y}^{\prime}\left(\delta-\epsilon\right)\epsilon_{0}^{2}\bar{\nu}_{L}\gamma^{\mu}\nu_{L}A_{\mu}^{\prime}\,,
\label{eq:GPnunu}
\end{align}
where $M_{0}=v\sqrt{g_{2}^{2}+g_{Y}^{\prime2}}/2\approx M_{Z}$, almost
identical to the original $Z$ boson mass in the electroweak theory
for small $\delta,\epsilon$, and we define $\epsilon_{0}=M_{\gamma^{\prime}}/M_{0}\approx M_{1}/M_{Z}$.

The decay width of the dark photon to neutrinos is given by
\begin{equation}
\Gamma_{\gamma^\prime \nu \bar \nu} \sim \frac{g_Y^{\prime 2} (\delta -\epsilon)^2 \epsilon_0^4}{48\pi} M_{\gamma^\prime}\,.
\end{equation}
For a 100~keV dark photon where we have $\epsilon_0 \sim 10^{-6}$, take the largest value of $|\delta-\epsilon|\sim 10^{-12}$,
one estimates that $\Gamma_{\gamma^\prime \nu \bar \nu} \sim 10^{-55}$~GeV,
which is much smaller than the dark photon to three photon decay width, given by
\begin{equation}
\Gamma_{\gamma^\prime \to 3\gamma} \sim \frac{(\delta-\epsilon)^2 \alpha^4 M_{\gamma^\prime}^9}{m_e^8}\,.\label{eq:Dw3ph}
\end{equation}
The above decay width gives rise to the lifetime of the dark photon to be roughly $10^{18}$~second,
greater than the age of the Universe.
While we are going to see in Section~\ref{sec:EoM} that
even the dark photon is stable during the age of the Universe,
its couplings to the SM particles will be further constrained by the
isotropic diffuse photon background (IDPB) data.

The interactions of $A^{\prime},A^{\gamma},Z$ to the dark fermion
$\chi$ can be written as
\begin{equation}
-\mathcal{L}_{\chi}=g_{X}Q_{\chi}\left(\mathcal{R}_{11}A_{\mu}^{\prime}+\mathcal{R}_{12}A_{\mu}^{\gamma}+\mathcal{R}_{13}Z_{\mu}\right)\bar{\chi}\gamma^{\mu}\chi\,.\label{eq:GBcDF}
\end{equation}
After the mixing, the dark fermion $\chi$ will couple to the photon with
\begin{equation}
-\mathcal{L}_{\chi}^{{\rm em}}=Q_{\epsilon} \bar{\chi}\gamma^{\mu}\chi A_{\mu}^{\gamma}
=\epsilon g_{X}Q_{\chi}\cos\theta\,\bar{\chi}\gamma^{\mu}\chi A_{\mu}^{\gamma}\,,\label{eq:millicharge}
\end{equation}
where $Q_{\epsilon}\equiv\epsilon g_{X}Q_{\chi}\cos\theta$ is the
millicharge carried by $\chi$.

One can see clearly from the Eq.~(\ref{eq:millicharge})
that the millicharge carried by the dark fermion $\chi$ is generated by the mass mixing effect,
and thus only depends on the mass mixing parameter~$\epsilon$.
Once the mass mixing is turned off ($\epsilon=0$),
no millicharge will be generated.
Thus the generation of millicharge
has nothing to do with the kinetic mixing between
$U(1)_{X}$ and the hypercharge gauge field.

Finally we summarize the order of the couplings induced by the kinetic mixing and the mass mixing:
\begin{gather}
A^\gamma i\bar{i} \sim e Q_i\,,\qquad A^\gamma \chi\bar{\chi} \sim \epsilon g_X\,, \label{Phcoup} \\
Z i\bar{i} \sim g_2\,,\qquad Z \chi\bar{\chi} \sim \delta g_X\,, \label{Zcoup} \\
A^\prime i\bar{i} \sim |\delta-\epsilon| g_Y\,, 
\qquad A^\prime \chi\bar{\chi} \sim g_X\,. \label{DPcoup}
\end{gather}

\subsubsection{$U(1)_{X}$ is broken via a dark Higgs}\label{sec:mixingEH}

If the $U(1)_{X}$ is broken at some scale and achieves a mass via
the dark $U(1)_{X}$ Higgs $\phi$, the dark $U(1)_{X}$ breaking
scale is important. Prior to the $U(1)_{X}$ breaking, the massless
$U(1)_{X}$ mixed with the massless hypercharge gauge field via gauge
kinetic terms. After the mixing the dark fermion $\chi$ carrying the
$U(1)_{X}$ charge, carries a milli-hypercharge proportional to the
kinetic mixing parameter $\delta$, as described in Section~\ref{sec:mixing2U1}. As
the temperature of the universe drops down, and considering the case
that the $U(1)_{X}$ breaking scale is higher than the electroweak
phase transition, in the stage of $U(1)_{X}$ is broken while before
the electroweak phase transition, the massive $U(1)_{X}$ mixed with
the massless hypercharge via kinetic terms. As discussed in Section~\ref{sec:mixing2U10M},
the dark particles carry zero hypercharge. When both of the $U(1)_{X}$
and electroweak symmetry are broken, the $U(1)_{X}$ gauge boson will
mix with the electroweak sector and the mixing matrices become $3\times3$,
c.f., Eqs.~(\ref{eq:darkHCD}) and (\ref{eq:HCD})
\begin{equation}
\mathcal{K}=\left(\begin{array}{ccc}
1 & \delta & 0\\
\delta & 1 & 0\\
0 & 0 & 1
\end{array}\right)\,,\qquad M^{2}=\left(\begin{array}{ccc}
g_{X}^{2}u^{2}+\frac{1}{4}y^{2}g_{Y}^{2}v^{2} & \frac{1}{4}yg_{Y}^{2}v^{2} & -\frac{1}{4}yg_{2}g_{Y}v^{2}\\
\frac{1}{4}yg_{Y}^{2}v^{2} & \frac{1}{4}g_{Y}^{2}v^{2} & -\frac{1}{4}g_{2}g_{Y}v^{2}\\
-\frac{1}{4}yg_{2}g_{Y}v^{2} & -\frac{1}{4}g_{2}g_{Y}v^{2} & \frac{1}{4}g_{2}^{2}v^{2}
\end{array}\right)\,.
\end{equation}
For the case of a general gauged $U(1)_{y+X}$ theory, where $y$
is a very small fraction of the hypercharge, studied in for example~\cite{Zhang:2022nnh},
the dark particles can carry a tiny amount of hypercharge proportional to $y$.
A simultaneous diagonalization of the above two matrices gives rise
to the following transformation to the gauge bosons
\begin{equation}
\left(\begin{array}{c}
C\\
B\\
A_{3}
\end{array}\right)=\mathcal{R}^{H}\cdot \left(\begin{array}{c}
A^{\prime}\\
A^{\gamma}\\
Z
\end{array}\right)=\left(\begin{array}{ccc}
\times & 0 & \times\\
\times & \times & \times\\
\times & \times & \times
\end{array}\right)\left(\begin{array}{c}
A^{\prime}\\
A^{\gamma}\\
Z
\end{array}\right)\,,
\end{equation}
where we only show the $\{12\}$-component of the rotation matrix $\mathcal{R}^{H}$,
relevant for generating the mixing between the photon and dark fermions.
The photon and the dark fermion interaction is then given by
\begin{align}
-\mathcal{L} & =g_{X}Q_{\chi}\mathcal{R}_{12}^{H}A_{\mu}^{\gamma}\bar{\chi}\gamma^{\mu}\chi+g_{Y}yA_{\mu}^{\gamma}\bar{\chi}\gamma^{\mu}\chi
=\frac{y\,e}{\cos\theta_{W}}A_{\mu}^{\gamma}\bar{\chi}\gamma^{\mu}\chi\,,
\end{align}
where the dark particles can carry a millicharge $Q_{\epsilon}=ye/\cos\theta_{W}$ proportional to $y$
and is not originated from the mixing effect.
In this case, neither the kinetic mixing nor the mass mixing plays any role of generating the millicharge.

The other possibility we do not discuss, is that
the SM gauge bosons also acquire tiny masses from the dark Higgs.
In this scenario, the dark Higgs mass as well as its couplings to the SM particles
are severely constrained.

\subsubsection{Kinetic mixing between a massless $U(1)$ and the hypercharge gauge field}\label{sec:mixingmlU1}

In this subsection we discuss the case that the kinetic mixing between
a {\it massless} $U(1)_{X}$ and the hypercharge gauge field,
which can generate a millicharge proportional to the kinetic mixing parameter.
Now the kinetic mixing matrix together with the mass matrix corresponding to
the gauge eigenbasis $V^{T}=\left(B,C,A_{3}\right)$ are given by
\begin{equation}
\mathcal{K}=\left(\begin{array}{ccc}
1 & \delta & 0\\
\delta & 1 & 0\\
0 & 0 & 1
\end{array}\right)\,,\qquad M^{2}=\left(\begin{array}{ccc}
\frac{1}{4}g_{Y}^{2}v^{2} & 0 & -\frac{1}{4}g_{2}g_{Y}v^{2}\\
0 & 0 & 0\\
-\frac{1}{4}g_{2}g_{Y}v^2 & 0 & \frac{1}{4}g_{2}^{2}v^{2}
\end{array}\right)\,.
\end{equation}
Using the formalism we have discussed in the previous subsection,
diagonalizing both of the above two matrices simultaneously,
is equivalent to diagonalizing a single matrix $K^{T}M^{2}K$,
with matrix $K$ given in Eq.~(\ref{eq:33KMnuTR}).
In this case, this is done by using an orthogonal transformation $O$ given by
\begin{equation}
O=\left(\begin{array}{ccc}
\frac{-\sqrt{1-\delta^{2}}g_{2}}{\sqrt{\left(1-\delta^{2}\right)g_{2}^{2}+g_{Y}^{2}}} & 0 & \frac{g_{Y}}{\sqrt{\left(1-\delta^{2}\right)g_{2}^{2}+g_{Y}^{2}}}\\
0 & 1 & 0\\
\frac{-g_{Y}}{\sqrt{\left(1-\delta^{2}\right)g_{2}^{2}+g_{Y}^{2}}} & 0 & \frac{-\sqrt{1-\delta^{2}}g_{2}}{\sqrt{\left(1-\delta^{2}\right)g_{2}^{2}+g_{Y}^{2}}}
\end{array}\right)\,,
\end{equation}
such that $\left(KO\right)^{T}M^{2}KO$ is diagonal
\begin{equation}
\left(KO\right)^{T}M^{2}KO=\left(\begin{array}{ccc}
0 & 0 & 0\\
0 & 0 & 0\\
0 & 0 & \frac{v^{2}}{4}\left(g_{2}^{2}+\frac{1}{1-\delta^{2}}g_{Y}^{2}\right)
\end{array}\right)\,,
\end{equation}
which gives rise to only one nonzero eigenvalue for the $Z$ boson
mass-square.

The full transformation matrix is then computed as
\begin{equation}
\mathcal{R}^{0}=KO
=\left(\begin{array}{ccc}
-\frac{g_{2}}{\sqrt{\left(1-\delta^{2}\right)g_{2}^{2}+g_{Y}^{2}}} & 0 & \frac{g_{Y}}{\sqrt{1-\delta^{2}}\sqrt{\left(1-\delta^{2}\right)g_{2}^{2}+g_{Y}^{2}}}\\
\frac{g_{2}\,\delta}{\sqrt{\left(1-\delta^{2}\right)g_{2}^{2}+g_{Y}^{2}}} & 1 & -\frac{g_{Y}\,\delta}{\sqrt{1-\delta^{2}}\sqrt{\left(1-\delta^{2}\right)g_{2}^{2}+g_{Y}^{2}}}\\
-\frac{g_{Y}}{\sqrt{\left(1-\delta^{2}\right)g_{2}^{2}+g_{Y}^{2}}} & 0 & -\frac{g_{2}\sqrt{1-\delta^{2}}}{\sqrt{\left(1-\delta^{2}\right)g_{2}^{2}+g_{Y}^{2}}}
\end{array}\right)\,.
\end{equation}
The interaction of three neutral gauge bosons to the dark fermion
$\chi$ is thus given by
\begin{equation}
-\mathcal{L}_{\chi}=g_{X}Q_{\chi}\left(\mathcal{R}^0_{21}A_{\mu}^{\gamma}+\mathcal{R}^0_{22}A_{\mu}^{\prime}+\mathcal{R}^0_{23}Z_{\mu}\right)\bar{\chi}\gamma^{\mu}\chi\,.
\end{equation}
Specifically, the coupling of the dark fermion $\chi$ to the photon
reads
\begin{equation}
-\mathcal{L}_{\chi}^{\mathrm{em}}  =g_{X}Q_{\chi}\mathcal{R}^0_{21}A_{\mu}^{\gamma}\bar{\chi}\gamma^{\mu}\chi
\approx \frac{\delta g_{2}g_{X}Q_{\chi}}{\sqrt{g_{2}^{2}+g_{Y}^{2}}} A_{\mu}^{\gamma}\bar{\chi}\gamma^{\mu}\chi\,,
\end{equation}
generating a millicharge proportional to the kinetic mixing parameter, carried by the dark fermion.

\section{Coupled Boltzmann equations with multiple temperatures}\label{sec:BEMT}

\begin{figure}
\begin{center}
\includegraphics[scale=0.5]{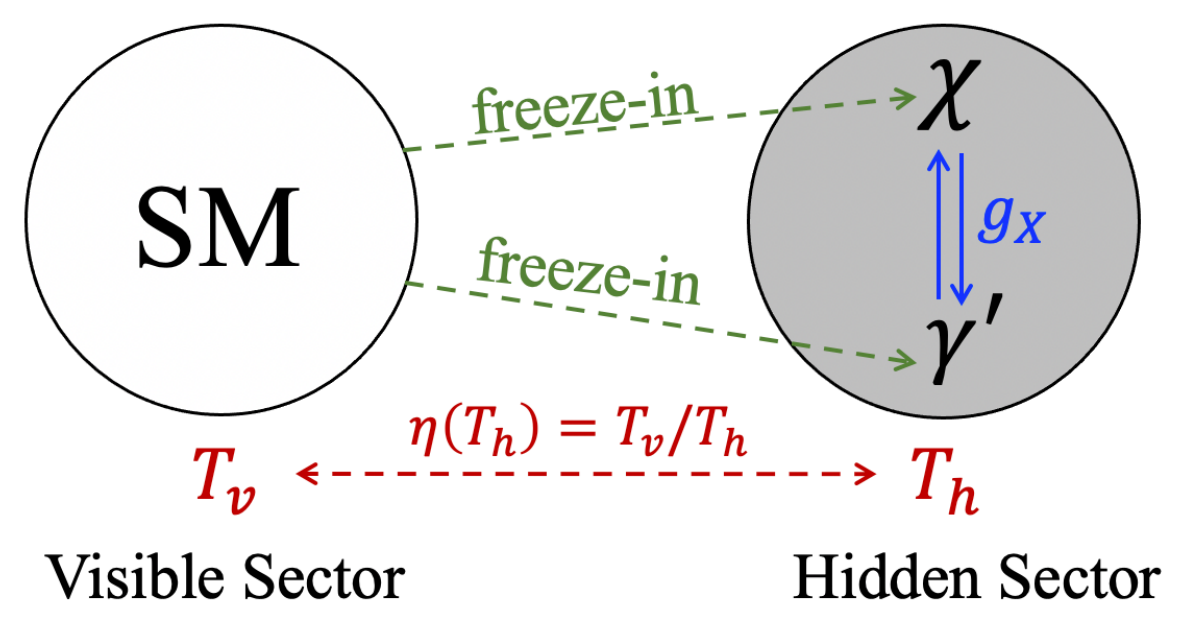}
\caption{A graphic illustration of the model we discuss.
The $U(1)_X$ hidden sector connects to the SM with tiny kinetic and mass mixings,
and thus it has only feeble interactions with SM particles.
The hidden sector possesses a different temperature $T_h$,
related to the visible sector temperature $T_v$ with the function $\eta(T_h)$.
The dark fermion $\chi$ and the dark photon $\gamma^\prime$
are produced through the freeze-in of the SM particles.
Inside the hidden sector, $\chi$ and $\gamma^\prime$
can have either strong or weak (maybe ultraweak) interactions
such as $\chi\bar{\chi} \leftrightarrow \gamma^\prime\gamma^\prime$.
The hidden sector interactions and the freeze-in productions
occur simultaneously and are related to $T_h$ and $T_v$ respectively,
and must be addressed jointly.}
\label{fig:U1Model}
\end{center}
\end{figure}

In Fig.~\ref{fig:U1Model}, we graphically illustrate the model we discuss.
Initially there is negligible amount of matter in the hidden sector.
The dark fermion $\chi$ and the dark photon $\gamma^\prime$ are produced
through the freeze-in production of the SM particles.
Once the matter inside the hidden sector accumulated,
hidden sector interactions start to take place,
such as $\chi\bar{\chi} \leftrightarrow \gamma^\prime \gamma^\prime $
which will change the number densities of $\chi$  and $\gamma^\prime$.
The hidden sector interactions occur simultaneously with the freeze-in processes.
The hidden sector possesses a different temperature $T_h$,
which is related to the visible sector temperature $T_v$
(the temperature of the observed Universe, and we will omit the subscript $_v$ from here),
with the function $\eta(T_h)$.
When the dark photon mass is greater than twice of the electron mass,
the dark photon will eventually decay to (mostly) electron-positron pairs.
When the dark photon mass is less than twice of electron mass,
it can only decay into two neutrinos and to three photons and must be made stable
through the age of the Universe, see discussions in Section~\ref{sec:mixingSt}.

As shown in~\cite{Aboubrahim:2020lnr,Aboubrahim:2021ycj,Aboubrahim:2021dei},
for the cases that SM extended by one or multiple
hidden sectors interacting feebly with the visible sector,
the hidden sectors will evolve by themselves,
especially they can possess temperatures distinct from the visible sector temperature.
In this section we will review the temperature-dependent coupled
Boltzmann equations of the millicharge dark particle $\chi$ and the
dark photon $\gamma^{\prime}$ for the model described in Section~\ref{sec:mixingSt},
as well as the evolution equation of $\eta(T_h)=T/T_{h}$, with $T\,(T_{h})$
being the visible (hidden) sector temperature.
By solving these coupled differential equations simultaneously, one obtains the evolution of
hidden sector particle number densities and the hidden sector temperature.
In the analysis we notice that the entropies in the hidden and the visible sectors
are not separately preserved, but their sum is conserved, which is imposed in the analysis.
Since we are mostly interested
in the evolution of the hidden sector, we will use the hidden sector
temperature $T_{h}$ as the clock and the temperature in the visible
sector is related to the hidden sector via the function $\eta$.

\subsection{Derivation of the evolution of hidden sector particles}

In this subsection we review the derivation of the evolution of the hidden sector temperature~\cite{Aboubrahim:2020lnr}.
We start from the two Friedmann equations for a flat universe
\begin{align}
H^{2} & =\frac{8\pi G_{N}}{3}\rho\,,\label{FR1}\\
\frac{\ddot{a}}{a} & =-\frac{4\pi G_{N}}{3}(\rho+3p)\,,\label{FR2}
\end{align}
where $G_{N}$ is Newton's gravitational constant, $\rho$ and $p$
are respectively the energy density and pressure, giving rise to the
continuity equation
\begin{equation}
\frac{{\rm d}\rho}{{\rm d}t}+3H(\rho+p)=0\,.\label{FR3}
\end{equation}
Using $T_{h}$ as the clock, we can then obtain the following relation
from Eq.~(\ref{FR3})
\begin{equation}
\frac{{\rm d}T_{h}}{{\rm d}t}=-\frac{4\zeta\rho}{{\rm d}\rho/{\rm d}T_{h}}H\,,\label{FR4}
\end{equation}
where $\zeta=\frac{3}{4}(1+p/\rho)$. Here $\zeta=1$ is for the radiation
dominated era and $\zeta=3/4$ for the matter dominated universe.
We assume that the hidden sector interacts with the visible sector
feebly and the continuity equations for the two sectors are modified
to be
\begin{align}
\frac{{\rm d}\rho_{h }}{{\rm d}t}+3H\left(\rho_{h }+p_{h }\right) & =j_{h }\,,\label{EOCh}\\
\frac{{\rm d}\rho_{v}}{{\rm d}t}+3H\left(\rho_{v}+p_{v}\right) & =-j_{h}\,,\label{EOCv}
\end{align}
where the subscripts $v$ and $h$ correspond to the visible and hidden
sectors, respectively, and $j_{h}$ is the source term in the hidden
sector and arises from freeze-in. 
Using the chain rule and together with Eqs. (\ref{FR4}) and (\ref{EOCh}), one obtains
\begin{equation}
\rho\frac{{\rm d}\rho_{h}}{{\rm d}T_{h}}=\left(\frac{\zeta_{h}}{\zeta}\rho_{h}-\frac{j_{h}}{4H\zeta}\right)\frac{{\rm d}\rho}{{\rm d}T_{h}}\,,\label{a7}
\end{equation}
where $\zeta_{h}=\frac{3}{4}(1+p_{h}/\rho_{h})$
and $\zeta_{h}=1$ for radiation dominance and $\zeta_{h}=3/4$
for matter dominance in the hidden sector.
The total energy density is given by the sum of visible sector contribution and hidden sector contribution, i.e.,
$\rho=\rho_{v}+\rho_{h}$, which gives rise to
${\rm d}\rho/{\rm d}T_{h}={\rm d}\rho_{v}/{\rm d}T_{h}+{\rm d}\rho_{h}/{\rm d}T_{h}$.
Together with Eq.~(\ref{a7}),
${\rm d}\rho_{v}/{\rm d}T_{h}$ can be solved
in terms of ${\rm d}\rho_{h}/{\rm d}T_{h}$ as
\begin{equation}
\frac{{\rm d}\rho_{v}}{{\rm d}T_{h}}=\frac{\zeta\rho_{v}+\rho_{h}(\zeta-\zeta_{h})+j_{h}/(4H)}{\zeta_{h}\rho_{h}-j_{h}/(4H)}\frac{{\rm d}\rho_{h}}{{\rm d}T_{h}}\,.\label{a10}
\end{equation}

In the analysis, we use the constraint that the total entropy $S=sR^{3}$
is conserved which gives ${\rm d}s/{\rm d}t+3Hs=0$  with
\begin{align}
s & =\frac{2\pi^{2}}{45}\big(h_{{\rm eff}}^{v}T^{3}+h_{{\rm eff}}^{h}T_{h}^{3}\big)\,,\label{entropy}
\end{align}
where $h_{{\rm eff}}^{v}\,(h_{{\rm eff}}^{h})$ is the visible (hidden) effective entropy degrees of freedom.
The Hubble parameter now depends on both $T$ and $T_{h}$
\begin{equation}
H^{2}=\frac{8\pi G_{N}}{3}\big[\rho_{v}(T)+\rho_{h}(T_{h})\big]\,,\label{hubble}
\end{equation}
where $\rho_{v}(T)$ and $\rho_{h}(T_{h})$ are the energy density in the
visible sector and in the hidden sector respectively and are given by
\begin{align}
\rho_{v} & =\frac{\pi^{2}}{30}g_{{\rm eff}}^{v}T^{4}\,,~~\rho_{h}=\frac{\pi^{2}}{30}g_{{\rm eff}}^{h}T_{h}^{4}\,.\label{rho-1}
\end{align}
$g_{{\rm eff}}^{v},h_{{\rm eff}}^{v}$ are functions of $T$ and we
use the fits given in~\cite{Kolb:1990vq,Gondolo:1990dk,Gelmini:1990je}
to parameterize them while $g_{{\rm eff}}^{h},h_{{\rm eff}}^{h}$
are functions of $T_{h}$ and we use temperature dependent integrals
given in~\cite{Hindmarsh:2005ix} to parameterize them.

Combining equations discussed above, the evolution of the ratio of temperatures for two sectors $\eta=T/T_{h}$ can be derived as
\begin{equation}
\frac{{\rm d}\eta}{{\rm d} T_{h}}=-\frac{A_{v}}{B_{v}}+
\frac{\zeta\rho_{v}+\rho_{h}(\zeta-\zeta_{h})+j_{h}/(4H)}{B_{v} [ \zeta_{h}\rho_{h}-j_{h}/(4H) ]}
\frac{{\rm d}\rho_{h}}{{\rm d}T_{h}}\,,\label{y30}
\end{equation}
with
\begin{align}
A_{v} & =\frac{\pi^{2}}{30}\Big(\frac{{\rm d}g_{{\rm eff}}^{v}}{{\rm d}T}\eta^{5}T_{h}^{4}+4g_{{\rm eff}}^{v}\eta^{4}T_{h}^{3}\Big)\,,\label{y4}\\
B_{v} & =\frac{\pi^{2}}{30}\Big(\frac{{\rm d}g_{{\rm eff}}^{v}}{{\rm d}T}\eta^{4}T_{h}^{5}+4g_{{\rm eff}}^{v}\eta^{3}T_{h}^{4}\Big)\,.\label{y5}
\end{align}

Now $\rho_{h}$ and $p_{h}$, which enter in the definition of $\zeta_{h}$,
are determined in terms of $\rho_{\gamma'},p_{\gamma'},\rho_{\chi},p_{\chi}$
so that $\rho_{h}=\rho_{\gamma^{\prime}}+\rho_{\chi}$ and $p_{h}=p_{\gamma^{\prime}}+p_{\chi}$,
where $\rho_{\gamma'}$ and $p_{\gamma'}$ are given by
\begin{align}
\rho_{\gamma^{\prime}} & =\frac{g_{\gamma^{\prime}}T_{h}^{4}}{2\pi^{2}}\int_{x_{\gamma^{\prime}}}^{\infty}\frac{x^{3}{\rm d}x}{e^{x}-1}\,,~~~\text{and}~~~p_{\gamma^{\prime}} =\frac{g_{\gamma^{\prime}}T_{h}^{4}}{6\pi^{2}}\int_{x_{\gamma^{\prime}}}^{\infty}\frac{\left(x^{2}-x_{\gamma^{\prime}}^{2}\right){\rm d}x}{e^{x}-1}\,,\\
\rho_{\chi} & =\frac{g_{\chi}T_{h}^{4}}{2\pi^{2}}\int_{x_{\chi}}^{\infty}\frac{x^{3}{\rm d}x}{e^{x}+1}\,,~~~\text{and}~~~p_{\chi} =\frac{g_{\chi}T_{h}^{4}}{6\pi^{2}}\int_{x_{\chi}}^{\infty}\frac{\left(x^{2}-x_{\chi}^{2}\right){\rm d}x}{e^{x}+1}\,,\label{rhop}
\end{align}
where $g_{\gamma^{\prime}}=3,g_{\chi}=4$ are degrees of freedom of
dark photon and dark fermion respectively, and $x_{\gamma^{\prime}}=M_{\gamma^{\prime}}/T_{h},x_{\chi}=m_{\chi}/T_{h}$.
The $H$ and $s$ are given by
\begin{align}
s & =\frac{2\pi^{2}}{45}\left(h_{\mathrm{eff}}^{v}\eta^{3}+h_{\mathrm{eff}}^{h}\right)T_{h}^{3}\,,\\
H^{2} & =\frac{\pi^{2}}{90M_{\mathrm{Pl}}^{2}}\left(g_{\mathrm{eff}}^{v}\eta^{4}+g_{\mathrm{eff}}^{h}\right)T_{h}^{4}\,,
\end{align}
and the ratio $s/H$ can be written as
\begin{align}
\frac{s}{H} & =\frac{2\sqrt{2}\pi}{\sqrt{45}}\frac{h_{{\rm eff}}}{\sqrt{g_{{\rm eff}}}}M_{{\rm Pl}}T_{h}\,,\label{sH}
\end{align}
where $M_{\mathrm{Pl}}\equiv\sqrt{\frac{1}{8\pi G_{N}}}=2.4\times10^{18}~\mathrm{GeV}$.

In the dark sector, the effective degrees of freedom include those
for the dark photon and for the dark fermion
\begin{align}
g_{{\rm eff}}^{h} & =g_{{\rm eff}}^{\gamma^{\prime}}+\frac{7}{8}g_{{\rm eff}}^{\chi}\,,~~~\text{and}~~~h_{{\rm eff}}^{h}=h_{{\rm eff}}^{\gamma^{\prime}}+\frac{7}{8}h_{{\rm eff}}^{\chi}\,.
\end{align}
At temperature $T_{h}$, $g_{{\rm eff}}$ and $h_{{\rm eff}}$ for
the particles $\gamma^{\prime}$ and $\chi$ are given by
\begin{equation}
\begin{aligned}g_{{\rm eff}}^{\gamma^{\prime}} & =\frac{45}{\pi^{4}}\int_{x_{\gamma^{\prime}}}^{\infty}\frac{\sqrt{x^{2}-x_{\gamma^{\prime}}^{2}}}{e^{x}-1}x^{2}{\rm d}x\,,~~~\text{and}~~~h_{{\rm eff}}^{\gamma^{\prime}}=\frac{45}{4\pi^{4}}\int_{x_{\gamma^{\prime}}}^{\infty}\frac{\sqrt{x^{2}-x_{\gamma^{\prime}}^{2}}}{e^{x}-1}(4x^{2}-x_{\gamma^{\prime}}^{2}){\rm d}x\,,\\
g_{{\rm eff}}^{\chi} & =\frac{60}{\pi^{4}}\int_{x_{\chi}}^{\infty}\frac{\sqrt{x^{2}-x_{\chi}^{2}}}{e^{x}+1}x^{2}{\rm d}x\,,~~~\,\text{and}~~~h_{{\rm eff}}^{\chi}=\frac{15}{\pi^{4}}\int_{x_{\chi}}^{\infty}\frac{\sqrt{x^{2}-x_{\chi}^{2}}}{e^{x}+1}(4x^{2}-x_{\chi}^{2}){\rm d}x\,.
\end{aligned}
\label{hdof}
\end{equation}
When temperatures go above the mass of the particles,
the particles become relativistic and their effective degrees of freedom take
constant values: $g_{{\rm eff}}^{\gamma^{\prime}}=h_{{\rm eff}}^{\gamma^{\prime}}=3$
for $T_{h}\gg M_{\gamma^{\prime}}$ and $g_{{\rm eff}}^{\chi}=h_{{\rm eff}}^{\chi}=4$
for $T_{h}\gg m_{\chi}$.

\subsection{Coupled Boltzmann equations for two sectors}\label{sec:CBE}

With the above preparation, we are now ready to write down the Boltzmann
equations for the number densities of the dark fermions $\chi$ and
of the dark photons $\gamma^{\prime}$, which govern the evolution of the hidden sector.
For the model we discuss,
we have the following Boltzmann equations for number densities of
the dark fermion $\chi$ and the dark photon $\gamma^{\prime}$:
\begin{align}
\frac{{\rm d}n_{\chi}}{{\rm d}t}+3Hn_{\chi}=\sum_{i\in{\rm SM}} & \Big\{
{\color{Green} (n_{\chi}^{{\rm eq}})^{2}\langle\sigma v\rangle_{\chi\bar{\chi}\to i\bar{i}}^{T}+n_{\gamma^{*}}\langle\Gamma\rangle_{\gamma^{*}\rightarrow\chi\bar{\chi}}^{T}} \nonumber \\
 & +\theta(M_{\gamma^{\prime}}-2m_{\chi})\big[-n_{\chi}^{2}\langle\sigma v\rangle_{\bar{\chi}\chi\rightarrow\gamma^{\prime}}^{T_{h}}+n_{\gamma^{\prime}}\langle\Gamma\rangle_{\gamma^{\prime}\to\chi\bar{\chi}}^{T_{h}}\big]
\nonumber \\
 &  -n_{\chi}^{2}\langle\sigma v\rangle_{\chi\bar{\chi}\to\gamma^{\prime}\gamma^{\prime}}^{T_{h}}+n_{\gamma^{\prime}}^{2}\langle\sigma v\rangle_{\gamma^{\prime}\gamma^{\prime}\to\chi\bar{\chi}}^{T_{h}} \Big\}\,,\\
\frac{{\rm d}n_{\gamma^{\prime}}}{{\rm d}t}+3Hn_{\gamma^{\prime}}=\sum_{i\in{\rm SM}} & \Big\{ n_{\chi}^{2}\langle\sigma v\rangle_{\chi\bar{\chi}\to\gamma^{\prime}\gamma^{\prime}}^{T_{h}}-n_{\gamma^{\prime}}^{2}\langle\sigma v\rangle_{\gamma^{\prime}\gamma^{\prime}\to\chi\bar{\chi}}^{T_{h}}\nonumber \\
 & +\theta(M_{\gamma^{\prime}}-2m_{\chi})\big[n_{\chi}^{2}\langle\sigma v\rangle_{\bar{\chi}\chi\rightarrow\gamma^{\prime}}^{T_{h}}-n_{\gamma^{\prime}}\langle\Gamma\rangle_{\gamma^{\prime}\to\chi\bar{\chi}}^{T_{h}}\big]\nonumber \\
 & + {\color{Green} \theta(M_{\gamma^{\prime}}-2m_{i})\big[n_{i}^{2}\langle\sigma v\rangle_{i\bar{i}\rightarrow\gamma^{\prime}}^{T}}-n_{\gamma^{\prime}}\langle\Gamma\rangle_{\gamma^{\prime}\to i\bar{i}}^{T_{h}}\big] \nonumber \\
 & + {\color{Green} n_{i}^{2}\langle\sigma v\rangle_{i\bar{i}\rightarrow\gamma^{\prime}\gamma}^{T}+2n_{i}n_{\gamma^{\prime}}^{{\rm eq}}\langle\sigma v\rangle_{i\gamma^{\prime}\rightarrow i\gamma}^{T}} \Big\}\,,
\end{align}
where the upper subscript is the visible sector temperature $T$ or
hidden sector temperature $T_{h}$ that the corresponding process
is depending on.
The freeze-in processes are marked by {\color{Green} green color} to be distinguishable to various hidden sector processes.
Later we will use $T_{h}$ as the reference temperature
and replace $T$ by $T_{h}\eta$. In the above equations, $i$ and
$\bar{i}$ represent SM fermions and anti-fermions, $\gamma^{*}$
denotes the plasmon which expresses photon with effective in-medium thermal
mass in the early universe, c.f., Appendix~\ref{sec:APPPlasmon} for a brief review.
All the terms in the above equations with subscript $^{{\rm eq}}$
indicate that we are calculating the corresponding process reversely for computational simplicity.
In such calculations, we use the
equilibrium values of comoving number densities (depending on the visible sector temperature $T$)
\begin{equation}
Y_{a}^{{\rm eq}}=\frac{n_{a}^{\mathrm{eq}}}{s}=\frac{g_{a}}{2\pi^{2}s}m_{a}^{2}TK_{2}\left(\frac{m_a}{T}\right)\,,
\qquad{\rm for\,particle\,species}\,a\,,
\end{equation}
which are usually not in equilibrium in the thermal bath. For example,
the freeze-in process $i\bar{i}\to\chi\bar{\chi}$ is computed via
the inverse process $\chi\bar{\chi}\to i\bar{i}$ using the comoving
number density of $\chi$ as if it is in equilibrium in the thermal bath,
which is much easier to calculate compared to the original forward process.
The theta functions indicate whether the three-point decay or combination processes can occur.
If the mass of dark photon is greater than twice of the dark fermion
mass, i.e., $M_{\gamma^{\prime}}>2m_{\chi}$, three-point processes
$\gamma^{\prime}\leftrightarrow\chi\bar{\chi}$
are much stronger than the four-point processes
$\chi\bar{\chi}\leftrightarrow\gamma^{\prime}\gamma^{\prime}$
which can be safely ignored.
However, we can conclude from our analysis in Section~\ref{sec:EHSP}:
for the freeze-in production of the dark photon,
the four-point freeze-in channels {\it must be considered}
even the three-point freeze-in channels are present.
The additional factor of 2 corresponding
to the process $i\gamma\rightarrow i\gamma^{\prime}$ also counts
the anti-fermion contribution $\bar{i}\gamma\rightarrow\bar{i}\gamma^{\prime}$.

\begin{figure}
\begin{center}
\includegraphics[scale=0.66]{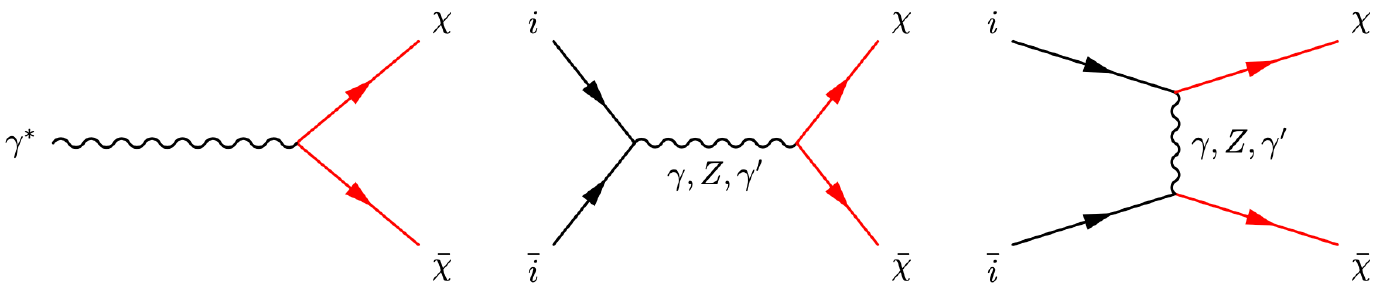}
\caption{
A display of production channels for the dark fermion $\chi$,
including SM fermions annihilation $i \bar{i} \to \chi \bar{\chi}$ through $s$ channel,
as well as the plasmon decay $\gamma^* \to \chi \bar{\chi}$. }
\label{fig:FeyChiFI}
\end{center}
\end{figure}

\begin{figure}
\begin{center}
\includegraphics[scale=0.42]{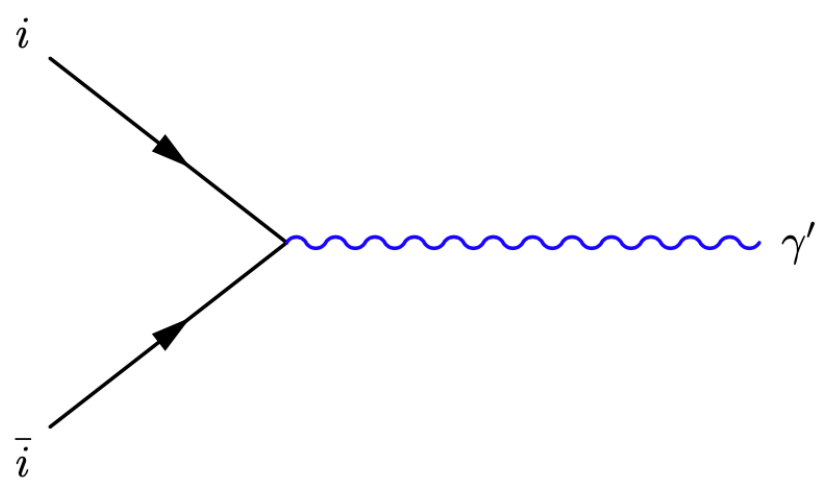}
\caption{
A display of dark photon three-point production channel, when the dark photon mass is heavier than at least twice of electrons.
As the mass of the dark photon increases, more combination channels involving SM fermions will become accessible.}
\label{fig:FeydpFI}
\end{center}
\end{figure}

\begin{figure}
\begin{center}
\includegraphics[scale=0.7]{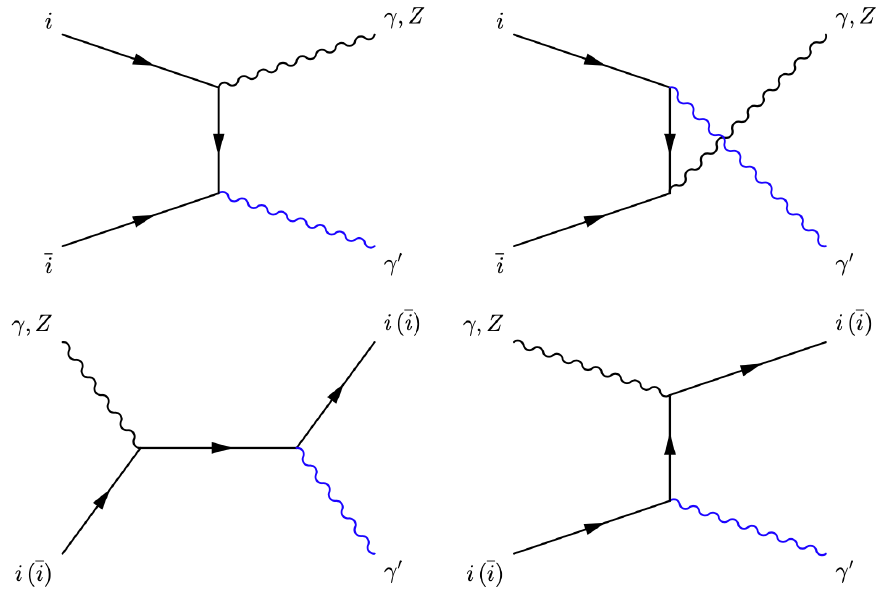}
\caption{A display of dark photon four-point production channels, including $ i \bar{i} \to \gamma^\prime + \gamma,Z$
and $ i (\bar{i}) + \gamma,Z \to i (\bar{i})  \gamma^\prime$, although interactions involving the $Z$ boson are suppressed.
Crucially, we noticed in our analysis that
even the three-point productions shown in Fig.~\ref{fig:FeydpFI} are available,
the contribution from four-point production channels are still quite significant
and should not be neglected.}
\label{fig:FeydpFI4}
\end{center}
\end{figure}

\begin{figure}
\begin{center}
\includegraphics[scale=0.3]{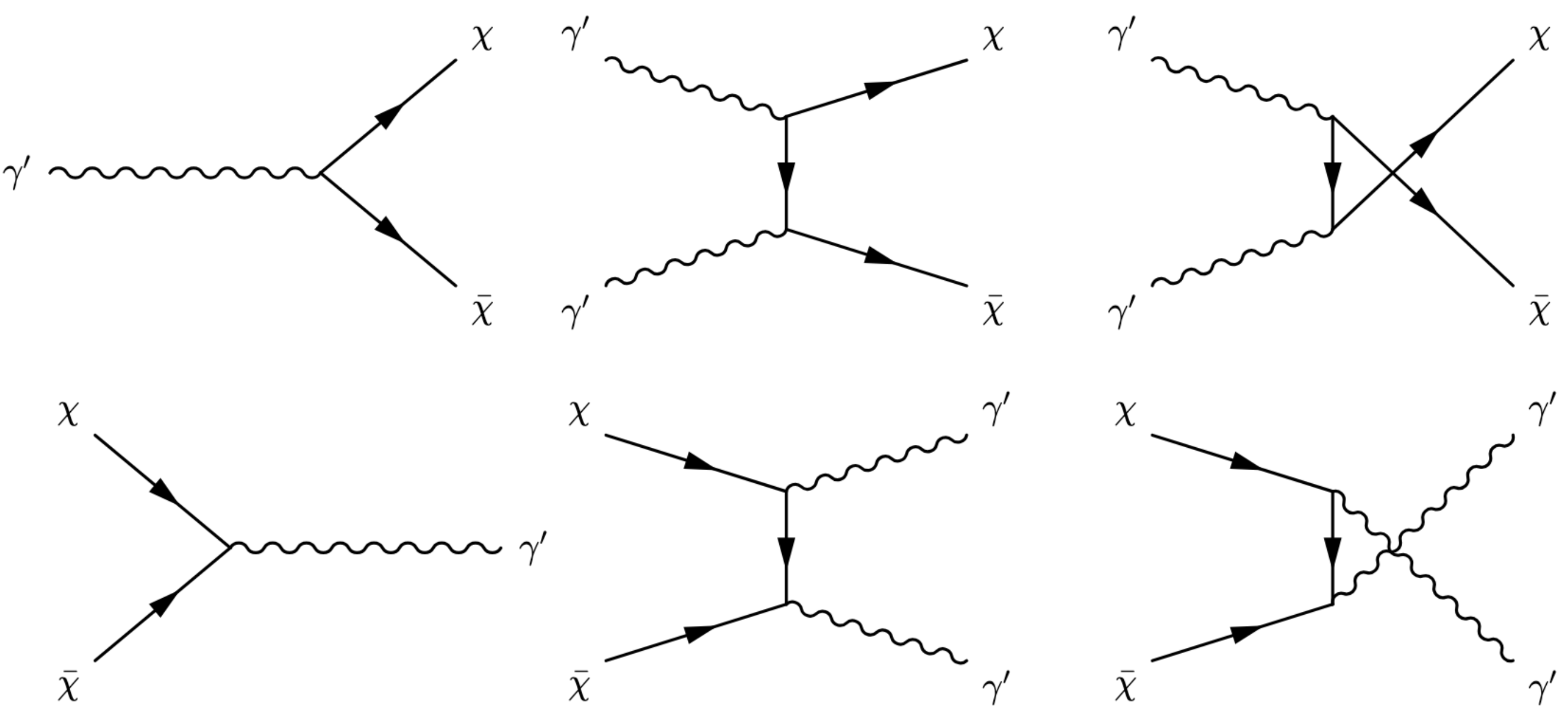}
\caption{
A display of various hidden sector self-interactions with particle number changing,
including: $\chi \bar{\chi} \leftrightarrow \gamma^\prime$,
$\chi \bar{\chi} \leftrightarrow \gamma^\prime \gamma^\prime$.
The hidden sector can encounter dark matter freeze-out depending on the masses of the dark particles,
discussed explicitly in Section~\ref{sec:EoM} for six different cases.}
\label{fig:FeyHSSelf}
\end{center}
\end{figure}

All relevant freeze-in processes for the dark fermion $\chi$ are shown in Fig.~\ref{fig:FeyChiFI}.
The freeze-in processes for the dark photon $\gamma^\prime$ are shown in Fig.~\ref{fig:FeydpFI},
if $M_{\gamma^\prime}$ is greater than twice of electron so that the three-point processes
via SM fermion combination $i\bar{i} \to \gamma^\prime$ can occur.
If $M_{\gamma^\prime}<2m_e$, since the partial decay widths of dark photon to neutrinos are highly suppressed,
c.f., Eq.~(\ref{eq:GPnunu}),
the feasible production channels can only be the four-point interactions, shown in Fig.~\ref{fig:FeydpFI4}.
We notice here that the production channel of $ i \bar{i} \to \gamma^\prime \gamma^\prime$ are suppressed,
compared to $ i \bar{i} \to \gamma\gamma^\prime$.
Four-point freeze-in production via interactions
$ i \bar{i} \to Z \gamma^\prime$,
$i Z \rightarrow i\gamma^{\prime}$ and $\bar{i} Z \rightarrow\bar{i}\gamma^{\prime}$
are also suppressed compared with the same interactions with $Z$ replaced by the photon.
We also collect all the hidden sector self-interactions in Fig.~\ref{fig:FeyHSSelf},
depending on the masses of the dark fermion and the dark photon.

We then analyze the evolution of $n_{\chi}$, $n_{\gamma^{\prime}}$
and $\eta$ as a function of $T_{h}$. For the computation of the
relic density, it is more convenient to deal directly with the comoving
number densities defined by the number density of a species $a$ divided
by the entropy $s$, i.e., $Y_{a}=n_{a}/s$.
We assume there is no initial abundance for the dark particles $\chi,\gamma^{\prime}$
and they are produced through freeze-in processes from SM particles.
The hidden sector particles can
have rather strong interactions such as $\chi\bar{\chi}\leftrightarrow\gamma^{\prime}\gamma^{\prime}$
within the hidden sector.
Thus for the case that $U(1)_{X}$ hidden sector interacts with the visible sector
with feeble couplings and evolves almost independently with the temperature $T_{h}=T/\eta$,
the evolution equations of the dark fermion $\chi$ and the dark photon $\gamma^{\prime}$
are given by the coupled Boltzmann equations for comoving number densities
$Y_{\chi}$ and $Y_{\gamma'}$ and $\eta$:
\begin{align}
\frac{{\rm d}Y_{\chi}}{{\rm d}T_{h}}= & -\frac{s}{H}\frac{{\rm d}\rho_{h}/{\rm d}T_{h}}{4\rho_{h}-j_{h}/H}\sum_{i\in{\rm SM}}\Big\{
{\color{Green} {(Y_{\chi}^{{\rm eq}})^{2}\langle\sigma v\rangle_{\chi\bar{\chi}\to i\bar{i}}^{T_{h}\eta}+\frac{1}{s}Y_{\gamma^{*}}\langle\Gamma\rangle_{\gamma^{*}\rightarrow\chi\bar{\chi}}^{T_{h}\eta}}}\nonumber \\
 & +\theta(M_{\gamma^{\prime}}-2m_{\chi})\big[-Y_{\chi}^{2}\langle\sigma v\rangle_{\bar{\chi}\chi\rightarrow\gamma^{\prime}}^{T_{h}}+\frac{1}{s}Y_{\gamma^{\prime}}\langle\Gamma\rangle_{\gamma^{\prime}\to\chi\bar{\chi}}^{T_{h}}\big]
 \nonumber \\
 &  -Y_{\chi}^{2}\langle\sigma v\rangle_{\chi\bar{\chi}\to\gamma^{\prime}\gamma^{\prime}}^{T_{h}}+Y_{\gamma^{\prime}}^{2}\langle\sigma v\rangle_{\gamma^{\prime}\gamma^{\prime}\to\chi\bar{\chi}}^{T_{h}}\Big\}\,,\label{eq:YchiBol}\\
\frac{{\rm d}Y_{\gamma^{\prime}}}{{\rm d}T_{h}}= & -\frac{s}{H}\frac{{\rm d}\rho_{h}/{\rm d}T_{h}}{4\rho_{h}-j_{h}/H}\sum_{i\in{\rm SM}}\Big\{ Y_{\chi}^{2}\langle\sigma v\rangle_{\chi\bar{\chi}\to\gamma^{\prime}\gamma^{\prime}}^{T_{h}}-Y_{\gamma^{\prime}}^{2}\langle\sigma v\rangle_{\gamma^{\prime}\gamma^{\prime}\to\chi\bar{\chi}}^{T_{h}}\nonumber \\
 & +\theta(M_{\gamma^{\prime}}-2m_{\chi})\big[Y_{\chi}^{2}\langle\sigma v\rangle_{\bar{\chi}\chi\rightarrow\gamma^{\prime}}^{T_{h}}-\frac{1}{s}Y_{\gamma^{\prime}}\langle\Gamma\rangle_{\gamma^{\prime}\to\chi\bar{\chi}}^{T_{h}}\big]\nonumber \\
 & + {\color{Green} \theta(M_{\gamma^{\prime}}-2m_{i})\big[Y_{i}^{2}\langle\sigma v\rangle_{i\bar{i}\rightarrow\gamma^{\prime}}^{T_{h}\eta}}-\frac{1}{s}Y_{\gamma^{\prime}}\langle\Gamma\rangle_{\gamma^{\prime}\to i\bar{i}}^{T_{h}}\big] \nonumber \\
 & + {\color{Green} Y_{i}^{2}\langle\sigma v\rangle_{i\bar{i}\rightarrow\gamma^{\prime}\gamma}^{T_{h}\eta}+2Y_{i}Y_{\gamma^{\prime}}^{{\rm eq}}\langle\sigma v\rangle_{i\gamma^{\prime}\rightarrow i\gamma}^{T_{h}\eta}} \Big\}\,,\label{eq:YDPBol}\\
\frac{{\rm d}\eta}{{\rm d}T_{h}}= & -\frac{A_{v}}{B_{v}}+\frac{\rho_{v}+j_{h}/(4H)}{\rho_{h}-j_{h}/(4H)}\frac{\frac{{\rm d}\rho_{h}}{{\rm d}T_{h}}}{B_{v}}\,, \label{eq:etaBol}
\end{align}
where the energy transfer density (the source term of the hidden sector)
is given by
\begin{align}
j_{h}= & \sum_{i\in{\rm SM}}\Big[2Y_{i}^{2}s^{2}J_{i\bar{i}\rightarrow\chi\bar{\chi}}^{T}+2Y_{\gamma^{*}}sJ_{\gamma^{*}\rightarrow\chi\bar{\chi}}^{T}+Y_{i}^{2}s^{2}J_{i\bar{i}\rightarrow\gamma^{\prime}\gamma}^{T}+2Y_{i}Y_{\gamma^{\prime}}^{{\rm eq}}s^{2}J_{i\gamma^{\prime}\rightarrow i\gamma}^{T}\nonumber \\
 & +\theta(M_{\gamma^{\prime}}-2m_{i})\big(Y_{i}^{2}s^{2}J_{i\bar{i}\rightarrow\gamma^{\prime}}^{T}-Y_{\gamma^{\prime}}sJ_{\gamma^{\prime}\rightarrow i\bar{i}}^{T_{h}}\big)\Big]\,,
\end{align}
and
\begin{align}
Y_{a}^{\mathrm{eq}}(T)Y_{b}^{\mathrm{eq}}(T)s^{2}J(ab\rightarrow cd)(T) & =\frac{g_{a}g_{b}}{128\pi^{4}}T \int_{s_{0}}^{\infty}ds\sigma_{cd\rightarrow ab}s\left(s-s_{0}\right)K_{2}(\sqrt{s}/T)\,,\\
Y_{a}J_{a\rightarrow bc} & =Y_{a}m_{a}\Gamma_{a\rightarrow bc}\,.
\end{align}
The cross-section is given by
\begin{align}
	\sigma_{ab\to cd} & =\frac{1}{g_{a}g_{b}}\frac{1}{2E_{a}2E_{b}\left|v_{a}-v_{b}\right|}\int\mathrm{d}\Pi_{c}\mathrm{d}\Pi_{d}\left|M\right|_{ab\to cd}^{2}\left(2\pi\right)^{4}\delta^{4}\left(p_{a}+p_{b}-p_{c}-p_{d}\right)\,,
\end{align}
where $\left|M\right|_{ab\to cd}^{2}$ is {\it summed over} initial and final spins,
the momentum phase space element is defined as
\begin{equation}
	\mathrm{d}\Pi_{i}\equiv\frac{\mathrm{d}^{3}\mathbf{p}_{i}}{\left(2\pi\right)^{3}2E_{i}}\,,
\end{equation}
and the thermally averaged cross-section reads
\begin{align}
	\left\langle \sigma v\right\rangle _{ab\to cd}^{T}\equiv & \frac{\int\mathrm{d}\Pi_{a}\mathrm{d}\Pi_{b}\,\sigma ve^{-E_{a}/T}e^{-E_{b}/T}}{\int\mathrm{d}\Pi_{a}\mathrm{d}\Pi_{b}e^{-E_{a}/T}e^{-E_{b}/T}}\,.
\end{align}
We now inspect some specific details concerning coefficients that require attention
when solving the Boltzmann equations.
\begin{itemize}
	\item For processes involving SM fermions,
  one needs to consider all types of fermions along with their corresponding anti-fermions.
  Take the process $i\bar{i}\to\gamma^{\prime}$ as an example,
  in the calculation we need the value of $n_{q_{i}}n_{\bar{q}_{i}}$ which is derived as
	\begin{align}
		n_{q_{i}}n_{\bar{q}_{i}}= 3g_{q}g_{\bar{q}}\left[\frac{m_{i}^{2}}{2\pi^{2}}TK_{2}\left( \frac{m_{i}}{T}\right)\right]^{2}
		=\left(3\times2\times2\right)\times\left[\frac{m_{i}^{2}}{2\pi^{2}}TK_{2}\left( \frac{m_{i}}{T} \right)\right]^{2}\,,
	\end{align}
	where the factor 3 indicates three colors of quarks annihilating with anti-quarks carrying the same color,
for all six flavors of quarks.
This calculation is analogous for leptons without accounting for the color factor.
	\item When calculating the energy density and pressure, contributions from
	both $\chi$ and $\bar{\chi}$ should be considered. Hence in Eq.~(\ref{rhop}) we have
	\begin{equation}
		g_{\text{\ensuremath{\chi}}}^{\mathrm{total}}=g_{\chi}+g_{\bar{\chi}}=4\,,
	\end{equation}
  and we omit the subscript ``total'' for simplicity.
	\item Since we consider a model with symmetric component of dark matter,
  we have $Y_\chi = Y_{\bar{\chi}}$ with $g_{\chi}=g_{\bar{\chi}}=2$.
  The evolution of $\bar{\chi}$ is identical to the evolution of $\chi$ in the hidden sector,
  and thus the total dark fermion number density in a comoving volume is $2Y_\chi$.
\end{itemize}
All types of thermally averaged cross-sections for various processes
are summarized below:
\begin{enumerate}
\item Fermion annihilation $a\bar{a}\rightarrow bc$. SM fermions or hidden
sector fermions can annihilate to a pair of fermion and anti-fermion,
or to two gauge bosons, which can be universally expressed as
\begin{equation}
\langle\sigma v\rangle_{a\bar{a}\rightarrow bc}^{T}=\frac{1}{8m_{a}^{4}TK_{2}^{2}\left(m_{a}/T\right)}\int_{4m_{a}^{2}}^{\infty}
\mathrm{d} s\,\sigma(s)\sqrt{s}\left(s-4m_{a}^{2}\right)K_{1}(\sqrt{s}/T)\,.
\end{equation}
\item Fermion elastic scattering $ab\rightarrow cd$. A fermion scatters
off a gauge boson, producing a fermion with a gauge boson
\begin{equation}
\begin{aligned}\langle\sigma v\rangle_{ab\rightarrow cd}^{T} & =\frac{1}{8m_{a}^{2}m_{b}^{2}TK_{2}\left(m_{a}/T\right)K_{2}\left(m_{b}/T\right)}\times\\
 & \quad\quad\quad\int_{\left(m_{a}+m_{b}\right)^{2}}^{\infty} \mathrm{d}s\,\sigma(s)\sqrt{s}\left[s-\left(m_{a}+m_{b}\right)^{2}\right]K_{1}\left(\sqrt{s}/T\right)\,.
\end{aligned}
\end{equation}
\item Neutral gauge boson decay $a\rightarrow b\bar{b}$. A neutral gauge
boson ($Z,\gamma^{\prime}$ and plasmon $\gamma^{*}$) can decay into
a pair of fermion and anti-fermion
\begin{equation}
\left\langle \Gamma\right\rangle _{a\rightarrow b\bar{b}}^{T}=\Gamma_{a\rightarrow b\bar{b}}\frac{K_{1}\left(m_{a}/T\right)}{K_{2}\left(m_{a}/T\right)}\,.
\end{equation}
\item Fermion pair combination $b\bar{b}\rightarrow a$.
A pair of fermion and anti-fermion combines into a massive gauge boson
\begin{equation}
\langle\sigma v\rangle_{b\bar{b}\rightarrow a}^{T}=\frac{\sigma^{0}_{b\bar{b}\to a}
(m_{a}^{2})m_{a}\left(m_{a}^{2}-4m_{b}^{2}\right)K_{1}\left(m_{a}/T\right)}{8m_{b}^{4}TK_{2}^{2}\left(m_{b}/T\right)}\,,
\end{equation}
with $\sigma^{0}_{b\bar{b}\to a}$ representing the part without the delta function in the cross-section $\sigma_{b\bar{b}\rightarrow a}(s)$.
\end{enumerate}
In the above equations, $K_{1}$ and $K_{2}$ are the modified Bessel
function of the second kind and degrees one and two, respectively.
The cross-sections of all relevant processes discussed in this work
are given in Appendix~\ref{sec:APPScatter}.

Given a set of values $\{m_\chi, M_{\gamma^\prime}, g_X, \delta, \epsilon\}$ and the initial value of $\eta = T/T_h$,
one can calculate the full evolution of the hidden sector particles as well as the change of the hidden sector temperature.
The initial value of $\eta = T/T_h$ will not modify the final abundance so much
but will alter the details of the dark particles evolution,
as discussed in~\cite{Aboubrahim:2020lnr,Li:2023nez}.
A complete analysis for
the evolution of a $U(1)$ hidden sector interacting feebly with the SM
from the reheating stage, including the determination of the initial value of $\eta$,
is discussed in a companion paper~\cite{FandZh}.

In the calculation, the expansion of non-relativistic thermally averaged cross-section is useful, and is given in~\cite{Gondolo:1990dk} as
\begin{equation}
\langle\sigma v\rangle =a^{(0)}+\frac{3}{2}a^{(1)}x^{-1} + \frac{15}{8} a^{(2)}  x^{-2} + \cdots
\end{equation}
at low temperatures,
where $x=m/T$, and $a^{(n)}$ indicates the $n$th derivative of $\sigma v_{\rm lab}$ with respect to
$\kappa = (s-4m^2)/4m^2$  ($m$ is the mass of the incoming particle) evaluated at $\kappa=0$.
Similarly for the thermally averaged decay width, one has the expansion at low temperatures,
\begin{equation}
\langle\Gamma\rangle_{a\rightarrow bc}=\Gamma_{a\rightarrow bc}\left(1-\frac{3}{2}x^{-1}+\frac{15}{8}x^{-2}+\cdots\right)\,.
\end{equation}

\section{Evolution of matter in visible and hidden sectors }\label{sec:EoM}

\subsection{$U(1)_{X}$ sub-GeV dark matter via freeze-in}\label{sec:FI}

In this section, we discuss the production of $U(1)_{X}$ sub-GeV
millicharge dark matter $\chi$ and dark photon $\gamma^{\prime}$ through freeze-in mechanism.
In some of the cases, the dark photon can also be stable and serve as a dark matter candidate,
although we will see its abundance is usually tiny. As discussed in Section~\ref{sec:mixing},
millicharge dark matter can be produced via Stueckelberg mass mixing,
with a millicharge proportional to $\epsilon=M_{2}/M_{1}$. In addition
to the freeze-in production processes, there are also quite strong
interactions within the hidden sector like $\chi\bar{\chi}\leftrightarrow\gamma^{\prime}\gamma^{\prime}$,
and $\chi\bar{\chi}\leftrightarrow\gamma^{\prime}$ if the dark photon
is heavier than twice of $\chi$ mass. Since the couplings between
the visible sector and the hidden sector are feeble, the hidden sector
temperature $T_{h}$ evolves almost independently from the visible
sector temperature $T$ (the temperature of the Universe). As the
freeze-in processes and the hidden sector interactions (depending
on the hidden sector temperature $T_{h}$) occur simultaneously, we
use the formalism discussed in Section~\ref{sec:BEMT} to calculate the full evolution
of the hidden sector particles as well as the hidden sector temperature.

We will now explore six distinct scenarios upon the masses of dark
matter $\chi$ and the dark photon $\gamma^{\prime}$, each exhibiting
markedly unique characteristics in the evolution of the hidden sector.
The common feature of all these scenarios is that they have the same
freeze-in production channel:
\begin{itemize}
\item \textbf{Dark fermion $\chi$ freeze-in production channels:} SM fermion
annihilation $i\bar{i}\to\chi\bar{\chi}$, plasmon decay $\gamma^{*}\to\chi\bar{\chi}$.
The $Z$ boson decay channel $Z\to\chi\bar{\chi}$ is suppressed,
since the abundance of $Z$ in the early universe is Boltzmann suppressed
at the temperature when the dark fermion $\chi$ is majorly produced
(around the mass of $\chi$ in the sub-GeV region).
\item \textbf{Dark photon $\gamma^{\prime}$ freeze-in production channels:}
For $M_{\gamma^{\prime}}>2m_{e}$, the major production channel is the
three-point channels $i\bar{i}\to\gamma^{\prime}$
and four-point channels $i\bar{i}\to\gamma\gamma^{\prime}$,
$i\gamma\to i\gamma^{\prime}$, $\bar{i}\gamma\to\bar{i}\gamma^{\prime}$.
As found in our calculation, the four-point freeze-in channels are
as important as the three-point processes, and {\it must be taken into account at all times.}
For $M_{\gamma^{\prime}}<2m_{e}$, the production channels are $i\bar{i}\to\gamma\gamma^{\prime}$,
$i\gamma\to i\gamma^{\prime}$, $\bar{i}\gamma\to\bar{i}\gamma^{\prime}$.
We emphasize here again that in the model the couplings of the dark
photon to neutrinos are highly suppressed, c.f., Eq.~(\ref{eq:GPnunu}), and thus
either the decay of dark photon to neutrinos or the freeze-in production
via neutrinos combination can be safely neglected.
\end{itemize}
We firstly discuss the case that the dark photon mass is greater than twice
of the electron mass, i.e., $M_{\gamma^{\prime}}>2m_{e}$. In this
case, the dark fermion $\chi$ is millicharged and serves as the dark
matter candidate. The dark photons, due to the mixing of $U(1)_{X}$
and the SM, will ultimately undergo decay into electron and positron
pairs, and are thus not dark matter candidates. The dark photon also
has to decay before BBN for a valid model. Within this category we
discuss three different cases:

\paragraph*{Case 1: $m_{\chi}>M_{\gamma^{\prime}}>2m_{e}$, dark matter $\chi$}

In this case hidden sector particles can interchange through the interactions
$\chi\bar{\chi}\leftrightarrow\gamma^{\prime}\gamma^{\prime}$ (dark
matter self-interactions such as $\chi\bar{\chi}\leftrightarrow\chi\bar{\chi}$
and $\chi\chi\leftrightarrow\chi\chi$ will not change the number
densities of dark fermion and dark photon and we will not discuss
at this point). When the temperature drops down, the hidden sector
encounters the dark fermion $\chi$ freeze-out. The dark photon will
eventually decay into SM fermions through $\gamma^{\prime}\to i\bar{i}$ (mostly
to electron and positron pairs).

\paragraph*{Case 2: $2m_{\chi}>M_{\gamma^{\prime}}>m_{\chi}>2m_{e}$, dark matter
$\chi$}

In this case hidden sector particles interchange through the interactions
$\gamma^{\prime}\gamma^{\prime}\leftrightarrow\chi\bar{\chi}$. When
the temperature drops down, the hidden sector encounters the dark
photon $\gamma^{\prime}$ freeze-out. The leftover of the dark photon
finally decay to SM fermions through $\gamma^{\prime}\to i\bar{i}$.

\paragraph*{Case 3: $M_{\gamma^{\prime}}>2m_{\chi}>2m_{e}$, dark matter $\chi$}

In this case hidden sector particles interchange mostly through the
three-point interactions $\gamma^{\prime}\leftrightarrow\chi\bar{\chi}$.
Because of the rather strong interactions among the hidden sector,
the dark photon will soon decay to $\chi\bar{\chi}$ when the temperature
drops down. Four-point interactions
$\chi\bar{\chi} \leftrightarrow \gamma^{\prime}\gamma^{\prime}$ can be neglected in this case.
In addition, the decay channels of dark photon to SM fermions can also
be neglected since the decay of $\gamma^{\prime}\to\chi\bar{\chi}$
is so much stronger.\\

We then discuss the case that the dark photon mass is less than twice of
the electron mass, i.e., $M_{\gamma^{\prime}}<2m_{e}$. Again the
dark fermion $\chi$ is millicharged and serves as the dark matter
candidate. Since the decay channel $\gamma^{\prime}\to e\bar{e}$
is forbidden, the dark photon can only decay to two neutrinos and
to three photons which are usually quite rare, c.f., Fig.~\ref{fig:FeyZpDecay}.
Focusing on these two
interactions, by tuning the $U(1)_{X}$ gauge coupling as well as
the kinetic mixing parameter $\delta$ and mass mixing parameter $\epsilon$,
one can make the photon stable throughout the entire age of the Universe,
establishing the dark photon as a viable dark matter candidate. However,
although the dark photon's lifetime is extended beyond the age of
the Universe, it can still undergo decay, even in minuscule amounts.
This decay contributes to the isotropic diffuse photon background
(IDPB), which can be precisely measured. The IDPB imposes an even
more stringent constraint on the couplings between the dark
photon and SM fermions. We incorporate this constraint into our analysis
and discover that it further diminishes the $\delta,\epsilon$ when
compared to the constraints posed by dark photon decay.

Within the category $M_{\gamma^{\prime}}<2m_{e}$ we also discuss
three different cases:\footnote{For the models with $M_{\gamma^{\prime}}<2m_{e}$ and $M_{\gamma^{\prime}}<2m_{\chi}$, one cannot make
the dark photon unstable and decaying before BBN considering all the
relevant experimental constraints.}

\paragraph*{Case 4: $2m_{e}>m_{\chi}>M_{\gamma^{\prime}}$, dark matter $\chi,\gamma^{\prime}$}

The hidden sector evolves through the interactions $\chi\bar{\chi}\leftrightarrow\gamma^{\prime}\gamma^{\prime}$,
and the freeze-out of $\chi$ will occur when the temperature drops.
The dark photon $\gamma^{\prime}$ remains stable throughout the age of the Universe
and receives stringent constraints from both its decay width as well as the IDPB.

\paragraph*{Case 5: $2m_{e}>2m_{\chi}>M_{\gamma^{\prime}}>m_{\chi}$, dark matter
$\chi,\gamma^{\prime}$}

The hidden sector evolves through the interactions $\gamma^{\prime}\gamma^{\prime}\leftrightarrow\chi\bar{\chi}$,
and the freeze-out of $\gamma^{\prime}$ will occur when the temperature
drops. The dark photon $\gamma^{\prime}$ is stable and the leftover
of $\gamma^{\prime}$ cannot decay. The dark photon $\gamma^{\prime}$
again receives stringent constraints from both its decay width as
well as the IDPB.

\paragraph*{Case 6: $2m_{e}>M_{\gamma^{\prime}}>2m_{\chi}$, dark matter $\chi$}

For this case the hidden sector evolves (mostly) via three-point interactions $\gamma^{\prime}\leftrightarrow\chi\bar{\chi}$.
With a relatively large $U(1)_{X}$ gauge coupling constant $g_{X}$,
the decay $\gamma^{\prime}\to\chi\bar{\chi}$ is rapid.
When the temperature drops down, all the dark photon will decay into dark fermions.
The evolution of the dark sector in this case is almost identical to Case 3.\\

Upon all the above six distinct cases, the interactions among the dark particles
play very important roles in the hidden sector evolution.
As the dark particles $\chi$ and $\gamma^\prime$ are accumulating via the freeze-in production,
at some moment the hidden sector particles may reach thermal equilibrium within the hidden sector,
since the hidden sector (gauge) coupling constant is much stronger than the freeze-in couplings.
Depending on the masses of  $\chi$ and $\gamma^\prime$,
the three-point interactions $\chi\bar{\chi}\leftrightarrow\gamma^{\prime}$ and
the four-point interactions $\chi\bar{\chi}\leftrightarrow\gamma^{\prime}\gamma^{\prime}$ become active.
For Case 3 and Case 6, the three-point interactions $\chi\bar{\chi}\leftrightarrow\gamma^{\prime}$ are allowed,
which overwhelm the four-point interactions $\chi\bar{\chi}\leftrightarrow\gamma^{\prime}\gamma^{\prime}$.
Eventually, the dark photon will entirely decay to $\chi \bar{\chi}$, leaving the dark fermion $\chi$ as the dark matter.
For Case 1 and Case 4, in which $m_\chi > M_{\gamma^\prime}$,
the dark freeze-out for $\chi$ will occur in the hidden sector.
As the interaction $\chi\bar{\chi}\to\gamma^{\prime}\gamma^{\prime}$ goes out of equilibrium,
the dark freeze-out ended and the amount of the dark fermion $\chi $ freezes at the final value.
While for Case 2 and Case 5, in which $2m_\chi > M_{\gamma^\prime} > m_\chi $,
the dark freeze-out for $\gamma^\prime$ will occur in the hidden sector.
The dark freeze-out will finish when the interaction $\gamma^{\prime}\gamma^{\prime} \to \chi\bar{\chi}$
goes out of equilibrium, the leftover of the dark photon will either decay to electron-positron pair if $M_{\gamma^\prime} > 2m_e$ (Case 2)
or will remain as another component of dark matter if $M_{\gamma^\prime} < 2m_e$ (Case 5)
and in the latter case its lifetime exceeds the age of the Universe.

\subsection{Constraints, benchmark models and phenomenology}\label{sec:pheno}

\begin{figure}
\begin{center}
\includegraphics[scale=0.5]{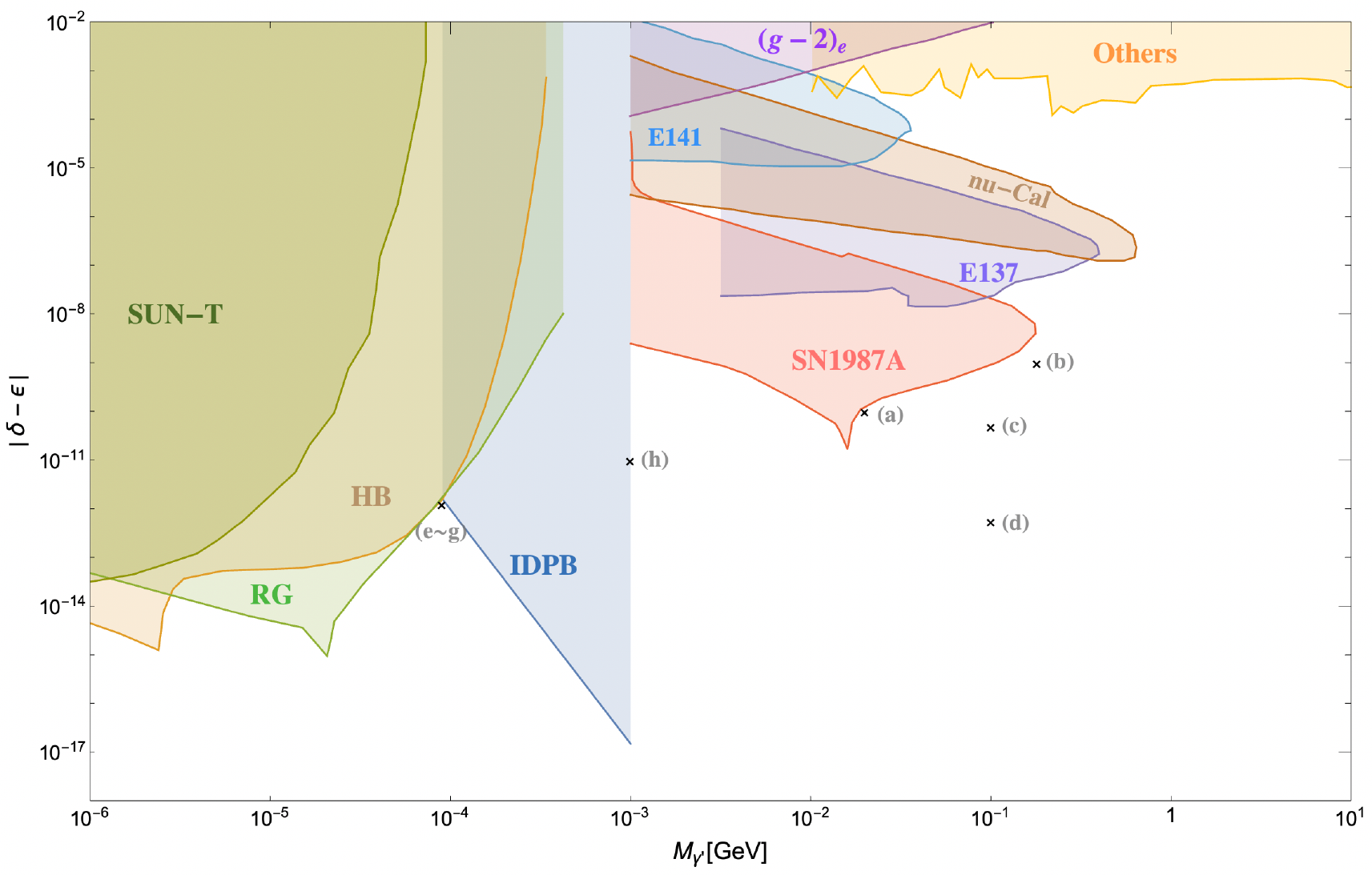}
\caption{[color online] A display of current constraints (colored regions) on the absolute value of
the kinetic mixing parameter minus the mass mixing parameter $|\delta - \epsilon|$,
translated from the current constraints on the simplified kinetic mixing model people usually referenced,
reviewed in Section~\ref{sec:mixing2U1}, using Eq.~(\ref{eq:dmevsd}).
The limits for $M_{\gamma^{\prime}}\gtrsim$~1~MeV come from supernova (SN1987A)~\cite{Chang:2016ntp}, $(g-2)_{e}$~\cite{Pospelov:2008zw} and from various beam dump experiments: E141~\cite{Riordan:1987aw}, E137~\cite{Bjorken:1988as,Batell:2014mga,Marsicano:2018krp}, $\nu$-Cal~\cite{Blumlein:2011mv,Blumlein:2013cua}.  The light orange area in the upper right labeled ``Others'' includes some collider/fixed target experiments:  CMS~\cite{CMS:2019kiy}, BaBar~\cite{BaBar:2014zli}, LHCb~\cite{LHCb:2019vmc}, KLOE~\cite{KLOE-2:2011hhj,KLOE-2:2012lii,KLOE-2:2014qxg,KLOE-2:2016ydq} and A1~\cite{Merkel:2014avp}.
For the dark photon mass in the 1 keV -- 1 MeV window, the shaded regions depict the existing constraints from solar lifetime (Sun-T), red giants (RG), horizontal branches (HB)~\cite{Redondo:2013lna,An:2013yfc,Hardy:2016kme} and from isotropic diffuse photon background (IDPB)~\cite{Yuksel:2007dr,Redondo:2008ec,An:2014twa}. All our benchmark models are also shown in the plot.}
\label{fig:debound}
\end{center}
\end{figure}

First of all we discuss all the relevant experimental constraints for the models we consider in this work.
In the literature, people usually assume a kinetic mixing between an additional $U(1)_X$ and the $U(1)_{\rm em}$
with a mixing parameter $\boldsymbol{\delta}$,
which is different from the kinetic mixing parameter $\delta$ in our work indicating the kinetic mixing between
$U(1)_X$ and the hypercharge gauge field $U(1)_Y$.
As experiments setting constraints on the kinetic mixing parameter mostly from the dark photon interaction with
electron and positron pair, we will have a translated relation of the current experimental constraints on our kinetic mixing parameter $\delta$.
The coupling of the dark photon to electron and positron pair through the Lagrangian from Eqs.~(\ref{eq:2U1kinM}) and (\ref{eq:2U1kinMint})
is computed to be
\begin{align}
     \mathcal{L}_{\gamma^{\prime}e\bar{e}}&=-\varepsilon eJ^{\mu}A_{\mu}^{\prime}
     =-\frac{g_{2}g_{Y}}{\sqrt{g_{2}^{2}+g_{Y}^{2}}}\boldsymbol{\delta} \bar{e}\gamma^{\mu}eA_{\mu}^{\prime}\,.
     \label{eq:OLDgpee}
\end{align}
While for the case that $U(1)_X$ mixed with $U(1)_{Y}$ via kinetic terms and also the Stueckelberg mass terms,
the coupling of the dark photon to electron and positron pair is given by Eq.~(\ref{eq:StNGBc})
\begin{equation}
     \mathcal{L}_{\gamma^{\prime}e\bar{e}}=-\frac{1}{2}\bar{e}\gamma^{\mu}\left(v_{e}^{\prime}-a_{e}^{\prime}\gamma^{5}\right)eA_{\mu}^{\prime}
     =-\frac{g_{2}^{2}g_{Y}^{\prime}}{g_{2}^{2}+g_{Y}^{\prime2}}\left(\delta-\epsilon\right)\bar{e}\gamma^{\mu}eA_{\mu}^{\prime}\,.
     \label{eq:NEWgpee}
\end{equation}
Comparing Eqs. (\ref{eq:OLDgpee}) and (\ref{eq:NEWgpee}), one finds the relation
\begin{align}
    \left(\delta-\epsilon\right)
    =\frac{g_{Y}\left(g_{2}^{2}+g_{Y}^{\prime2}\right)}{g_{2}g_{Y}^{\prime}\sqrt{g_{2}^{2}+g_{Y}^{2}}}\boldsymbol{\delta}
  \approx\frac{\sqrt{g_{2}^{2}+g_{Y}^{2}}}{g_{2}}\boldsymbol{\delta}\,,
  \label{eq:dmevsd}
\end{align}
where $g_{Y}^{\prime} \approx g_{Y}$ for small $\delta$,\,$\epsilon$.
Thus the experimental constraints set on $\boldsymbol{\delta}$
(parameter for kinetic mixing between $U(1)_X$ and the $U(1)_{\rm em}$
in the simplified model discussed in Section~\ref{sec:mixing2U1}) can be translated to
a constraint on the parameter $|\delta - \epsilon|$ for the model we consider in this work, discussed in Section~\ref{sec:mixingSt}.
We plot the current experimental constraints on the kinetic mixing parameter $\delta$ minus mass mixing parameter $\epsilon$ in Fig.~\ref{fig:debound},
with all our benchmark models exhibited.

As discussed in Section~\ref{sec:mixingSt},
when the dark photon is less massive than twice of electron mass,
dark photons can only decay to three photons, and
to pairs of neutrino and anti-neutrino,
as shown in Fig.~\ref{fig:FeyZpDecay}.
Within the parameter values we employ for our benchmark models,
the decay of the dark photon to three photons is dominant over the neutrino decay channel.

Photons from the decay of the dark photon (even its lifetime exceeds the age of the Universe)
will affect the observed $\gamma$-ray
background and consequently impose stringent constraints on the parameter space.
Various bounds on the decaying dark
matter with a wide range of mass have been explored in details in~\cite{Chen:2003gz,Pospelov:2008jd,DeLopeAmigo:2009dc,Ando:2015qda,Liu:2016ngs,Blanco:2018esa},
including 2-body and 3-body decays.
Model-independent constraints were established in~\cite{Yuksel:2007dr}
by demanding that the photon flux from $\gamma^{\prime}\to3\gamma$ decays must not surpass
the measurements of the diffuse $\gamma$-ray background.
Specifically, constraints on dark photon dark matter in the keV-MeV mass range
were explored in~\cite{Redondo:2008ec},
requiring that the $\gamma$-ray flux from decaying dark photon
does not exceed the total observed flux of $\gamma$-ray:
\begin{equation}
	\frac{\mu\,\tau}{\mathrm{GeV\,s}}\lesssim10^{27}\left(\frac{\omega}{\rm GeV}\right)^{1.3}\left(\frac{\Omega h^{2}}{0.1}\right)\,,
\end{equation}
with $\mu,\,\tau,\,\omega$ representing mass, lifetime and the energy
of the decaying dark photon respectively. The lifetime $\tau$ can
be calculated using the decay width $\Gamma_{\gamma^{\prime}\to3\gamma}$
given by Eq.~(\ref{eq:Dw3ph}), and the corresponding constraint is plotted
in Fig.~\ref{fig:debound} indicated by ``IDPB''.

\begin{figure}[]
\begin{center}
\includegraphics[scale=0.5]{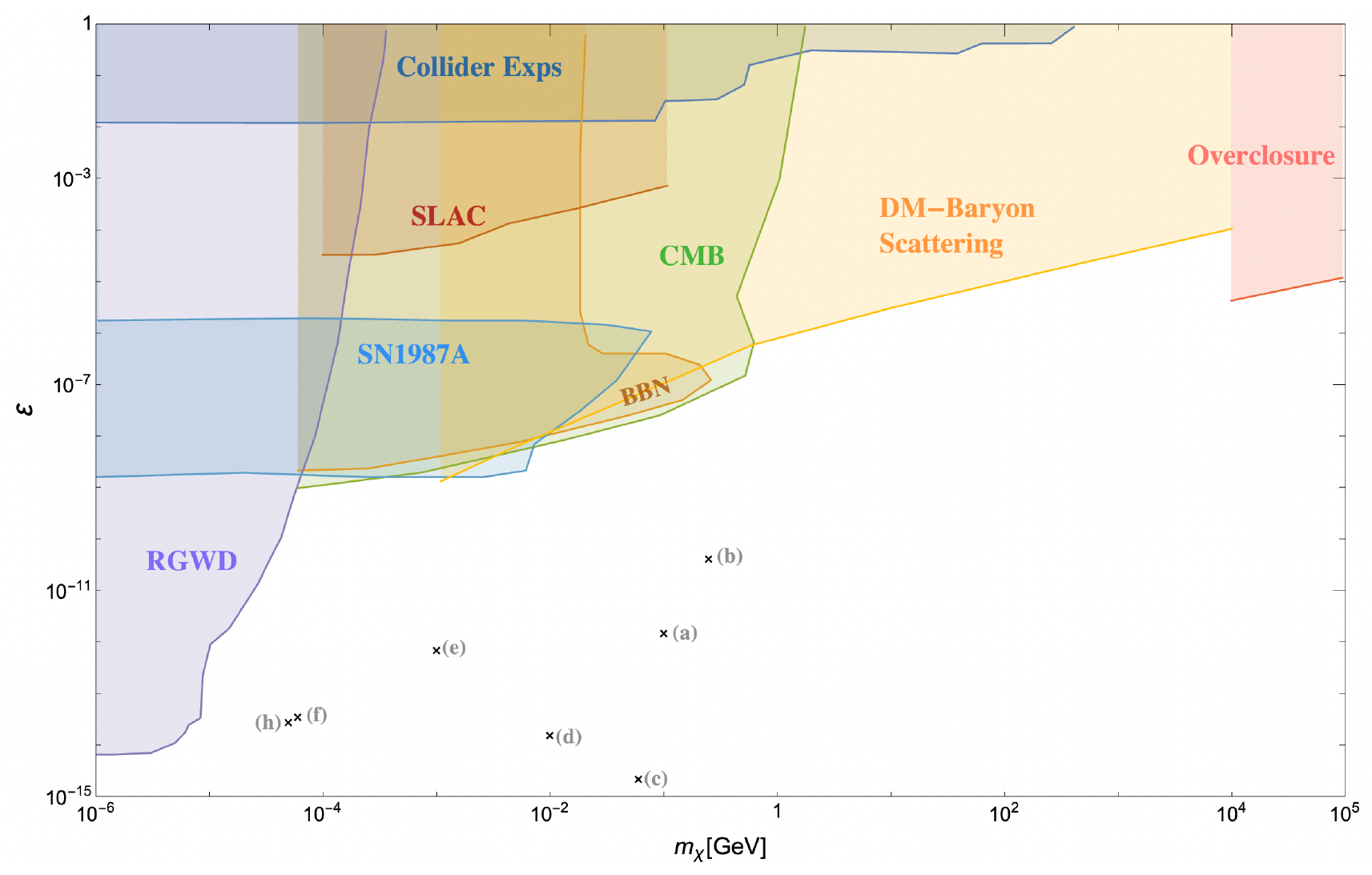}
\caption{[color online] A display of current constraints on the millicharge carried by the dark particle versus its mass.
The constraints arise from: red giants and white dwarfs (RGWD)~\cite{Vogel:2013raa,Fung:2023euv}, supernova (SN1897A)~\cite{Chang:2018rso}, SLAC milliQ experiment~\cite{Prinz:1998ua}, $N_{\rm eff}$ during BBN and CMB~\cite{Vogel:2013raa,Vinyoles:2015khy}, overclosure of the Universe~\cite{Davidson:1991si,Jaeckel:2012yz} and multifarious collider experiments such as LEP,  LHC,  etc~\cite{Davidson:2000hf,Jaeckel:2012yz}. We also show the limit on the elastic scattering  between baryons and millicharge dark matter given by~\cite{Xu:2018efh}.
Our benchmark models are also shown in the plot.}
\label{fig:millibound}
\end{center}
\end{figure}

People also searched millicharge particles as dark matter candidates and imposed constraints
from laboratory, astrophysical and cosmological experiments for a wide mass range of the millicharge particle.
Bounds from colliders including LEP, LHC, etc~\cite{Davidson:2000hf,Jaeckel:2012yz},
cover a wide mass range but are less restrictive compared to other constraints.
Astrophysical constraints involving red giants, horizontal-branch stars, white dwarfs and supernova 1987A,
are caused by new energy-loss channels which will lead to observational alterations to the standard stellar evolution process~\cite{Mohapatra:1990vq,Davidson:1991si,Davidson:1993sj,Davidson:2000hf,Badertscher:2006fm,Vogel:2013raa,Chang:2018rso}.
Additionally, particles carrying tiny electric charges have the possibility to interact with the plasma and get thermally excited in the early universe.
Both dark photons and millicharge particles in the hidden sector may
contribute to $N_{\rm eff}$, leading to deviations from the standard value,
and will thus receive constraints from cosmological observations~\cite{Steigman:1986nh,Davidson:1991si,Jaeckel:2008fi,Vogel:2013raa,Vinyoles:2015khy}.
Another important cosmological bound is from the overclosure of the Universe,
demanding the relic density of the millicharge dark matter to be less than the critical density today $\left(\Omega<1\right)$~\cite{Davidson:1991si,Jaeckel:2012yz}.
Limits on the elastic scattering cross-section between sub-GeV millicharge dark matter and baryons are given by~\cite{Xu:2018efh},
following the treatment of previous works~\cite{Cyburt:2002uw,Dvorkin:2013cea,Tashiro:2014tsa,Munoz:2015bca,Catena:2015uha,Gluscevic:2017ywp}
and using CMB temperature and polarization data~\cite{Planck:2015fie}
together with Lyman-$\alpha$ flux power spectrum data~\cite{SDSS:2004kjl}.
The most relevant experimental constraints on the millicharge carried by the dark particles are plotted in Fig.~\ref{fig:millibound},
with all our benchmark models exhibited in the allowed region.

We mention in passing that the (stable) dark photon will not contribute excessively to effective relativistic degree of freedom in the Universe in Case 4 and Case 5 of our benchmark models.
The dark photon contribution to $\Delta N_{\rm eff}$ is given by
\begin{equation}
\Delta N_{\rm eff} = \frac{8}{7} g_{\gamma^\prime} \left(\frac{T_h }{T_\gamma}\right)^4\,.
\end{equation}
At low temperatures, $g_{\gamma^\prime} = 3$ and ${T_h }/{T_\gamma} \ll 1$ and the above contribution is negligible.

Recall that the simplified model discussed in Section~\ref{sec:mixing2U1},
which considers the kinetic mixing between a (massless) $U(1)_X$ and the $U(1)_{\rm em}$.
In the mass eigenbasis the dark particle carries a millicharge $g_X Q_{\chi} \delta /\sqrt{1-\delta^2}$, c.f., Eq.~(\ref{eq:Int1}),
where $g_X$ is the $U(1)_X$ gauge coupling and $Q_\chi$ is the $U(1)_X$ charge carried by the dark fermion.

In the literature,  people usually takes the above expression for the millicharge carried by the dark fermion
even the $U(1)_X$  is not massless.
As we discussed in depth in Section~\ref{sec:mixing}, this treatment is inappropriate and indeed,
there will be {\it no millicharge generated at all}
for the case of the massless photon field mixing with an additional {\it massive} $U(1)$ via only kinetic terms.
Not mentioning considering  $U(1)_X$ mixing directly with the photon field,
which is a linear combination of $\{B,A_3\}$ in the electroweak theory,
is not rigorous in the first place.
Nevertheless, experiments can still set bounds directly to the millicharge $Q_{\boldsymbol{\epsilon}} $ (if there is any)
carried by the dark particle,  given by the expression
\begin{equation}
Q_{\boldsymbol{\epsilon}} = g_X Q_\chi \delta /\sqrt{1-\delta^2} \equiv \boldsymbol{\epsilon} e\,,
\end{equation}
for the simplified model discussed in Section~\ref{sec:mixing2U1}, and
in the literature bounds were set directly to the parameter $\boldsymbol{\epsilon}$.

Considering the mixing effects within the full electroweak theory,
where the $U(1)_X$ connects with $U(1)_Y$ through
both kinetic mixing and mass mixing,
the dark matter carries a millicharge given by Eq.~(\ref{eq:millicharge}),
\begin{equation}
Q_\varepsilon =  \epsilon g_X Q_\chi \cos \theta  \equiv \varepsilon e\,,
\label{eq:Qeandmm}
\end{equation}
giving rise to $\varepsilon =  \epsilon g_X Q_\chi \cos \theta /e$.
The experimental constraints set on $\varepsilon$ can be then
transformed into the constraints on the mass mixing parameter $\epsilon$ in our models.

The current experimental constraints on the millicharge carried by the dark particle
with respect to the dark particle mass is plotted in Fig.~\ref{fig:millibound}.
In the model we discuss, the millicharge is proportional to the mass mixing parameter $\epsilon$,
and the constraints on $\epsilon$ can be obtained directly from Eq.~(\ref{eq:Qeandmm}).
We also mark our benchmark models on Fig.~\ref{fig:millibound} indicating the millicharge carried by the dark particles in the respective models.

\begin{table}
	\centering
	\resizebox{\textwidth}{22mm}{
		\begin{tabular}{|c|c|c|c|c|c|c|c|c|c|c|c|}
			\hline
			\multicolumn{2}{|c|}{Case} &
			\multicolumn{1}{c|}{Model} &
			\multicolumn{1}{c|}{$M_{\gamma^\prime}$} &
			\multicolumn{1}{c|}{$m_\chi$} &
			\multicolumn{1}{c|}{$g_X$} &
			\multicolumn{1}{c|}{$\delta$} &
			\multicolumn{1}{c|}{$\epsilon$} &
			\multicolumn{1}{c|}{$\varepsilon$} &
			\multicolumn{1}{c|}{$\Omega_\chi h^2$} &
			\multicolumn{1}{c|}{$\Omega_{\gamma^\prime}h^2$} &
			\multicolumn{1}{c|}{$\tau_{\gamma^\prime}$}\\ \hline
			
			\multicolumn{1}{|c|}{\multirow{2}{*}{1}} &
			\multicolumn{1}{c|}{\multirow{2}{*}{$m_\chi>M_{\gamma^\prime}>2m_e$}} &
			$a$ & $20$ & $100$ &
			$0.0054$ & $1\times10^{-13}$ & $1\times10^{-10}$ &$1.57\times10^{-12}$& $0.120$ &
			$0$ & $0.616$ \\ \cline{3-12}
			
			\multicolumn{1}{|c|}{} &
			\multicolumn{1}{c|}{} &
			$b$ & $180$ & $250$ &
			$0.015$ & $1\times10^{-12}$ & $1\times10^{-9}$ &$4.36\times10^{-11}$ & $0.120$ &
			$0$ & $6.82\times10^{-4}$ \\ \hline
			
			\multicolumn{1}{|c|}{2}&
			\multicolumn{1}{c|}{$2m_\chi>M_{\gamma^\prime}>m_\chi>2m_e$} &
			$c$ & $100$ & $60$ &
			$1.59\times10^{-5}$ & $1\times10^{-14}$ & $5\times10^{-11}$  & $2.29\times10^{-15}$& $0.120$ &
			$0$ & $0.491$ \\ \hline
			
		    \multicolumn{1}{|c|}{3}&
			\multicolumn{1}{c|}{$M_{\gamma^\prime}>2m_\chi>2m_e$} &
			$d$ & $100$ & $10$ &
			$0.01$ & $1\times10^{-14}$ & $5.6\times10^{-13}$  & $1.62\times10^{-14}$& $0.120$ &
			$0$ & $2.48\times10^{-19}$ \\ \hline
			
			\multicolumn{1}{|c|}{4} &
			\multicolumn{1}{c|}{$2m_e>m_\chi>M_{\gamma^\prime}$} &
			\cellcolor{blue!10}{$e$} & \cellcolor{blue!10}{$0.09$}& \cellcolor{blue!10}{$1$}&
			\cellcolor{blue!10}{$0.20$}& \cellcolor{blue!10}{$1\times10^{-14}$} & \cellcolor{blue!10}{$1.27\times10^{-12}$} & \cellcolor{blue!10}{$7.34\times10^{-13}$} & \cellcolor{blue!10}{$7.42\times10^{-12}$ }&\cellcolor{blue!10}{$4.43\times10^{-3}$}&\cellcolor{blue!10}{$2.67\times10^{30}$ }  \\ \hline			
			
			\multicolumn{1}{|c|}{\multirow{2}{*}{5}} &
			\multicolumn{1}{c|}{\multirow{2}{*}{$2m_e>2m_\chi>M_{\gamma^\prime}>m_\chi$}} &
			\cellcolor{blue!10}{$f$} & \cellcolor{blue!10}{$0.09$}& \cellcolor{blue!10}{$0.06$}&
			\cellcolor{blue!10}{$0.01$}& \cellcolor{blue!10}{$1\times10^{-14}$} & \cellcolor{blue!10}{$1.27\times10^{-12}$} & \cellcolor{blue!10}{$3.67\times10^{-14}$}  & \cellcolor{blue!10}{$5.94\times10^{-3}$ }&\cellcolor{blue!10}{$2.58\times10^{-9}$}&\cellcolor{blue!10}{$2.67\times10^{30}$ }  \\ \cline{3-12}
			
			\multicolumn{1}{|c|}{} &
			\multicolumn{1}{c|}{} &
			\cellcolor{blue!10}{$g$} & \cellcolor{blue!10}{$0.09$}& \cellcolor{blue!10}{$0.06$}&
			\cellcolor{blue!10}{$1.5\times10^{-4}$}& \cellcolor{blue!10}{$1\times10^{-14}$} & \cellcolor{blue!10}{$1.27\times10^{-12}$}& \cellcolor{blue!10}{$5.50\times10^{-16}$}  & \cellcolor{blue!10}{$1.85\times10^{-3}$ }&\cellcolor{blue!10}{$3.03\times10^{-3}$}&\cellcolor{blue!10}{$2.67\times10^{30}$ }  \\ \hline
						
			\multicolumn{1}{|c|}{6} &
			\multicolumn{1}{c|}{$2m_e>M_{\gamma^\prime}>2m_\chi$} &
			$h$ & $1$ & $0.05$ &
			$0.001$ & $4\times10^{-13}$ & $1\times10^{-11}$& $2.9\times10^{-14}$ & $0.120$ &
			$0$ & $6.20\times10^{-18}$  \\ \hline
	\end{tabular}}
	\caption{The benchmark models we consider in this work for six different types of models. We compute the full evolution of all above benchmark models and calculate the corresponding dark matter relic density for each model. One cannot achieve a full occupation of the dark matter relic density for models in the shaded blue region.
	The lifetimes (in the unit of seconds) of the dark photon for each model are listed in the last column.
    $M_{\gamma^\prime}$ and $m_\chi$ are in MeVs.}
	\label{TableBench}
\end{table}

We consider the following benchmark models, shown in Table~\ref{TableBench},
to illustrate our formalism of computing the hidden sector evolution.
In the models we consider,  after the mixing,
the dark fermion carried the millicharge proportional to $\epsilon$,
and $Z$ boson couples to the dark fermion with strength proportional to $\delta$, c.f., Eqs.~(\ref{Phcoup}) and (\ref{Zcoup}).
These couplings play important roles in calculating the evolution of the dark particles.
Specifically, the couplings of dark photon to SM fermions are proportional to $|\delta - \epsilon|$, c.f., Eq.~(\ref{DPcoup}).
Our investigation shows for each set of the $\{ M_{\gamma^\prime}, m_\chi , g_X \}$ value,
the evolution of the hidden sector temperature and the dark particles
makes no difference for the two cases  $\delta \gg \epsilon$ and $\epsilon \gg \delta$, given the same value of $|\delta - \epsilon|$.
The same value of  $|\delta - \epsilon|$
results in almost identical shape of the plots for the two cases of  $\delta \gg \epsilon$ and $\epsilon \gg \delta$,
although there is a separate constraint on $\epsilon$, c.f., Fig.~\ref{fig:millibound}.
In this paper we will not discuss the fine-tuned case that both of $\delta$ and $\epsilon$ are large but
their difference takes a tiny value, i.e., $\delta\sim\epsilon\gg |\delta - \epsilon| $.
For the following plots we will only show the case of $\epsilon \gg \delta$ for a chosen value of $|\delta - \epsilon|$,
displayed in Fig.~\ref{fig:debound},
which leads to a relatively larger millicharge carried by the dark fermion.

\subsection{Evolution of hidden sector particles}\label{sec:EHSP}

In this subsection, we discuss in depth the evolution of the hidden sector particles
for eight benchmark models covering six distinct cases summarized in the previous subsection.
We calculate the coupled Boltzmann equations given by Eqs.~(\ref{eq:YchiBol})-(\ref{eq:etaBol})
and plot the evolution for the dark fermion $\chi$ and the dark photon $\gamma^\prime$,
as well as interaction rates for various hidden sector interactions related to the dark freeze-out, versus Hubble.
In the calculation we choose an initial value of $\eta=10$ at the temperature $T=10^5$~GeV.
As discussed in~\cite{Aboubrahim:2020lnr,Li:2023nez},
the initial value of $\eta = T/T_h$ will not modify the final abundance so much
but will alter the details of the dark particles evolution.
A comprehensive analysis for the evolution of a $U(1)$ hidden sector interacting feebly with the SM
from the reheating stage, including the determination of the initial value of $\eta$,
is discussed in a companion paper~\cite{FandZh}.

\begin{figure}
	\centering
	\subfigure[]{
		\includegraphics[scale=0.47,trim=34 0 52 0,clip]{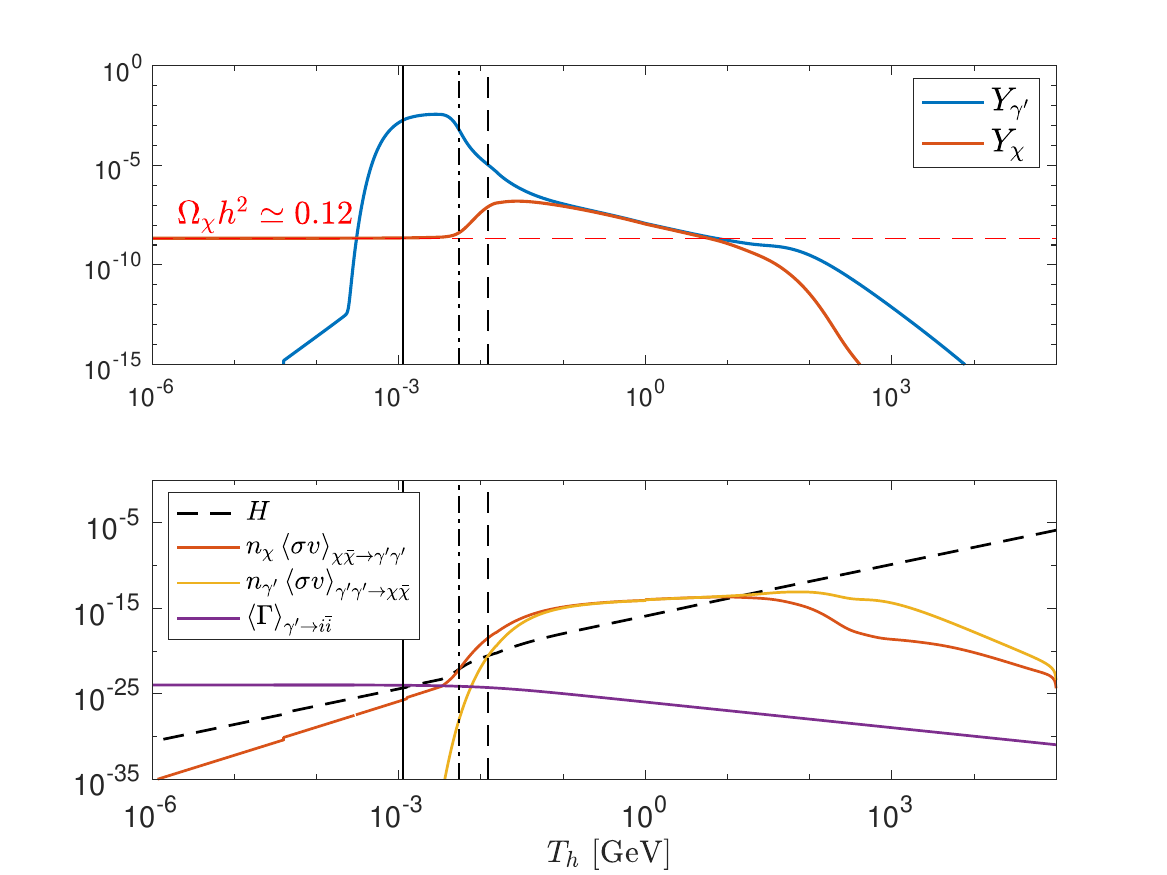} }
	\subfigure[]{
		\includegraphics[scale=0.47,trim=34 0 52 0,clip]{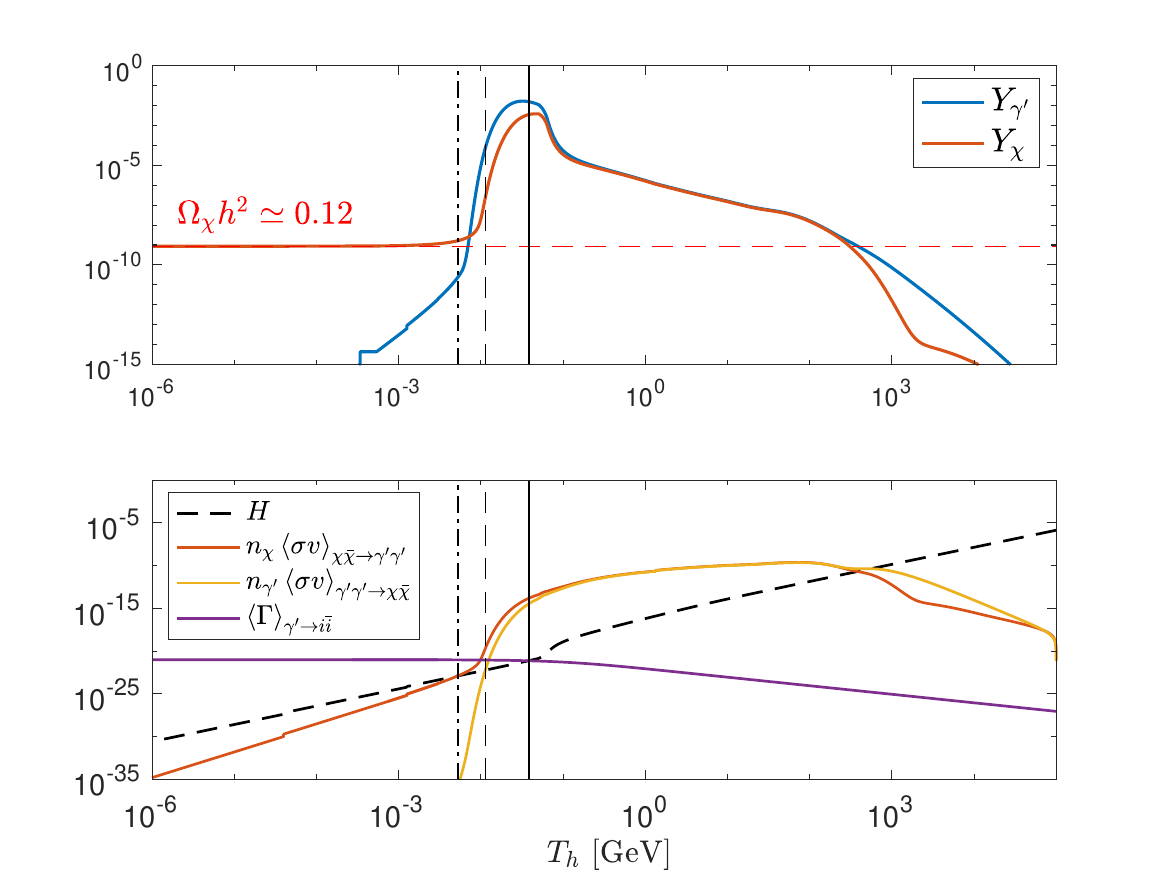} }
	\caption{ \label{Case1} [Color online] The exhibition of the hidden sector evolution for Case 1 (models a,b).
		For each model, the above figure shows the evolution of the comoving number densities ($Y$) of the dark fermion $\chi$
		and the dark photon $\gamma^\prime$.
		The red horizontal dash line denotes the observed dark matter relic density with dark matter mass equal to $m_\chi$.
		In the lower figure we trace the interaction rates of three important processes compared to the Hubble line.
		When the interaction rate is above the Hubble line, the corresponding interaction takes place very rapidly and
        particles participated in the interaction are in chemical equilibrium.
		The black vertical solid line shows at this temperature  the thermally averaged decay width of
		the dark photon to SM fermions (mostly to the electron-positron pairs) going above Hubble,
		indicating the decay of dark photon becomes rapid.
        The black vertical dash line corresponds to the process of $\gamma^\prime \gamma^\prime \to \chi\bar{\chi}$ becoming inactive.
        The black vertical dash-dotted line corresponds to the process of $\chi\bar{\chi} \to \gamma^\prime \gamma^\prime$
		becoming inactive.}
\end{figure}

\paragraph{Case 1 (dark matter $\chi$):}

As we can see from Fig.~\ref{fig:debound}, for the above 1~MeV region,
the dark photon receives the strongest constraint at the mass around 17 MeV,
corresponding to a constraint $|\delta -\epsilon| \lesssim 2\times 10^{-11}$.
Using this lower bound value, the dark photon decays not so rapidly and may spoil the BBN constraint.
Thus we choose the dark photon mass to be 20 MeV,  a little off the 17 MeV point in the benchmark model a,
corresponding to the lifetime of the dark photon less than 1 second.
When the dark photon mass is far from 20 MeV, there would less restrictions on $|\delta -\epsilon|$, c.f., Fig.~\ref{fig:debound}.
In model b, we take a larger value of $|\delta -\epsilon|$, resulting in a larger coupling
between the visible sector and the hidden sector.
We show benchmark models a,b for Case 1 in Fig.~\ref{Case1}.
In both models,  the dark fermion can make up the whole dark matter constituent.

For each benchmark model, we plot the evolution of the dark fermion $\chi$ and the dark photon $\gamma^\prime$,
and with the same temperature scales we also plot the interaction rates of
hidden sector particle evolutions versus the Hubble line.
When the interaction rate is above the Hubble scale at a certain temperature,
the corresponding interaction is quite intense and
particles involving in this interaction are in thermal equilibrium.
For model a:
The black vertical dash line corresponds to the process of
$\gamma^\prime \gamma^\prime \to \chi\bar{\chi}$ becoming inactive,
indicating the start of the dark freeze-out,
represented by the number density of $\chi$ about to drop;
The black vertical dash-dotted line corresponds to the process of
$\chi\bar{\chi} \to \gamma^\prime \gamma^\prime$
becoming inactive, indicating the finish of the dark freeze-out,
represented by the number density of $\chi$ tending to level off;
The black vertical solid line shows that at this temperature the thermally averaged decay width of
the dark photon to SM fermions (mostly to the electron-positron pairs) goes above Hubble,
indicating the decay of the dark photon becoming rapid,
which corresponds to a steep drop for the dark photon number density.

For model b, the constraint on  $|\delta -\epsilon|$ is less restrictive and can be taken to be a larger value compared to model a.
Thus the freeze-in rate is much larger and more dark particles are produced.
In the plots, one can see that even the purple line showing the dark photon decay becomes active in a quite early stage,
the robust freeze-in production still generates a massive amount of
both the dark fermion $\chi$ and the dark photon $\gamma^\prime$.
The evolutions of $\chi$ and $\gamma^\prime$ are determined by the combined effects of freeze-in,
hidden sector interactions and the dark photon decay to SM particles.
The dark photon finally diminishes to vanish before BBN.

\begin{figure}
	\centering
	\addtocounter{subfigure}{+2}
	\subfigure[]{
    \includegraphics[scale=0.47,trim=34 0 52 0,clip]{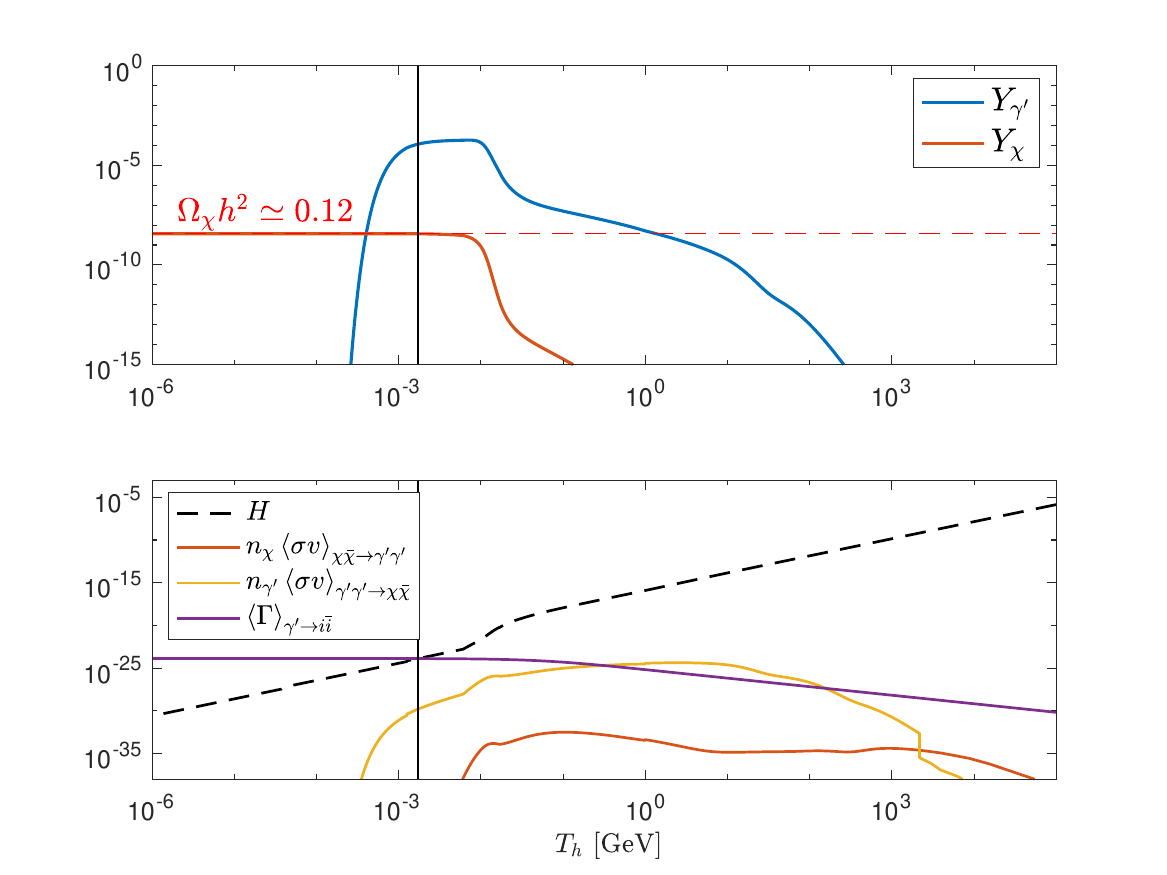}}
	\caption{ \label{Case2} [Color online] The exhibition of the hidden sector evolution for Case 2 (model c).
		The above figure shows the evolution of the comoving number densities of the dark fermion $\chi$
		and the dark photon $\gamma^\prime$, with
		the red horizontal line denoting the observed dark matter relic density with dark matter mass equal to $m_\chi$.
		The lower figures show the interplay between the hidden sector interaction rates versus Hubble.}
\end{figure}


\paragraph{Case 2 (dark matter $\chi$):}

For case 2 we study benchmark model c, and the full evolution is shown in Fig.~\ref{Case2}.
In this case, for a relatively large $g_X$, the hidden sector will have strong freeze-out effect
via the process $\gamma^\prime \gamma^\prime\to \chi\bar{\chi}$
and render an overproduction of the dark matter $\chi$.
The only way to achieve the observed dark matter relic density
is to weaken the hidden sector interaction $\gamma^\prime \gamma^\prime\to \chi\bar{\chi}$
with a rather small value of $g_X$,
such that less $\gamma^\prime$ will transform to the dark fermion $\chi$,
noticing that $\gamma^\prime \gamma^\prime\to \chi\bar{\chi}$ dominates over
$\chi\bar{\chi}\to \gamma^\prime \gamma^\prime$ as seen from the lower figure in Fig.~\ref{Case2}.
The upper plot in Fig.~\ref{Case2} shows the evolution of the hidden sector particles,
whereas the lower plot exhibits the hidden sector interaction rates are below the Hubble line,
corresponding to the hidden sector interactions occurring not rapid.
However, hidden sector interactions, even ultraweak,
play crucial roles generating the correct amount of dark matter relic density.
Later, when the dark photon decay began to dominate, shown by the solid black vertical line,
the number density of the dark photon immediately drops down to zero.
The abundance of the dark fermion, however, gradually accumulates without reduction,
and can finally reach the current observed value of the dark matter relic density,
given by suitable parameters we choose.

\begin{table}
\begin{center}
\begin{tabular}{|c|c|}
\hline
Comparison of different calculations & $\Omega_{\chi}h^{2}$\tabularnewline
\hline
All freeze-in processes included & 0.1195\tabularnewline
\hline
Plasmon decay process excluded & 0.1195\tabularnewline
\hline
Four-point $\gamma^{\prime}$ freeze-in processes excluded & 0.0643\tabularnewline
\hline
Pure freeze-in production of $\chi$ & $10^{-9}$ \tabularnewline
\hline
\end{tabular}
\end{center}
\caption{A comparison of different calculations of the dark matter relic density for the benchmark model c.}
\label{TableBMc}
\end{table}

Since in this case the parameters are chosen in such a way that
the hidden sector interactions  $\chi\bar{\chi} \leftrightarrow \gamma^\prime \gamma^\prime$
never achieved thermal equilibrium,
it is a good occasion to show the contribution of the four-point freeze-in processes of the dark photon
which are usually ignored in the literature, given by the last line in Eq.~(\ref{eq:YDPBol}),
as long as the three-point freeze-in productions of the dark photon ($i \bar i \to \gamma^\prime$) are present.
In Table~\ref{TableBMc} we compare several different computations with or without some important freeze-in processes.
The first result corresponds to the calculated relic density for the dark fermion $\chi$
including all relevant processes given by Eqs.~(\ref{eq:YchiBol}) and (\ref{eq:YDPBol}).
The second result shows the $\chi$ relic density not including the contribution from the plasmon decay,
which is almost identical to the first result including all the processes,
which agrees with the previous findings that the production of the millicharge particle from
plasmon decay becomes important only when the millicharge particle is lighter than the mass of the electron~\cite{Dvorkin:2019zdi,Chang:2019xva}.
The third result shows the $\chi$ relic density without including all the four-point freeze-in processes of the dark photon.
The last result is computed from the pure freeze-in production of $\chi$,
ignoring the freeze-in production of the dark photon and dropping out all hidden sector interactions,
which is a much smaller value compared with the previous one.
This can be understood as follows:
the freeze-in production of $\chi$ depends on the interaction $i\bar{i}\to \chi\bar{\chi}$,
which always has an additional $g_X^2$ suppression compared to the dark photon freeze-in production,
c.f., the rough estimation of the couplings in Eqs.~(\ref{Phcoup})-(\ref{DPcoup}).
In model c we choose a rather small $g_X\sim 10^{-5}$ and thus the
pure freeze-in production of $\chi$ is quite rare.

It is commonly stated that when the interaction rate exceeds the Hubble scale,
this interaction is considered to be ``in equilibrium''.
In other words, this interaction occurs rapidly at the corresponding Hubble scale.
While our results indicate that
even the rate of an interaction never reached the Hubble scale in the early universe,
this interaction definitely still took place and its imprints can be remarkable.
In this case, the abundance of dark fermions accumulates primarily through the dark photon freeze-out,
while the dark photon itself decays into SM fermions, eventually disappearing without trace.

We conclude: (1) Although the interaction rates of $\chi\bar{\chi} \leftrightarrow \gamma^\prime \gamma^\prime$
are always below Hubble in this case,
the freeze-in production of the dark photon plays a crucial role in generating the abundance of the dark fermion $\chi$.
One should not ignore the hidden sector interactions, even they are ultraweak,
and compute the dark matter relic density through ``pure freeze-in'' production.
(2) The four-point freeze-in production of the dark photon is as important as the three-point freeze-in production channels,
and must be taken into account at all times.

\begin{figure}
	\centering
	\addtocounter{subfigure}{+3}
	\subfigure[]{
		\includegraphics[scale=0.47,trim=34 0 52 0,clip]{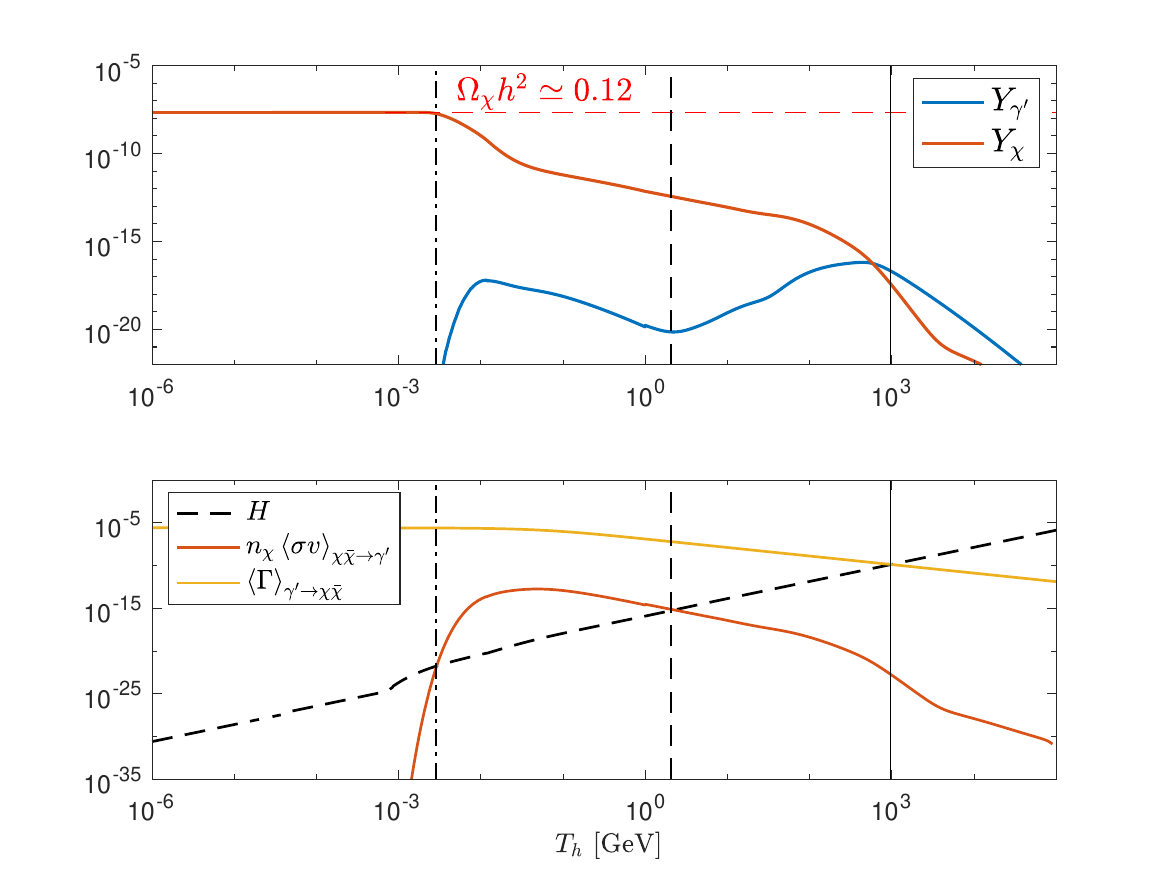} }
	\caption{ \label{Case3} [Color online] The exhibition of the hidden sector evolution for Case 3 (model d).
		The top figure shows the evolution of the comoving number densities of the dark fermion $\chi$
		and the dark photon $\gamma^\prime$, with
		the red horizontal line denoting the observed dark matter relic density with dark matter mass equal to $m_\chi$.
		In the lower figure we trace the hidden sector interaction rates compared to the Hubble line.
		The black vertical solid line shows at this temperature the thermally averaged decay width of $\gamma^\prime \to \chi\bar{\chi}$
		becoming active,
        and the corresponding interaction rate is extremely strong.
        The black vertical dash line shows the process $\chi\bar{\chi} \to \gamma^\prime$ becoming active and
		the black vertical dash-dotted line corresponds to the same process
        becoming insignificant 		and all of the dark photon will eventually decay to dark fermions. }
\end{figure}

\paragraph{Case 3 (dark matter $\chi$):}

In this case the dark photon mass is larger than twice of the dark fermion mass
and thus a vigorous decay of the dark photon dominates all the interaction channels.
Especially, the decay of the dark photon to SM fermion pairs $\gamma^\prime \to i \bar i$ can be completely neglected
compared to the process  $\gamma^\prime \to \chi \bar \chi$ occurring at the same time.
As can be seen from Fig.~\ref{Case3},
the abundance of the dark fermion $\chi$ always increases before it reaches the final value.
The first drop in the dark photon abundance is due to $\gamma^\prime \to \chi\bar{\chi}$ becoming active,
indicated by the black vertical solid line.
When the interaction rate of the process $\chi \bar{\chi} \to \gamma^\prime$ goes above the Hubble line,
indicated by the black vertical dash line,
the abundance of the dark photon soon increases.
The black vertical dash-dotted line corresponds to the process of $\chi\bar{\chi} \to \gamma^\prime$
becoming insignificant and
finally all dark photon decay into $\chi \bar{\chi} $ and its abundance becomes zero.

\begin{figure}
	\centering
	\addtocounter{subfigure}{+4}
	\subfigure[]{
		\includegraphics[scale=0.47,trim=34 0 52 0,clip]{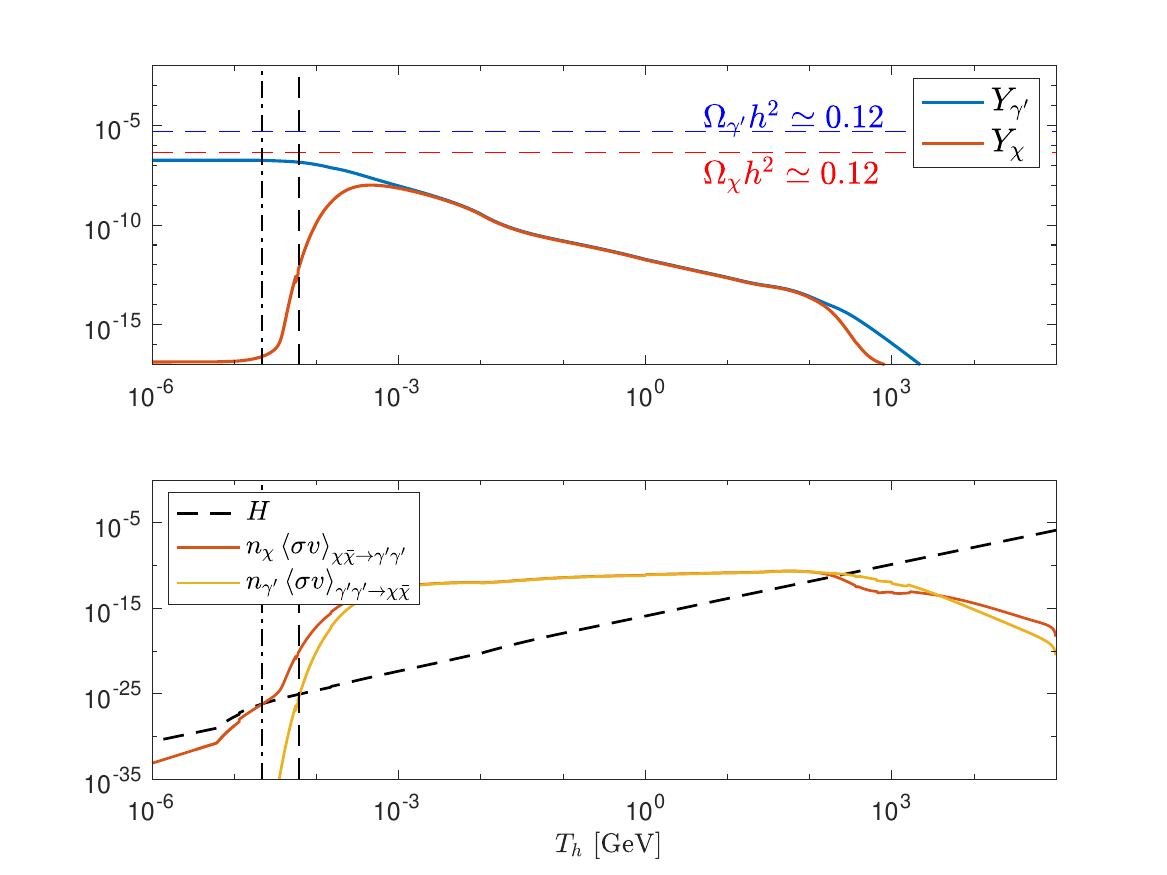} }\\
		\subfigure[]{
		\includegraphics[scale=0.47,trim=34 0 52 0,clip]{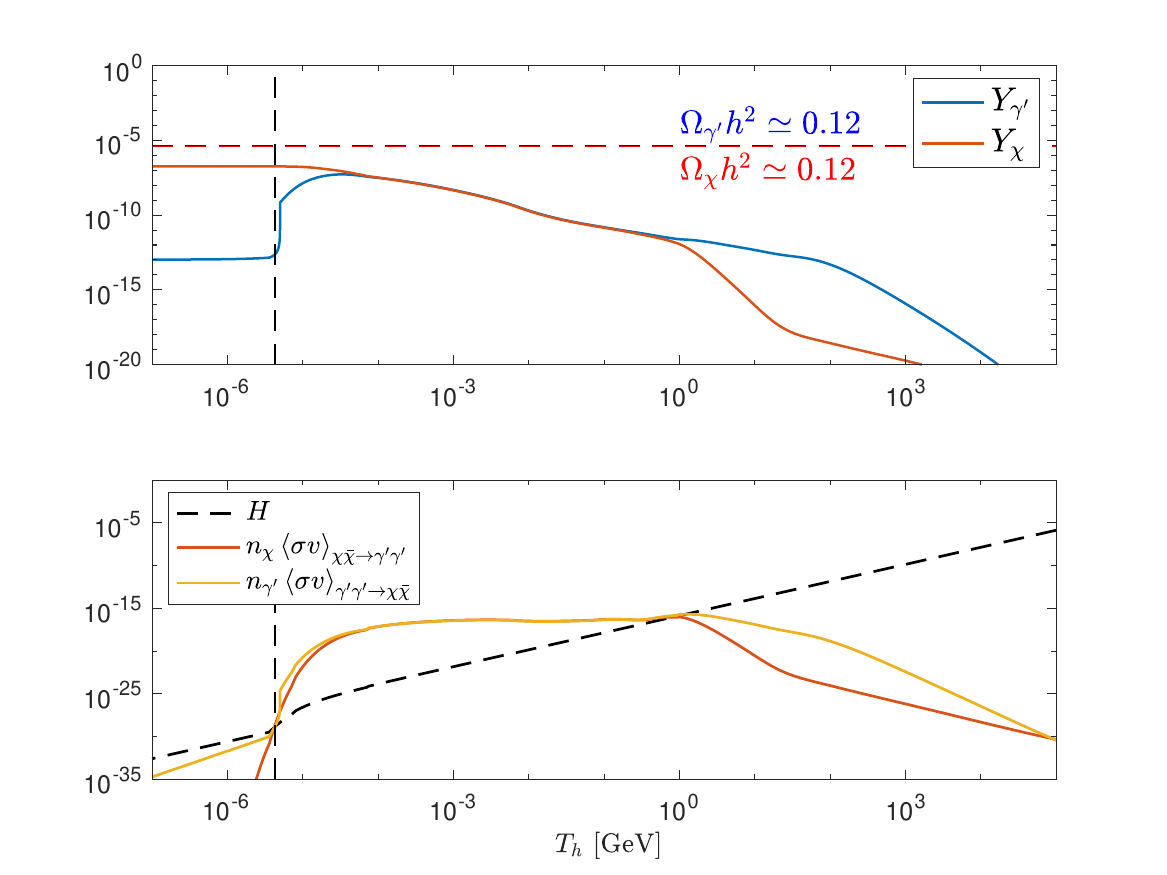} }
	\subfigure[]{
		\includegraphics[scale=0.47,trim=34 0 52 0,clip]{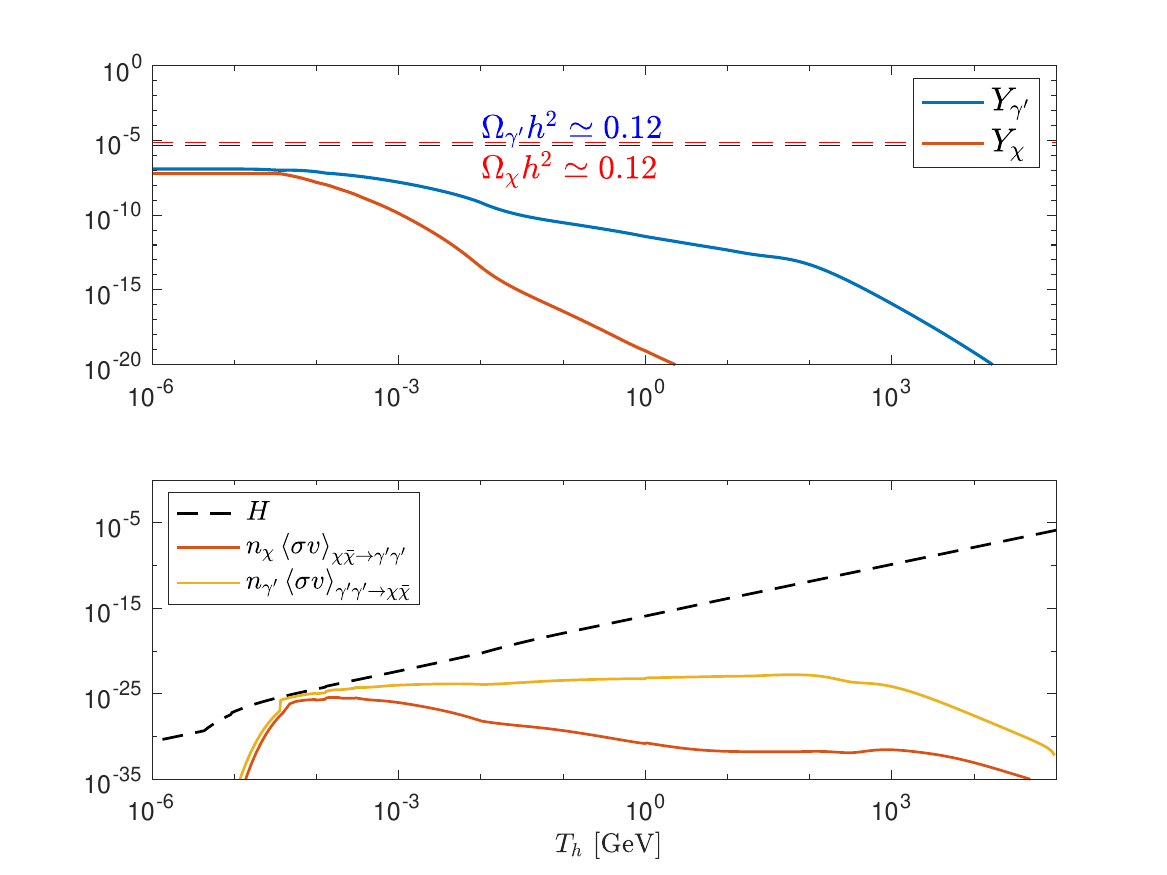} }
  \caption{ \label{Case5} [Color online] The exhibition of the hidden sector evolution for  Case 4 (model e) and Case 5 (models f,g).
	For each model, the above figure shows the evolution of the comoving number densities of the dark fermion $\chi$
	and the dark photon $\gamma^\prime$.
	The red horizontal line denoting the observed dark matter relic density if the dark matter is totally made up by the dark fermion $\chi$.
	whereas the blue horizontal line shows the observed dark matter relic density if the dark matter is totally made up by the dark photon $\gamma^\prime$.
	In the lower figures for each models,
	we trace the interaction rates of two processes among the hidden sector compared to the Hubble.
The IDPB constraint is so strong that one can only take a much smaller value for both $\delta$ and $\epsilon$ compared with other cases.
As a result, in Case~4 and Case~5, one cannot obtain a large enough value for the dark matter relic abundance.
In model g, we choose a certain $g_X$ value such that the hidden sector interaction rate is about tangent to the Hubble line,
and we obtain the largest relic abundance for the dark matter $\gamma^\prime$ and $\chi$.}
\end{figure}

\paragraph{Case 4 and Case 5 (dark matter $\chi,\gamma^\prime$):}

In these two cases the dark photon mass is less than twice of the electron mass and thus
the dark photon can be stable throughout the history of the Universe,
as discussed in Section~\ref{sec:mixingSt}.
As noted earlier, the IDPB constraint is stronger than
the experimental constraints on both the kinetic mixing and the millicharge carried by the dark matter,
and thus restrict the $|\delta - \epsilon|$ value to be much smaller than Case 1,2,3 and 6.
In these two cases, both the dark fermion $\chi$ and the dark photon $\gamma^\prime$ are dark matter candidates.
We conclude, for a single extra $U(1)$ mixed with the SM,
one {\it cannot} achieve a full occupation of the dark matter relic density
with the dark photon dark matter (at most about $5\%$).
For a full occupation of the dark photon dark matter,
one needs at least two extra $U(1)$ hidden sectors and was studied in depth in~\cite{Aboubrahim:2021ycj}.

In Case 4,
we choose the benchmark model e such that they can produce
the highest abundance of the dark matter,
corresponding to a rather large hidden sector gauge coupling $g_X$.
In this case one can see clearly a dark freeze-out occurring right after the interaction rates of
$\gamma^\prime \gamma^\prime \to \chi \bar \chi $
and
$\chi \bar \chi \to \gamma^\prime \gamma^\prime $
start to deviate.
The dark freeze-out ended at the interaction $\chi \bar \chi \to \gamma^\prime \gamma^\prime $
becomes inactive, indicated by the black vertical dash-dotted line.

For Case 5 we choose two types of benchmark models:
one type (model f) with larger $U(1)_X$ gauge coupling and the hidden sector interactions are active
and the dark freeze-out for the dark photon takes place via the interaction
$\gamma^\prime \gamma^\prime \to \chi \bar \chi $,
where we see a quick drop in $Y_{\gamma^\prime}$ around $T_h \sim 10^{-2}$~MeV.
The dark freeze-out ended when $\gamma^\prime \gamma^\prime \to \chi \bar \chi $ becoming inactive
indicated by the black vertical dash line.
The $U(1)_X$ gauge coupling of the other type (models g) is chosen to be small
such that the hidden sector interactions $\chi \bar \chi \leftrightarrow \gamma^\prime \gamma^\prime$
never go beyond Hubble.
In model g, we find the largest relic density for the dark matter $\gamma^\prime$ and $\chi$.
Thus for the case that the SM extended by a single $U(1)$,
the $\mathcal{O}$(keV) dark photon dark matter can only occupy at most  $\sim 5 \%$ of the total observed dark matter relic density.

As the dark fermion mass goes below the mass of the electron,
the  freeze-in contribution from plasmon decay $\gamma^*\to \chi \bar{\chi}$
becomes important~\cite{Dvorkin:2019zdi,Chang:2019xva},
especially when the dark fermion mass is around 50~keV.
However, the dark fermion number density cannot be computed separately,
even the interactions inside the hidden sector are ultraweak,
as we already discovered in the analysis for the benchmark model c.
To calculate the final relic abundance of the dark fermion,
one needs to compute the full evolution of the entire hidden sector
including all of the hidden sector interactions.
The hidden sector interactions $\chi \bar{\chi} \leftrightarrow \gamma^\prime\gamma^\prime$
play crucial roles in determining the final relic abundance of both $\chi$ and $\gamma^\prime$.

\begin{table}
	\centering
\begin{tabular}{|c|c|c|c|c|}
\hline
\multirow{2}{*}{Benchmark models} & \multicolumn{2}{c|}{Plasmon decay included} & \multicolumn{2}{c|}{Plasmon contribution excluded}\tabularnewline
\cline{2-5} \cline{3-5} \cline{4-5} \cline{5-5}
 & $\Omega_{\chi}h^{2}$ & $\Omega_{\gamma^{\prime}}h^{2}$ & $\Omega_{\chi}h^{2}$  & $\Omega_{\gamma^{\prime}}h^{2}$\tabularnewline
\hline
Model e & $7.4193\times10^{-12}$ & $4.4344\times10^{-3}$ & $7.4191\times10^{-12}$ & $4.4327\times10^{-3}$\tabularnewline
\hline
Model f & $5.9415\times10^{-3} $ & $2.5842\times10^{-9}$ & $5.9397\times10^{-3}$ & $2.5807\times10^{-9}$ \tabularnewline
\hline
\end{tabular}
\caption{The results of
dark matter relic densities including or excluding the plasmon decay contribution for benchmark models e and f.
In these two benchmark models, both the dark fermion $\chi$ and the dark photon $\gamma^\prime$ are dark matter candidates.}
\label{TablePeff}
\end{table}

In Table~\ref{TablePeff} we list the final result of the relic abundance of $\chi$ and $\gamma^\prime$
with and without the inclusion of the plasmon decay.
We can see the contribution of the plasmon decay to the final relic densities of  $\chi$ and $\gamma^\prime$ are indeed not significant,
because of the interactions inside the hidden sector.

\begin{figure}[t!]
	\centering
	\addtocounter{subfigure}{+7}
	\subfigure[]{
		\includegraphics[scale=0.47,trim=34 0 52 0,clip]{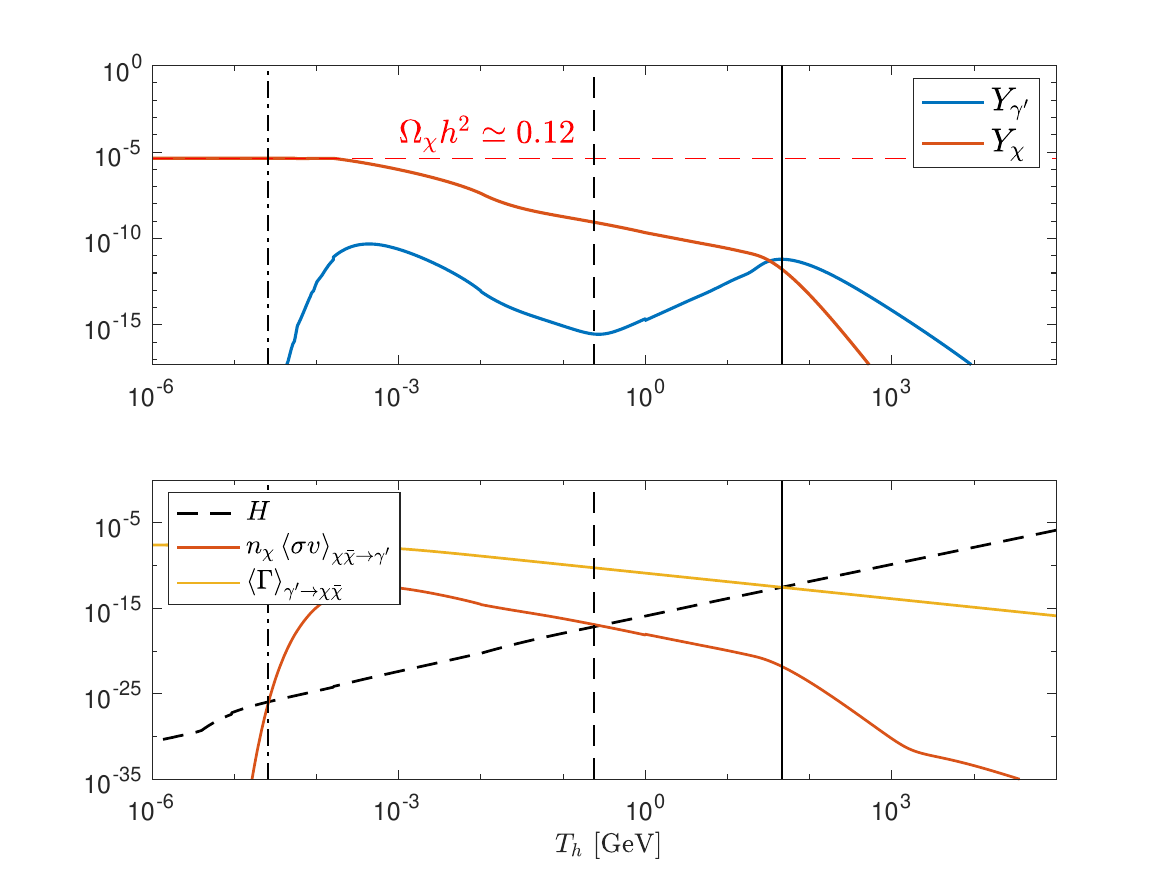} }
	\caption{ \label{Case6} [Color online] The exhibition of the hidden sector evolution for Case 6 (model h).
		The above figure shows the evolution of the comoving number densities of the dark fermion $\chi$
		and the dark photon $\gamma^\prime$, with
		the red horizontal line denoting the observed dark matter relic density with the dark matter being the dark fermion $\chi$.
		The lower figure shows the interaction rates of two processes among the hidden sector compared to the Hubble.
		In this case, even the mass of the dark photon is taken to be less than twice of electron mass,
		the massive decay of the dark photon to dark fermions plays the dominant role.
        In this case the dark matter candidate can only be the dark fermion $\chi$.}
\end{figure}

\paragraph{Case 6 (dark matter $\chi$):}

In this case, the intensive decay of the dark photon
through the channel $\gamma^\prime \to \chi \bar\chi $  plays the dominant role.
This case is similar to Case 3:
the abundance of the dark fermion $\chi$ always increases before it reaches the final value.
The drop in the dark photon abundance around the black vertical dash line
is due to the inverse interaction $\chi \bar \chi \to \gamma^\prime$ becoming active.
Finally all dark photon decay into $\chi \bar \chi$ and its abundance becomes zero.\\

\subsection{Summary of the computation results}\label{sec:SumRes}

In this section, we investigate eight benchmark models (a-h) from six distinct cases
depending on the masses of the dark photon and the dark fermion,
given by Table~\ref{TableBench}.
In models a,b,c,d,h, the dark matter candidate is the dark fermion $\chi$;
whereas in models e,f,g, the dark matter candidates are the dark fermion $\chi$ and the dark photon $\gamma^\prime$.
We have computed and plotted the full evolution of the dark fermion $\chi$ and the dark photon $\gamma^\prime$
as well as the hidden sector temperature,
governed by the coupled Boltzmann equations given in Section~\ref{sec:CBE}.
We also keep track of the hidden sector interaction rates versus Hubble
to see the intensity of the corresponding interaction.
In models a,b,d,e,f,h we find the hidden sector encountered apparent dark freeze-out.

We now succinctly outline the primary discoveries for the analysis of the eight benchmark models encompassing six distinct cases,
depending on the masses of the dark photon and the dark fermion, given by Table~\ref{TableBench}:
\begin{itemize}
\item An interaction that never reached equilibrium,
indicated by its interaction rate was always less than the Hubble scale,
does not imply that this interaction did not occur.
If the hidden sector particles are produced through the freeze-in mechanism,
as long as the hidden sector has self-interactions involving various dark particles,
the self-interactions within the hidden sector (even ultraweak)
play significant roles in the evolution of the hidden sector (see Table~\ref{TableBMc}, the fourth result).
  One must take full consideration of the contribution from hidden sector interactions at all times.
  As long as the hidden sector possesses interactions among hidden sector particles, even they are ultraweak,
  one should not ignore these hidden sector interactions and compute the dark matter relic density through ``pure freeze-in'' production.
\item For the extension of the SM with an extra $U(1)$ hidden sector, we find
the contribution from four-point freeze-in processes for the dark photon: $i\bar i \to \gamma\gamma^\prime$,
$i \gamma \to i \gamma^\prime$, $\bar i \gamma \to \bar i \gamma^\prime$,
which are usually ignored in the literature,
are as important as the three-point freeze-in processes $i\bar i \to \gamma^\prime$,
where $i$ denotes the SM fermions (see Table~\ref{TableBMc}, the third result).
We also expect that for more general freeze-in dark matter models,
the four-point freeze-in production channels are also crucial and should not be neglected,
even the three-point freeze-in channels for generating the same freeze-in particle are present.
\item The freeze-in production of the dark fermion $\chi$ from the plasmon decay,
however, does not play an important role in determining the final relic abundance of the $\chi$,
as long as the hidden sector self-interactions (in this case: $\chi \bar{\chi} \leftrightarrow \gamma^\prime\gamma^\prime$) are present,
see results from Table~\ref{TableBMc} and Table~\ref{TablePeff}.
  Though for a comprehensive analysis,
  it is advisable to incorporate the plasmon contribution into the computation.
\item In the analysis of Case 4 and Case 5 (models e,f,g),
  we find $\mathcal{O}$(keV) dark photon dark matter from a {\it single} $U(1)_X$ hidden sector with feeble couplings to the SM,
  can only occupy at most $\sim 5\%$ of the total observed dark matter relic density.
\end{itemize}

\section{Conclusion}\label{sec:Con}

In this paper we conduct a comprehensive study on the sub-GeV millicharge dark matter
produced through the freeze-in mechanism.
We discuss the SM extended by a $U(1)_X$ hidden sector,
and there exist tiny kinetic mixing and mass mixing between the $U(1)_X$ and the hypercharge gauge field.
The dark fermions carrying the $U(1)_X$ quantum numbers are naturally dark matter candidates.
After the mass mixing of the $U(1)_X$ and the hypercharge gauge field,
the dark fermion carries a millicharge.
We discuss in general how such sub-GeV millicharge dark matter can be generated
through the freeze-in mechanism from a theoretical perspective.
We highlight that there are several inappropriate treatments and misunderstandings in the literature regarding this subject.

The millicharge can be generated only in three ways:
\begin{enumerate}
  \item The dark particle carries a tiny amount of hypercharge as a prior,
  and one such possibility is discussed in Section~\ref{sec:mixingEH}.
  \item Theory involving kinetic mixing between a {\it massless} $U(1)$ and the hypercharge gauge field,
  and the generated millicharge is proportional to the kinetic mixing parameter, as discussed in Section~\ref{sec:mixingmlU1}.
  \item The mass mixing between two massive $U(1)$ (one may produce a massless $U(1)$ in the final mass eigenbasis,
  see the discussion in Section~\ref{sec:mixing2U12M}),
  {\it no matter the kinetic mixing is present or not},
  and the millicharge generated is proportional to the mass mixing parameter.
  In the electroweak theory, this can be done by
  a Stueckelberg mass mixing between the $U(1)_X$ and the hypercharge gauge field, as discussed in Section~\ref{sec:mixingSt}.
\end{enumerate}

We clarify some inappropriate treatments and misunderstandings that are presented in the literature:
\begin{enumerate}
  \item It is not appropriate to consider the kinetic mixing between the extra $U(1)$ and the photon field,
  i.e., the effective operator
  $\mathcal{L} \sim \frac{\boldsymbol{\delta}}{2}F_{\mu\nu}^{\rm em}F_{X}^{\mu\nu}$,
  especially from a theoretical perspective.
  The photon field is a linear combination of the hypercharge gauge field and the neutral $SU(2)$ gauge boson
  in the full electroweak theory.
  Thus the mixing between the extra $U(1)$ and the hypercharge gauge field
  may not generate a direct mixing between the extra $U(1)$ and the $U(1)_{\rm em}$~\cite{Feldman:2007wj}.
  It is indeed subtle to embed such effective term
  $\mathcal{L} \sim \frac{\boldsymbol{\delta}}{2}F_{\mu\nu}^{\rm em}F_{X}^{\mu\nu}$
  into a UV-completed theory,
  as discussed in depth in Section~\ref{sec:mixingSt}.
  \item Even from the effective field theory point of view,
the kinetic mixing between the massless photon field
and an extra massive $U(1)$ {\it cannot} generate a millicharge,
no matter how tiny the extra $U(1)$ mass is~\cite{Feldman:2007wj},
summarised in Section~\ref{sec:mixing2U10M}.
Taking the mass of the massive $U(1)$ to be extremely small (but still nonzero),
{\it cannot} transit to the case of kinetic mixing between two massless $U(1)$ fields,
which will generate a millicharge proportional to the kinetic mixing parameter.
  \item The kinetic mixing {\it does not} play any role in generating a millicharge
  as long as one of the two $U(1)$ fields gets massive.
  In this case only the mass mixing, e.g., Eq.~(\ref{eq:2U1massmix})
  can generate a millicharge.
\end{enumerate}

In the model exhibited in Fig.~\ref{fig:U1Model}, both the dark fermion $\chi$ and the dark photon $\gamma^\prime$
are produced via the freeze-in mechanism.
After the mass mixing, $\chi$ carries a millicharge in addition to the $U(1)_X$ quantum number.
$\chi$ is always stable and thus is the dark matter candidate,
whereas the dark photon can be made stable when its mass is less than twice of the electron mass.
Within the hidden sector, $\chi$ and $\gamma^\prime$
may simultaneously engage in strong or weak interactions
during the freeze-in production of $\chi$ and $\gamma^\prime$.
The interactions within the hidden sector alter the number densities of $\chi$ and $\gamma^\prime$ over time.
Hence, it is not possible to calculate the relic abundance of $\chi$ and $\gamma^\prime$
solely through pure freeze-in production.
With the formalism established in~\cite{Aboubrahim:2020lnr} and reviewed in Section~\ref{sec:BEMT},
we are capable of calculating the complete evolution of the hidden sector particles as well as the hidden sector temperature.
This calculation formalism is general and can be applied to models involving
any hidden sectors with feeble connections to the visible sector.
In these models, the hidden sector particles are produced through the freeze-in mechanism,
and exhibit strong or weak (even ultraweak) self-interactions within the hidden sector.

We perform a comprehensive analysis on eight benchmark models (a-h) encompassing six distinct cases,
depending on the masses of the dark photon and the dark fermion,
and determine the complete evolution of the $U(1)_X$ hidden sector.
In some of the models with intensive hidden sector interactions (models a,b,d,e,f,h),
we observe apparent dark freeze-out;
while in others with weak or even ultraweak hidden sector interactions (models c,g),
the dark freeze-out is not apparent,
but these interactions still play significant roles in determining the final relic abundance of hidden sector particles.
The intensities of the hidden sector interactions are characterized by the interaction rates,
and  are related to the hidden sector temperature which is evidently different from the visible sector temperature
(the temperature of our observed Universe).
The hidden sector temperature can be tracked in the calculation that
solving the coupled Boltzmann equations given in Section~\ref{sec:CBE}.
We have taken into account all of the freeze-in processes as well as the hidden sector self-interactions simultaneously.
We plot the evolution of the hidden sector particles
with an emphasize on the consequences of hidden sector interactions.
In addition to the dark fermion $\chi$, $\mathcal{O}$(keV) dark photon can also be dark matter candidate.
We find in models e,f,g that from a {\it single} $U(1)_X$ hidden sector with feeble couplings to the SM,
the dark photon dark matter can only occupy at most $\sim 5\%$ of the total observed dark matter relic density.

The  primary discoveries for the analysis of the eight benchmark models are summarised in Section~\ref{sec:SumRes}.
We tend to believe the following findings are universal and applicable to broader freeze-in scenarios:
\begin{enumerate}
  \item An interaction that never reached equilibrium,
indicated by its interaction rate was always less than the Hubble scale,
does not imply that this interaction did not occur.
On the contrary, this interaction definitely still took place and its imprints can be remarkable.
  \item If the hidden sector possesses self-interactions (even ultraweak),
  these interactions invariably occur simultaneously with the freeze-in production of hidden sector particles.
To accurately determine the relic abundance of the freeze-in dark matter,
it is essential to consider all pertinent freeze-in processes,
{\it as well as all\,} hidden sector interactions,
and solve the coupled Boltzmann equations while accounting for the temperature disparity between the visible and hidden sectors.
  One should not ignore the hidden sector interactions, even they are ultraweak,
  and compute the dark matter relic density through ``pure freeze-in'' production.
  \item For extensions of the SM, the four-point freeze-in processes are usually ignored,
especially when the three-point freeze-in production channels are present for the same freeze-in particle.
In our analysis we find that the four-point freeze-in production channels are {\it as important as} the three-point freeze-in channels,
and should not be neglected.
\end{enumerate}

\noindent \textbf{Acknowledgments: }

WZF acknowledges discussions with Xiaoyong Chu and Zuowei Liu.
WZF is supported by the National Natural Science Foundation of China under
Grant No. 11935009.

\appendix

\section{Kinetic and mass mixing in the Stueckelberg extension}\label{sec:APPKMMM}
\setcounter{equation}{0}
\renewcommand{\theequation}{A.\arabic{equation}}
In this Appendix we derive the mass eigenbasis of neutral gauge bosons
in the Stueckelberg extension of the SM with kinetic mixing and mass
mixing, which is firstly done in~\cite{Feldman:2007wj}.
We consider the Lagrangian given in Eqs.~(\ref{eq:KinMix}) and (\ref{eq:StMMix}),
resulting in the mixing matrices Eq.~(\ref{eq:33mixM}). We consider the case
that $M_{2}\neq0$ thus we have a tiny mass mixing parameterized by
$\epsilon=M_{2}/M_{1}$. In this case the physical photon in the mass
eigenbasis couples to hidden sector fermions with a strength proportional
to $\epsilon$, thus the hidden sector fermion is millicharged. A
simultaneous diagonalization of the kinetic matrix and mass matrix
can be done in the following steps. Firstly we rewrite the mass matrix
in terms of $\epsilon$ as
\begin{equation}
M_{{\rm St}}^{2}=\begin{pmatrix}M_{1}^{2} & M_{1}M_{2} & 0\\
M_{1}M_{2} & M_{2}^{2}+\frac{1}{4}v^{2}g_{Y}^{2} & -\frac{1}{4}v^{2}g_{2}g_{Y}\\
0 & -\frac{1}{4}v^{2}g_{2}g_{Y} & \frac{1}{4}v^{2}g_{2}^{2}
\end{pmatrix}=\begin{pmatrix}M_{1}^{2} & M_{1}^{2}\epsilon & 0\\
M_{1}^{2}\epsilon & M_{1}^{2}\epsilon^{2}+\frac{1}{4}v^{2}g_{Y}^{2} & -\frac{1}{4}v^{2}g_{2}g_{Y}\\
0 & -\frac{1}{4}v^{2}g_{2}g_{Y} & \frac{1}{4}v^{2}g_{2}^{2}
\end{pmatrix}\,.
\end{equation}
A transformation
\begin{equation}
K^{\prime}=KO^{\prime}=\begin{pmatrix}\frac{1}{\sqrt{1-\delta^{2}}} & 0 & 0\\
-\frac{\delta}{\sqrt{1-\delta^{2}}} & 1 & 0\\
0 & 0 & 1
\end{pmatrix}\begin{pmatrix}\cos\phi & -\sin\phi & 0\\
\sin\phi & \cos\phi & 0\\
0 & 0 & 1
\end{pmatrix}=\begin{pmatrix}\frac{1-\delta\epsilon}{\sqrt{\left(1-\delta^{2}\right)\left(1-2\delta\epsilon+\epsilon^{2}\right)}} & \frac{-\epsilon}{\sqrt{1-2\delta\epsilon+\epsilon^{2}}} & 0\\
\frac{\epsilon-\delta}{\sqrt{\left(1-\delta^{2}\right)\left(1-2\delta\epsilon+\epsilon^{2}\right)}} & \frac{1}{\sqrt{1-2\delta\epsilon+\epsilon^{2}}} & 0\\
0 & 0 & 1
\end{pmatrix}\,,
\end{equation}
with
\begin{equation}
\phi=\arctan\left(\frac{\epsilon\sqrt{1-\delta^{2}}}{1-\delta\epsilon}\right)\,,
\end{equation}
diagonalizes the kinetic matrix and change the mass matrix into the
form
\begin{equation}
K^{\prime T}M_{St}^{2}K^{\prime}=\begin{pmatrix}\frac{v^{2}g_{Y}^{\prime2}(\delta-\epsilon)^{2}+4M_{1}^{2}\left(1-2\delta\epsilon+\epsilon^{2}\right)}{4\left(1-\delta^{2}\right)} & \frac{v^{2}g_{Y}^{\prime2}(\epsilon-\delta)}{4\sqrt{1-\delta^{2}}} & \frac{v^{2}\mathrm{g_{2}}g_{Y}^{\prime}(\delta-\epsilon)}{4\sqrt{\left(1-\delta^{2}\right)}}\\
\frac{v^{2}g_{Y}^{\prime2}(\epsilon-\delta)}{4\sqrt{1-\delta^{2}}} & \frac{v^{2}g_{Y}^{\prime2}}{4} & -\frac{v^{2}\mathrm{g_{2}}g_{Y}^{\prime}}{4}\\
\frac{v^{2}\mathrm{g_{2}}g_{Y}^{\prime}(\delta-\epsilon)}{4\sqrt{\left(1-\delta^{2}\right)}} & -\frac{v^{2}\mathrm{g_{2}}g_{Y}^{\prime}}{4} & \frac{v^{2}\mathrm{g_{2}^{2}}}{4}
\end{pmatrix}\,,
\end{equation}
where we set $g_{Y}^{\prime}=g_{Y}/\sqrt{1-2\delta\epsilon+\epsilon^{2}}$.
Now the lower right $2\times2$ matrix is of the same form with the
SM $\{B,A_{3}\}$ mass mixing matrix. The rotation matrix
\begin{equation}
R_{1}^{\prime}=\begin{pmatrix}1 & 0 & 0\\
0 & \cos\theta & -\sin\theta\\
0 & \sin\theta & \cos\theta
\end{pmatrix}=\begin{pmatrix}1 & 0 & 0\\
0 & \frac{g_{2}}{\sqrt{g_{2}^{2}+g_{Y}^{\prime2}}} & \frac{-g_{Y}^{\prime}}{\sqrt{g_{2}^{2}+g_{Y}^{\prime2}}}\\
0 & \frac{g_{Y}^{\prime}}{\sqrt{g_{2}^{2}+g_{Y}^{\prime2}}} & \frac{g_{2}}{\sqrt{g_{2}^{2}+g_{Y}^{\prime2}}}
\end{pmatrix}\,,
\end{equation}
will transform the mass matrix into
\begin{equation}
\left(K^{\prime}R_{1}^{\prime}\right)^{T}M_{St}^{2}K^{\prime}R_{1}^{\prime}=\begin{pmatrix}\frac{v^{2}g_{Y}^{\prime2}(\delta-\epsilon)^{2}+4M_{1}^{2}\left(1-2\delta\epsilon+\epsilon^{2}\right)}{4\left(1-\delta^{2}\right)} & 0 & \frac{1}{4}v^{2}g_{Y}^{\prime}\left(\delta-\epsilon\right)\sqrt{\frac{\mathrm{g}_{2}^{2}+g_{Y}^{\prime2}}{1-\delta^{2}}}\\
0 & 0 & 0\\
\frac{1}{4}v^{2}g_{Y}^{\prime}\left(\delta-\epsilon\right)\sqrt{\frac{\mathrm{g}_{2}^{2}+g_{Y}^{\prime2}}{1-\delta^{2}}} & 0 & \frac{1}{4}v^{2}\left(\mathrm{g}_{2}^{2}+g_{Y}^{\prime2}\right)
\end{pmatrix}\,.
\end{equation}
A second rotation matrix
\begin{equation}
R_{2}^{\prime}=\begin{pmatrix}\cos\psi & 0 & \sin\psi\\
0 & 1 & 0\\
-\sin\psi & 0 & \cos\psi
\end{pmatrix}
\end{equation}
will fully diagonalize the above matrix with
\begin{equation}
\psi=\frac{1}{2}\arctan\frac{2v^{2}g_{Y}^{\prime}\left(\delta-\epsilon\right)\sqrt{\left(1-\delta^{2}\right)\left(\mathrm{g}_{2}^{2}+g_{Y}^{\prime2}\right)}}{-4M_{1}^{2}\left(1-2\delta\epsilon+\epsilon^{2}\right)+v^{2}\left[\mathrm{g}_{2}^{2}\left(1-\delta^{2}\right)+g_{Y}^{\prime2}\left(1-2\delta^{2}+2\delta\epsilon-\epsilon^{2}\right)\right]}\,.
\end{equation}
The full rotation matrix with mass mixing is then given by
\begin{align}
\mathcal{R}  =KO^{\prime}R_{1}^{\prime}R_{2}^{\prime}&=\begin{pmatrix}\frac{1}{\sqrt{1-\delta^{2}}} & 0 & 0\\
-\frac{\delta}{\sqrt{1-\delta^{2}}} & 1 & 0\\
0 & 0 & 1
\end{pmatrix}\begin{pmatrix}\cos\phi & -\sin\phi & 0\\
\sin\phi & \cos\phi & 0\\
0 & 0 & 1
\end{pmatrix}\begin{pmatrix}1 & 0 & 0\\
0 & \cos\theta & -\sin\theta\\
0 & \sin\theta & \cos\theta
\end{pmatrix}\begin{pmatrix}\cos\psi & 0 & \sin\psi\\
0 & 1 & 0\\
-\sin\psi & 0 & \cos\psi
\end{pmatrix}\nonumber \\
& =c\begin{pmatrix}\frac{\left(1-\delta\epsilon\right)\cos\psi}{\sqrt{1-\delta^{2}}}-\frac{g_{Y}^{\prime}\epsilon\sin\psi}{\sqrt{\mathrm{g}_{2}^{2}+g_{Y}^{\prime2}}} & -\frac{g_{2}\epsilon}{\sqrt{\mathrm{g}_{2}^{2}+g_{Y}^{\prime2}}} & \frac{g_{Y}^{\prime}\epsilon\cos\psi}{\sqrt{\mathrm{g}_{2}^{2}+g_{Y}^{\prime2}}}+\frac{\left(1-\delta\epsilon\right)\sin\psi}{\sqrt{1-\delta^{2}}}\\
\frac{(\epsilon-\delta)\cos\psi}{\sqrt{1-\delta^{2}}}+\frac{g_{Y}^{\prime}\sin\psi}{\sqrt{\mathrm{g}_{2}^{2}+g_{Y}^{\prime2}}} & \frac{g_{2}}{\sqrt{\mathrm{g}_{2}^{2}+g_{Y}^{\prime2}}} & -\frac{g_{Y}^{\prime}\cos\psi}{\sqrt{\mathrm{g}_{2}^{2}+g_{Y}^{\prime2}}}+\frac{\left(\epsilon-\delta\right)\sin\psi}{\sqrt{1-\delta^{2}}}\\
\frac{-g_{2}\sin\psi}{c\sqrt{\mathrm{g}_{2}^{2}+g_{Y}^{\prime2}}} & \frac{g_{Y}^{\prime}}{c\sqrt{\mathrm{g}_{2}^{2}+g_{Y}^{\prime2}}} & \frac{g_{2}\cos\psi}{c\sqrt{\mathrm{g}_{2}^{2}+g_{Y}^{\prime2}}}
\end{pmatrix}\,, \label{eq:RSt}
\end{align}
where the constant $c=1/\sqrt{1-2\delta\epsilon+\epsilon^{2}}\approx1$.

\section{Plasmon Effect}\label{sec:APPPlasmon}
\setcounter{equation}{0}
\renewcommand{\theequation}{B.\arabic{equation}}
An electromagnetic wave propagating through the plasma consists of
coherent vibrations and encompasses both transverse and longitudinal
components and travels at speeds slower than that of light. The quantization
of electromagnetic waves within a plasma yields a spin-1 particle
featuring two transverse and one longitudinal spin polarizations.
It is customary to refer to all three polarization states as \textquotedblleft plasmons\textquotedblright{}
in the literature, underscoring dispersion relations depend on the
properties of the plasma. Thus in the early universe, the photon acquires
an effective mass $m_{T}^{2}(k,T)=\omega_{T}^{2}(k,T)-k^{2}$, $m_{L}^{2}(k,T)=\omega_{L}^{2}(k,T)-k^{2}$,
for the transverse and longitudinal modes respectively, where $k$
is the momentum of plasmon. The massive photon, which we will refer
to as plasmon in this paper, can decay into light particle and anti-particle
pair carrying electric charges. The dark millicharged particles can
then be also produced by the plasmon $\gamma^{*}$ via the decay $\gamma^{*}\rightarrow\chi\bar{\chi}$.
In this appendix we review briefly the plasmon effect and the effective
plasmon frequency which is used for calculating the decay of the plasmon.
We follow the discussion of~\cite{Braaten:1993jw},
where the decay of plasmons to neutrinos was considered.

The typical electron velocity is defined by
\begin{equation}
v_{\star}=\omega_{1}/\omega_{p}\,,
\end{equation}
where the first mode frequency and the plasma frequency are given
by phase space integrals

\begin{align}
\omega_{1}^{2}(T) & =\frac{4\alpha}{\pi}\int_{0}^{\infty}\mathrm{d}p\frac{p^{2}}{E}\left(\frac{5}{3}v^{2}-v^{4}\right)\big[n_{e}(E,T)+\bar{n}_{e}(E,T)\big]\,,\\
\omega_{p}^{2}(T) & =\frac{4\alpha}{\pi}\int_{0}^{\infty}\mathrm{d}p\frac{p^{2}}{E}\left(1-\frac{v^{2}}{3}\right)\big[n_{e}(E,T)+\bar{n}_{e}(E,T)\big]\,,
\end{align}
where $n_{e},\bar{n}_{e}$ are respectively the number densities of
electron and positron with the Fermi distribution
\begin{gather}
n_{e}=\frac{1}{{\rm e}^{(E-\mu)/T}+1}\,,\qquad\bar{n}_{e}=\frac{1}{{\rm e}^{(E+\mu)/T}+1}\,.
\end{gather}

The electromagnetic polarization tensor of plasmon interacting with
electrons and positrons to the leading order is given by the thermal
integral
\begin{align}
\Pi^{\mu\nu}(K,T) & =16\pi\alpha\int\frac{{\rm d}^{3}p}{(2\pi)^{3}}\frac{1}{2E}\left(n_{e}(E,T)+\bar{n}_{e}(E,T)\right)\nonumber \\
 & \times\frac{P\cdot K\left(P^{\mu}K^{\nu}+K^{\mu}P^{\nu}\right)-K^{2}P^{\mu}P^{\nu}-(P\cdot K)^{2}g^{\mu\nu}}{(P\cdot K)^{2}-\left(K^{2}\right)^{2}/4}\,,\label{P}
\end{align}
where $P=(E,\vec{p})$ is the momentum of electrons and positrons
and $K=(\omega,\vec{k})$ is the plasmon momentum with $K^{2}=\omega^{2}-k^{2}$.
The transverse and longitudinal polarization functions are
\begin{align}
\Pi_{T}(\omega,k) & =\frac{1}{2}\left(\delta^{ij}-\hat{k}^{i}\hat{k}^{j}\right)\Pi^{ij}(\omega,\vec{k})\,,\\
\Pi_{L}(\omega,k) & =\Pi^{00}(\omega,\vec{k})\,,
\end{align}
where $\hat{k}$ is the normalized momentum vector. The effective
propagator can then be constructed in the Coulomb gauge ($\partial^{i}A_{i}=0$),
and the nonzero components are
\begin{align}
D_{T}^{ij}(\omega,\vec{k}) & =\frac{1}{\omega^{2}-k^{2}-\Pi_{T}(\omega,k)}\left(\delta^{ij}-\hat{k}^{i}\hat{k}^{j}\right)\,,\label{PT}\\
D_{L}(\omega,\vec{k}) & =\frac{1}{k^{2}-\Pi_{L}(\omega,k)}\,, \label{PL}
\end{align}
with the frequencies of the longitudinal and transverse modes defined
by the poles of effective propagators Eqs.~(\ref{PT}) and (\ref{PL})
\begin{align}
\omega_{T}(k)^{2} & =k^{2}+\Pi_{T}\left(\omega_{T}(k),k\right)\,,\\
\omega_{L}(k)^{2} & =\frac{\omega_{L}(k)^{2}}{k^{2}}\Pi_{L}\left(\omega_{L}(k),k\right)\,.
\end{align}
The dressed polarization vectors are given by
\begin{equation}
\tilde{\epsilon}_{L}^{\mu}(k)=\frac{\omega_{L}(k)}{k}\sqrt{Z_{L}(k)}\big(1,\vec{0}\big)^{\mu}\,,\qquad\tilde{\epsilon_{\pm}}^{\mu}(k)=\sqrt{Z_{T}(k)}\big(0,\vec{\epsilon}_{\pm}\big)^{\mu}\,,
\end{equation}
where the residue functions are
\begin{align}
Z_{L}(k) & =\left.\frac{k^{2}}{\omega^{2}}\left[-\frac{\partial\Pi_{L}}{\partial\omega^{2}}\left(\omega,k\right)\right]^{-1}\right|_{\omega\to\omega_{L}}\,,\\
Z_{T}(k) & =\left.\left[1-\frac{\partial\Pi_{T}}{\partial\omega^{2}}\left(\omega,k\right)\right]^{-1}\right|_{\omega\to\omega_{T}}\,.
\end{align}

The general dispersion relations valid at all temperatures and densities
up to first order in the electromagnetic fine structure constant $\alpha$
are given by \cite{Braaten:1993jw,Dvorkin:2019zdi},
\begin{align}
\omega_{T}^{2} & =k^{2}+\omega_{p}^{2}\frac{3\omega_{T}^{2}}{2v_{\star}^{2}k^{2}}\Big[1-\frac{\omega_{T}^{2}-v_{\star}^{2}k^{2}}{\omega_{T}^{2}}\frac{\omega_{T}}{2v_{\star}k}\ln\big(\frac{\omega_{T}+v_{\star}k}{\omega_{T}-v_{\star}k}\big)\Big]\,,\qquad0\leq k<\infty\\
\omega_{L}^{2} & =\omega_{p}^{2}\frac{3\omega_{L}^{2}}{v_{\star}^{2}k^{2}}\Big[\frac{\omega_{L}}{2v_{\star}k}\ln\big(\frac{\omega_{L}+v_{\star}k}{\omega_{L}-v_{\star}k}\big)-1\Big]\,,\qquad0\leq k<k_{{\rm max}}
\end{align}
where $k_{\mathrm{max}}$ is the maximum momentum of which the longitudinal
photon lies inside the light cone
\begin{align}
k_{{\rm max}} & =\frac{4\alpha}{\pi}\int_{0}^{\infty}{\rm d}p\frac{p^{2}}{E}\Big[\frac{1}{v}\ln\big(\frac{1+v}{1-v}\big)-1\Big][n_{e}(E)+\bar{n}_{e}(E)]\nonumber \\
 & =\omega_{p}\sqrt{\frac{3}{v_{\star}^{2}}\Big[\frac{1}{2v_{\star}}\ln\big(\frac{1+v_{\star}}{1-v_{\star}}\big)-1}\Big]\,,
\end{align}
where $v=p/E$ is the velocity of electrons and positrons.

The coupling to the electromagnetic current then gets renormalized
with
\begin{equation}
\tilde{\epsilon}_{T}^{\mu}=\sqrt{Z_{T}}\epsilon_{T}^{\mu}\,,\qquad\tilde{\epsilon}_{L}^{\mu}=\sqrt{Z_{L}}\epsilon_{L}^{\mu}\,,
\end{equation}
where
\begin{align}
Z_{T}(k) & =\frac{2\omega_{T}^{2}(\omega_{T}^{2}-v_{\star}^{2}k^{2})}{3\omega_{p}^{2}\omega_{T}^{2}+(\omega_{T}^{2}+k^{2})(\omega_{T}^{2}-v_{\star}^{2}k^{2})-2\omega_{T}^{2}(\omega_{T}^{2}-k^{2})}\,,\label{eq:ZTk}\\
Z_{L}(k) & =\frac{2(\omega_{L}^{2}-v_{\star}^{2}k^{2})}{3\omega_{p}^{2}-(\omega_{L}^{2}-v_{\star}^{2}k^{2})}\,.\label{eq:ZLk}
\end{align}

In this paper we will mostly deal with the \textbf{relativistic limit},
where we have $T\gg m_{e}$, $v_{\star}=1$, $k_{{\rm max}}\to\infty$.In
this case the dispersion relations reduce to
\begin{align}
\omega_{T}^{2} & =k^{2}+\omega_{p}^{2}\frac{3\omega_{T}^{2}}{2k^{2}}\Big[1-\frac{\omega_{T}^{2}-k^{2}}{\omega_{T}^{2}}\frac{\omega_{T}}{2k}\ln\big(\frac{\omega_{T}+k}{\omega_{T}-k}\big)\Big]\,,\qquad0\leq k<\infty\\
\omega_{L}^{2} & =\omega_{p}^{2}\frac{3\omega_{L}^{2}}{k^{2}}\Big[\frac{\omega_{L}}{2k}\ln\big(\frac{\omega_{L}+k}{\omega_{L}-k}\big)-1\Big]\,,\qquad0\leq k<\infty
\end{align}
with the plasma frequency simplified to be
\begin{equation}
\omega_{p,{\rm rel}}^{2}=\frac{4\alpha}{3\pi}\big(\mu^{2}+\frac{\pi^{2}T^{2}}{3}\big)\,.\label{eq:OmepRel}
\end{equation}
The renormalized factors $Z_{T}$ and $Z_{L}$, given by Eqs. (\ref{eq:ZTk})
and (\ref{eq:ZLk}), are also reduced with $v_{\star}=1$. In the
relativistic limit when $T$ is large, the number of net electrons
is much smaller than the total number of electron and positron pairs
and thus the chemical potential $\mu$ can be neglected. Hence, from
Eq. (\ref{eq:OmepRel}) we see a relation $\omega_{p,{\rm rel}}\sim0.1T$
in the relativistic limit. Recall the photon effective mass $m^{2}(k,T)=\omega^{2}(k,T)-k^{2}$,
the millicharge dark matter with mass $m$ can only be produced by
the plasmon if the temperature is greater than $\sim20\times m$.
In this paper we focus on the dark matter mass ranging from 50 keV
to 1 GeV, corresponding to a temperature $T>1$ MeV, which is safe
to take the relativistic limit (for the dark matter mass near 50 keV,
see discussion in~\cite{Dvorkin:2019zdi}, the relativistic limit is also valid).

We are interested in the plasmon decaying to the millicharge dark
matter via the channel $\gamma^{*}\rightarrow\chi\bar{\chi}$, and
the thermally averaged decay widths are given by
\begin{align}
\langle\Gamma\rangle_{\gamma_{L}^{*}\rightarrow\chi\bar{\chi}}n_{\gamma_{L}^{*}} & =\frac{Q_{\epsilon}}{(2\pi)^{3}}\int k^{2}{\rm d}k\frac{Z_{L}(k)\omega_{L}(k)\left(m_{L}(k)^{2}+2m_{\chi}^{2}\right)\sqrt{m_{L}(k)^{2}\left(m_{L}(k)^{2}-4m_{\chi}^{2}\right)}}{3m_{L}(k)^{4}\left(e^{\omega_{L}(k)/T}-1\right)}\,,\\
\langle\Gamma\rangle_{\gamma_{T}^{*}\rightarrow\chi\bar{\chi}}n_{\gamma_{T}^{*}} & =\frac{4Q_{\epsilon}}{(2\pi)^{3}}\int k^{2}{\rm d}k\frac{Z_{T}(k)\left(m_{T}(k)^{2}-m_{\chi}^{2}\right)\sqrt{m_{T}(k)^{2}\left(m_{T}(k)^{2}-4m_{\chi}^{2}\right)}}{3\omega_{T}(k)m_{T}(k)^{2}\left(e^{\omega_{T}(k)/T}-1\right)}\,,
\end{align}
in which $Q_{\epsilon}$ is the millicharge given in Eq.~(\ref{eq:millicharge}).

\section{Relevant scattering cross-sections}\label{sec:APPScatter}
\setcounter{equation}{0}
\renewcommand{\theequation}{C.\arabic{equation}}

The cross-sections for all relevant interactions are summarized in this Appendix.

\paragraph{1. $\chi\bar{\chi}\rightarrow\gamma,\gamma^{\prime},Z\rightarrow i\bar{i}$}

\begin{align}
\sigma_{\chi\bar{\chi} \rightarrow i\bar{i}}(s)&=\frac{g_{X}^{2}Q_{\chi}^{2}}{48\pi}\sqrt{\frac{s-4m_{i}^{2}}{s-4m_{\chi}^{2}}}\left(1+\frac{2m_{\chi}^{2}}{s}\right)\times \nonumber\\
 & \bigg\{\frac{\mathcal{R}_{13}^{2}\left[a_{i}^{2}\left(s-4m_{i}^{2}\right)+v_{i}^{2}\left(2m_{i}^{2}+s\right)\right]}{\left(s-M_{Z}^{2}\right)^{2}+M_{Z}^{2}\Gamma_{Z}^{2}}
 +\frac{\mathcal{R}_{11}^{2}\left[a_{i}^{\prime2}\left(s-4m_{i}^{2}\right)+v_{i}^{\prime2}\left(2m_{i}^{2}+s\right)\right]}{\left(s-M_{\gamma^{\prime}}^{2}\right)^{2}+M_{\gamma^{\prime}}^{2}\Gamma_{\gamma^{\prime}}^{2}}\nonumber\\
 & +\frac{2\mathcal{R}_{11}\mathcal{R}_{13}\left[a_{i}a_{i}^{\prime}\left(s-4m_{i}^{2}\right)+v_{i}v_{i}^{\prime}\left(2m_{i}^{2}+s\right)\right]}{\left[\left(s-M_{Z}^{2}\right)^{2}+M_{Z}^{2}\Gamma_{Z}^{2}\right]\left[\left(s-M_{\gamma^{\prime}}^{2}\right)^{2}+M_{\gamma^{\prime}}^{2}\Gamma_{\gamma^{\prime}}^{2}\right]}G \nonumber\\
 & +\frac{4g_{2}s_{\theta}Q_{i}v_{i}\mathcal{R}_{13}\mathcal{R}_{12}\left(2m_{i}^{2}+s\right)\left(s-M_{Z}^{2}\right)}{s\left[\left(s-M_{Z}^{2}\right)^{2}+M_{Z}^{2}\Gamma_{Z}^{2}\right]} \nonumber\\
 & +\frac{4g_{2}s_{\theta}Q_{i}v_{i}^{\prime}\mathcal{R}_{12}\mathcal{R}_{11}\left(2m_{i}^{2}+s\right)\left(s-M_{\gamma^{\prime}}^{2}\right)}{s\left[\left(s-M_{\gamma^{\prime}}^{2}\right)^{2}+M_{\gamma^{\prime}}^{2}\Gamma_{\gamma^{\prime}}^{2}\right]}\nonumber\\
 & +\frac{4(g_{2}s_{\theta}Q_{i}\mathcal{R}_{12})^{2}\left(2m_{i}^{2}+s\right)}{s^{2}}\bigg\}\,,
\end{align}
with
\begin{equation}
G=\left[\Gamma_{Z}M_{Z}\Gamma_{\gamma^{\prime}}M_{\gamma^{\prime}}+\left(s-M_{Z}^{2}\right)\left(s-M_{\gamma^{\prime}}^{2}\right)\right]\,.
\end{equation}

\paragraph{2. $\chi\bar{\chi}\rightarrow\gamma\gamma$}

\begin{equation}
\sigma_{\chi\bar{\chi}\rightarrow\gamma\gamma}(s) =\frac{g_{X}^{4}Q_{\chi}^{4}\mathcal{R}_{12}^{4}}{8\pi s\left(s-4m_{\chi}^{2}\right)}\left\{ -\frac{\sqrt{s-4m_{\chi}^{2}}}{\sqrt{s}}\left(s+4m_{\chi}^{2}\right)+\frac{\log A}{s}\left(s^{2}+4m_{\chi}^{2}s-8m_{\chi}^{4}\right)\right\} \,,
\end{equation}
with
\begin{equation}
A=\frac{\sqrt{s}+\sqrt{s-4m_{\chi}^{2}}}{\sqrt{s}-\sqrt{s-4m_{\chi}^{2}}}\,.
\end{equation}

\paragraph{3. $\gamma^{\prime}\rightarrow i\bar{i}$}

\begin{equation}
\Gamma_{\gamma^{\prime}\rightarrow i\bar{i}}=\frac{M_{\gamma^{\prime}}}{48\pi}\sqrt{1-\frac{4m_{i}^{2}}{M_{\gamma^{\prime}}^{2}}}\left[v_{i}^{\prime2}\left(1+\frac{2m_{i}^{2}}{M_{\gamma^{\prime}}^{2}}\right)+a_{i}^{\prime2}\left(1-\frac{4m_{i}^{2}}{M_{\gamma^{\prime}}^{2}}\right)\right]\,.
\end{equation}

\paragraph{4. $i\bar{i}\rightarrow\gamma^{\prime} $}

\begin{align}
\sigma_{i\bar{i}\rightarrow\gamma^{\prime}}(s)&=\sigma^0_{i\bar{i}\rightarrow\gamma^{\prime}}(s)\delta(s-M_{\gamma^{\prime}}^{2})\nonumber\\
&=\frac{\pi M_{\gamma^{\prime}}^{2}}{4s}\sqrt{\frac{s}{s-4m_{i}^{2}}}\left[v_{i}^{\prime2}\left(1+\frac{2m_{i}^{2}}{M_{\gamma^{\prime}}^{2}}\right)+a_{i}^{\prime2}\left(1-\frac{4m_{i}^{2}}{M_{\gamma^{\prime}}^{2}}\right)\right]\delta(s-M_{\gamma^{\prime}}^{2})\,.
\end{align}

\paragraph*{5. $i\bar{i}\to\gamma\gamma^{\prime}$, $i\gamma\to i\gamma^{\prime}$}~{}

\noindent
The squared scattering matrices summed over initial and final spins are
given by
	\begin{align}
		\sum_{\mathrm{spins}}\left|M\right|_{i\bar{i}\to\gamma\gamma^{\prime}}^{2}= & 2e^{2}Q_{i}^{2}\left\{ \mathcal{K}_{1}+\mathcal{K}_{2}\frac{1}{t_{i}u_{i}}+\mathcal{K}_{3}\left(\frac{1}{t_{i}^{2}}+\frac{1}{u_{i}^{2}}\right)+\mathcal{K}_{4}\left(\frac{1}{t_{i}}+\frac{1}{u_{i}}\right)+\mathcal{K}_{5}\left(\frac{u_{i}}{t_{i}}+\frac{t_{i}}{u_{i}}\right)\right\} \,,\protect\\
		\sum_{\mathrm{spins}}\left|M\right|_{i\gamma\to i\gamma^{\prime}}^{2}= & -2e^{2}Q_{i}^{2}\left\{ \mathcal{K}_{1}+\mathcal{K}_{2}\frac{1}{s_{i}u_{i}}+\mathcal{K}_{3}\left(\frac{1}{s_{i}^{2}}+\frac{1}{u_{i}^{2}}\right)+\mathcal{K}_{4}\left(\frac{1}{s_{i}}+\frac{1}{u_{i}}\right)+\mathcal{K}_{5}\left(\frac{u_{i}}{s_{i}}+\frac{s_{i}}{u_{i}}\right)\right\} \,,
	\end{align}
where $s_{i}=s-m_{i}^{2},t_{i}=t-m_{i}^{2},u_{i}=u-m_{i}^{2},$ and
\begin{align}
	\mathcal{K}_{1} & =4a_{i}^{\prime2}m_{i}^{2}/M_{\gamma^{\prime}}^{2}\,,\nonumber \\
	\mathcal{K}_{2} & =2\left[a_{i}^{\prime2}\left(8m_{i}^{4}-6m_{i}^{2}M_{\gamma^{\prime}}^{2}+M_{\gamma^{\prime}}^{4}\right)+v_{i}^{\prime2}\left(M_{\gamma^{\prime}}^{4}-4m_{i}^{4}\right)\right]\,,\nonumber \\
	\mathcal{K}_{3} & =-2m_{i}^{2}\left[a_{i}^{\prime2}\left(M_{\gamma^{\prime}}^{2}-4m_{i}^{2}\right)+v_{i}^{\prime2}\left(2m_{i}^{2}+M_{\gamma^{\prime}}^{2}\right)\right]\,,\nonumber \\
	\mathcal{K}_{4} & =-2\left[a_{i}^{\prime2}\left(M_{\gamma^{\prime}}^{2}-4m_{i}^{2}\right)+v_{i}^{\prime2}\left(2m_{i}^{2}+M_{\gamma^{\prime}}^{2}\right)\right]\,,\nonumber \\
	\mathcal{K}_{5} & =a_{i}^{\prime2}\left(2m_{i}^{2}/M_{\gamma^{\prime}}^{2}-1\right)+v_{i}^{\prime2}\,.
\end{align}
The corresponding cross-sections are
\begin{align}
	\sigma_{i\bar{i}\to\gamma\gamma^{\prime}}\left(s\right)= & \frac{e^{2}Q_{i}^{2}\left(s-M_{\gamma^{\prime}}^{2}\right)}{32\pi s\sqrt{s\left(s-4m_{i}^{2}\right)}}\left\{ \left(\mathcal{K}_{1}-2\mathcal{K}_{5}\right)+\mathcal{K}_{3}\frac{2s}{\left(s-M_{\gamma^{\prime}}^{2}\right)^{2}m_{i}^{2}}\right.\nonumber \\
	& \left.\quad\quad\quad\quad+\frac{2\sqrt{s}\left[\mathcal{K}_{2}-\mathcal{K}_{4}\left(s-M_{\gamma^{\prime}}^{2}\right)+\mathcal{K}_{5}\left(s-M_{\gamma^{\prime}}^{2}\right)^{2}\right]}{\left(s-M_{\gamma^{\prime}}^{2}\right)^{2}\sqrt{s\left(s-4m_{i}^{2}\right)}}\log B\right\} \,,\\
	\sigma_{i\gamma\to i\gamma^{\prime}}\left(s\right)= & \frac{-e^{2}Q_{i}^{2}D}{64\pi s\left(s-m_{i}^{2}\right)}\left\{ 2\mathcal{K}_{1}-\frac{C\mathcal{K}_{5}}{s}+\frac{2\mathcal{K}_{4}}{s-m_{i}^{2}}+\frac{2\mathcal{K}_{3}}{\left(s-m_{i}^{2}\right)^{2}}\left(1-\frac{4s^{2}}{C^{2}-D^{2}}\right)\right.\nonumber \\
	& \left.-\frac{2}{D\left(s-m_{i}^{2}\right)^{2}}\left[\mathcal{K}_{2}+\mathcal{K}_{4}\left(s-m_{i}^{2}\right)+\mathcal{K}_{5}\left(s-m_{i}^{2}\right)^{2}\right]\log\left(\frac{C+D}{C-D}\right)\right\} \,,\\
	\sigma_{\gamma\gamma^{\prime}\to i\bar{i}}\left(s\right)= & \frac{4s\left(s-4m_{i}^{2}\right)}{6\left(s-M_{\gamma^{\prime}}^{2}\right)^{2}}\cdot\sigma_{i\bar{i}\to\gamma\gamma^{\prime}}\left(s\right)\,,\\
	\sigma_{i\gamma^{\prime}\to i\gamma}\left(s\right)= & \frac{4\left(s-m_{i}^{2}\right)^{2}}{6D^{2}}\cdot\sigma_{i\gamma\to i\gamma^{\prime}}\left(s\right)\,,
\end{align}
with
\begin{align}
	B= & \left(\sqrt{s}+\sqrt{s-4m_{i}^{2}}\right)/\left(\sqrt{s}-\sqrt{s-4m_{i}^{2}}\right)\,,\nonumber\\
	C= & m_{i}^{2}-M_{\gamma^{\prime}}^{2}+s\,,\nonumber \\
	D= & \sqrt{s^{2}-2\left(m_{i}^{2}+M_{\gamma^{\prime}}^{2}\right)s+\left(m_{i}^{2}-M_{\gamma^{\prime}}^{2}\right)^{2}}\,.
\end{align}

\paragraph{6. $\chi\bar{\chi}\rightarrow\gamma^{\prime}\gamma^{\prime}$}

\begin{align}
\sigma_{\chi\bar{\chi}\rightarrow\gamma^{\prime}\gamma^{\prime}}(s) & =\frac{g_{X}^{4}Q_{\chi}^{4}\mathcal{R}_{11}^{4}}{8\pi s\left(s-4m_{\chi}^{2}\right)}\Bigg\{-\frac{\sqrt{\left(s-4M_{\gamma^{\prime}}^{2}\right)\left(s-4m_{\chi}^{2}\right)}}{M_{\gamma^{\prime}}^{4}+m_{\chi}^{2}\left(s-4M_{\gamma^{\prime}}^{2}\right)}\left[2M_{\gamma^{\prime}}^{4}+m_{\chi}^{2}\left(s+4m_{\chi}^{2}\right)\right]\nonumber \\
 & +\frac{\log E}{s-2M_{\gamma^{\prime}}^{2}}\left(s^{2}+4m_{\chi}^{2}s+4M_{\gamma^{\prime}}^{4}-8m_{\chi}^{4}-8m_{\chi}^{2}M_{\gamma^{\prime}}^{2}\right)\Bigg\}\,,
\end{align}
with
\begin{equation}
E=\frac{s-2M_{\gamma^{\prime}}^{2}+\sqrt{\left(s-4M_{\gamma^{\prime}}^{2}\right)\left(s-4m_{\chi}^{2}\right)}}{s-2M_{\gamma^{\prime}}^{2}-\sqrt{\left(s-4M_{\gamma^{\prime}}^{2}\right)\left(s-4m_{\chi}^{2}\right)}}\,.
\end{equation}

\paragraph{7. $\gamma^{\prime}\gamma^{\prime}\rightarrow\chi\bar{\chi}$}

\begin{equation}
9\left(s-4M_{\gamma^{\prime}}^{2}\right)\sigma_{\gamma^{\prime}\gamma^{\prime}\rightarrow\chi\bar{\chi}}(s)=8\left(s-4m_{\chi}^{2}\right)\sigma_{\chi\bar{\chi}\rightarrow\gamma^{\prime}\gamma^{\prime}}(s)\,.
\end{equation}

\paragraph{8. $\gamma^{\prime}\rightarrow\chi\bar{\chi}$}

\begin{equation}
\Gamma_{\gamma^{\prime}\rightarrow\chi\bar{\chi}}=\frac{g_{X}^{2}Q_{\chi}^{2}\mathcal{R}_{11}^{2}M_{\gamma^{\prime}}}{12\pi}
\sqrt{1-\frac{4m_{\chi}^{2}}{M_{\gamma^{\prime}}^{2}}}
\left(1+\frac{2m_{\chi}^{2}}{M_{\gamma^{\prime}}^{2}}\right)\,.
\end{equation}

\paragraph{9. $\chi\bar{\chi}\rightarrow\gamma^{\prime}$}

\begin{align}
\sigma_{\chi\bar{\chi}\rightarrow\gamma^{\prime}}(s) &=\sigma^0_{\chi\bar{\chi}\rightarrow\gamma^{\prime}}(s)\delta(s-M_{\gamma^{\prime}}^{2}) \nonumber \\
&=\frac{\pi g_{X}^{2}Q_{\chi}^{2}\mathcal{R}_{11}^{2}M_{\gamma^{\prime}}^{2}}{s}\sqrt{\frac{s}{s-4m_{\chi}^{2}}}\left(1+\frac{2m_{\chi}^{2}}{M_{\gamma^{\prime}}^{2}}\right)\delta(s-M_{\gamma^{\prime}}^{2})\,.
\end{align}

\end{document}